\documentclass[aps,pra, twocolumn, superscriptaddress]{revtex4-2}

\usepackage{gensymb}
\usepackage{graphicx}
\usepackage{amsmath}
\usepackage{textcomp}
\usepackage{subcaption}
\usepackage{caption}
\usepackage{nicefrac}
\usepackage{upgreek}
\usepackage{ctable}

\newcommand{\Yb}{\textsuperscript{171}Yb\textsuperscript{+}}
\newcommand{\Ybnext}{\textsuperscript{174}Yb\textsuperscript{+}}
\newcommand{\ket}[1]{\ensuremath{|{#1}\rangle}}

\newcommand{\state}[3]{\textsuperscript{#1}#2\textsubscript{#3}}
\newcommand{\fuzz}{FuzzButtons\textsuperscript{\textregistered}}
\newcommand{\um}[1]{{#1}~\ensuremath{\mu }m}

\newcolumntype{C}[1]{>{\centering\arraybackslash}p{#1}} 
\newcolumntype{R}[1]{>{\raggedleft\arraybackslash}p{#1}}                                                                  
\newcolumntype{L}[1]{>{\raggedright\arraybackslash}p{#1}}    
\begin{document}

\title{Engineering the Quantum Scientific Computing Open User Testbed (QSCOUT): Design details and user guide}

\author{Susan M. Clark}
\email[]{sclark@sandia.gov}
\homepage[]{https://qscout.sandia.gov}
\author{Daniel Lobser}
\author{Melissa Revelle}
\author{Christopher G. Yale}
\author{David Bossert}
\author{Ashlyn D. Burch}
\affiliation{Sandia National Laboratories, Albuquerque, New Mexico 87123, USA}
\author{Matthew N. Chow}
\affiliation{Sandia National Laboratories, Albuquerque, New Mexico 87123, USA}
\affiliation{Department of Physics and Astronomy, University of New Mexico, Albuquerque, New Mexico 87131, USA}
\affiliation{Center for Quantum Information and Control, CQuIC, University of New Mexico, Albuquerque, New Mexico 87131, USA}
\author{Craig W. Hogle}
\author{Megan Ivory}
\affiliation{Sandia National Laboratories, Albuquerque, New Mexico 87123, USA}
\author{Jessica Pehr}
\affiliation{Sandia National Laboratories, Albuquerque, New Mexico 87123, USA}
\affiliation{Currently at: IonQ, College Park, Maryland 20740, USA}
\author{Bradley Salzbrenner}
\author{Daniel Stick}
\author{William Sweatt}
\author{Joshua M. Wilson}
\author{Edward Winrow}
\affiliation{Sandia National Laboratories, Albuquerque, New Mexico 87123, USA}
\author{Peter Maunz}
\affiliation{Sandia National Laboratories, Albuquerque, New Mexico 87123, USA}
\affiliation{Currently at: IonQ, College Park, Maryland 20740, USA}

\begin{abstract}
The Quantum Scientific Computing Open User Testbed (QSCOUT) at Sandia National Laboratories is a trapped-ion qubit system designed to evaluate the potential of near-term quantum hardware in scientific computing applications for the US Department of Energy (DOE) and its Advanced Scientific Computing Research (ASCR) program.  Similar to commercially available platforms, most of which are based on superconducting qubits, it offers quantum hardware that researchers can use to perform quantum algorithms, investigate noise properties unique to quantum systems, and test novel ideas that will be useful for larger and more powerful systems in the future.  However, unlike most other quantum computing testbeds, QSCOUT uses trapped \Yb~ions as the qubits, provides full connectivity between qubits, and allows both quantum circuit and low-level pulse control access to study new modes of programming and optimization. The purpose of this manuscript is to provide users and the general community with details of the QSCOUT hardware and its interface, enabling them to take maximum advantage of its capabilities.
\end{abstract}

\maketitle

\section{Overview  and System Design }\label{overview}
Experimental quantum computing is beginning to realize its potential for computational speed-up through recent demonstrations of specialized algorithms with as few as 50 qubits~\cite{Arute2019,Zhong2020}. These machines are considered noisy intermediate-scale quantum (NISQ) computers~\cite{Preskill2018}, and though their ability to solve relevant, real-world problems is limited by their size and fidelity, they are very useful for investigating the best way to build and operate a future fault-tolerant quantum computer.  While access to NISQ hardware has become available over the last few years~\cite{IBM, Rigetti, IonQ}, it is still scarce and highly constrained to a small set of fixed physical operations.  We are developing the QSCOUT platform to reduce these two barriers in order to accelerate progress by the scientific community.  

QSCOUT consists of a small number of Ytterbium (\Yb)~ion qubits.  The first round of user experiments has three ions, with plans to expand to a linear chain of 32 ions in later rounds.  Trapped ions have long coherence times~\cite{Fisk1997}, high gate fidelities~\cite{Ballance2016}, and very low state preparation and measurement (SPAM) errors~\cite{Noek2013}, making them an ideal system to build a quantum processor.  The targets for QSCOUT are to provide up to 32 qubits with $>$~99.5~\% fidelity single-qubit gates, $>$~98~\% fidelity two-qubit gates, and $<$~0.5~\% crosstalk.  Despite being arranged in a linear chain, the qubits are fully connected through the vibrational modes of the chain, meaning a two-qubit gate can occur between any pair of ions in the chain without extra gates or swapping~\cite{Landsman2019}.  Single-qubit and two-qubit gates are performed using a pulsed laser to excite Raman transitions between the hyperfine states of the qubit.  In order to individually address the ions, one path of the pulsed-laser beam is split into 32 beams, with each beam passing through a distinct channel of a multichannel acoustic-optic modulator (AOM) before being focused onto a single ion.  This arrangement allows for independent frequency, amplitude, and phase control of the laser pulses applied to each ion.  

To trap a linear chain of ions, we are using an HOA-2.1 surface trap fabricated at Sandia National Labs~\cite{Maunz2016}, because it is capable of precisely controlling ion spacing and manipulating chains of ions.  To maintain a linear chain for enough time to perform many quantum operations, the trap must be under ultra-high vacuum to limit collisions that ``melt'' the chain or eject ions from the trap.  The vacuum chamber must have windows for optical access as well as electrical feedthroughs for rf and dc voltages to reach the trap.  The details of the vacuum chamber are described in Sec.~\ref{vacuum}. Section~\ref{cwlasers}  discusses the specific laser frequencies and locks needed for trapping, cooling, and detecting \Yb~ions, as well as our optomechanical system for minimizing vibrations and drift that degrade the quality of the delivered light.  Quantum gates are performed using a 355 nm pulse train~\cite{Islam2014}, which has tight spatial tolerances to achieve individual addressing of closely spaced ions.  Our technique for pulse train spacing compensation and optical delivery is outlined in Sec.~\ref{355laser}.  Distinguishable detection of ions is achieved by imaging light from each ion into a separate core of a multimode fiber array, with each core coupled to it own photomultiplier tube (PMTs) for counting photons.  Section~\ref{sec:imaging} describes this detection method in more detail.  We also developed new electrical hardware to control the timing, frequency, and amplitude of the rf pulses needed to modulate the optical signals that are delivered to the ions.  This device and corresponding firmware are described in Sec.~\ref{ControlHW}.  The high level programming language for users to interface with our hardware, Jaqal, has been described in~\cite{Landahl2020}, and design decisions in~\cite{Morrison2020}.  Finally, in Sec.~\ref{Results}, we describe the resulting ion performance achieved in preparation for the first round of user experiments. 

\section{Ultra-High Vacuum Chamber }\label{vacuum}
Because QSCOUT relies on the realization of multi-ion chains, the background gas pressure in the system is of paramount importance. While an individual ion lifetime in a surface-electrode trap is typically hours-long (even a few days), as we scale up to larger chains, that lifetime reduces as more ions are subject to background gas collisions~\cite{Pagano2018}. Since most collisions at room temperature are of sufficient energy to destabilize the ion chain, limiting collisions results in fewer ion losses, fewer quantum algoritum restarts, and faster processing time. To this end, we develop an ultra-high vacuum (UHV) system intended to realize long-lived ion chains incorporating vacuum practices from other fields to achieve lower background pressure.

UHV has been a standard of vacuum technology for decades. It has relied on the use of stainless steel components, copper gasketing, limiting organics to low-outgassing plastics, welding and brazing of components such as viewports, and utilizing ion pumps to maintain the necessary vacuum pressure. Additionally, a vacuum bake is a standard procedure to remove water vapor and other outgassing residues from the internal environment.

 \begin{figure*}[ht]
 \includegraphics[width=0.65\textwidth]{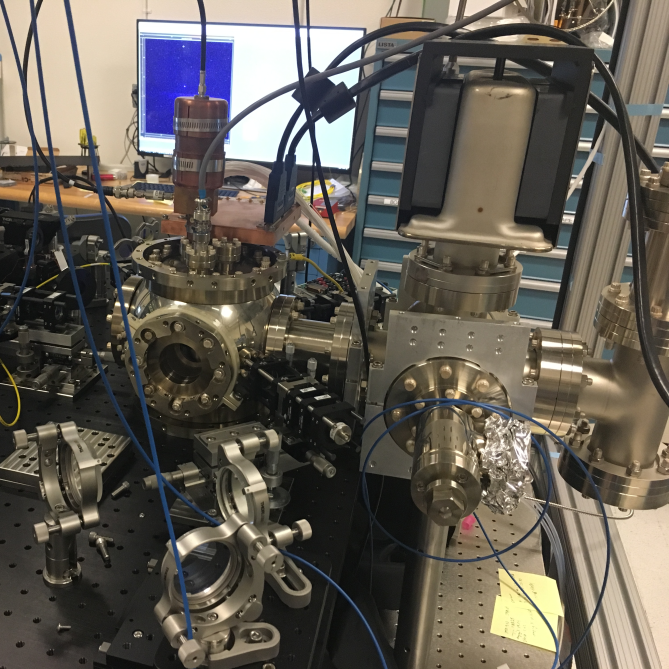}
 \caption{Ultra-high vacuum chamber for trapped ion quantum operations. The chamber consists of an experimental region (left) and a pumping region (right). The chamber has rf and dc feedthroughs that support trap potential generation as well as viewports for laser addressing of the ions. \label{chamber}}
 \end{figure*}

In trapped-ion systems, the vacuum chamber must support an ion trap, an ion source, and have viewports for laser access. In our chamber, the central experimental region is a Kimball Physics 6" CF Spherical Square (Fig.~\ref{chamber}). A feedthrough flange containing electrical and radio-frequency (rf) connections is attached at the top of the chamber. A stainless steel platform anchored to the feedthrough is suspended inside the chamber.  The platform and its various components are all machined from Grade 316L stainless steel and electropolished. The microfabricated surface-electrode ion trap is attached to this platform, facing downward to make it less likely dust particles will attach to the surface (Fig.~\ref{internals}a). The trap uses an rf signal to generate the trapping pseudopotential, and a number of dc electrodes shape and confine that potential to a particular region of the trap~\cite{House2008, Maunz2016, Revelle2020}. A thermal oven is used to create a neutral Yb flux, which is then photoionized to generate ions for trapping (more details in Sec.~\ref{leveldiagram}).  Imaging is performed from the bottom of the chamber and utlizes a re-entrant viewport to allow for the small working distance of the imaging lens assembly (more details in Sec.~\ref{sec:imaging}).  The spherical square has an octagonal structure, and viewports are affixed to seven of the eight sides to optically access the trap. One of the viewports which allows for optical access perpendicular to the trap axis is also a re-entrant viewport, visible in Fig.~\ref{chamber}. This is needed for individual ion addressing using lasers, described in Sec. \ref{355laser}. The transition to the pumping region of the chamber consists of a stainless steel vacuum cross and tee. This supports a 50 L/s ion pump (Agilent Varian VacIon Plus 55 StarCell), an ion gauge (Agilent Varian UHV-24P Nude Bayard-Alpert Ion Gauge with dual Ir-Th filament), a titanium sublimation pump (TSP), and all-metal bakeable valve. To prevent titanium film from sputtering on the trap surface, the TSP is positioned out of direct line-of-sight from the trap. For convenience, we constructed two chambers so that the backup, already baked with a trap, can be swapped as needed.

\subsection{Requirements}
The dc and rf wiring that convey the voltages for trap generation traditionally use organic materials in the form of insulation, multi-pin connectors, and circuit boards. The materials include polyether ether ketone (PEEK), polyimide films such as Kapton\textsuperscript{\textregistered} (DuPont), and RO4350B\texttrademark (Rogers Corporation). While these materials are typically classified as low-outgassing materials (total mass loss $<$ 1\%)~\cite{OutgassingNASA}, they can still contribute to the background gas pressure present in the chamber. In addition, the bulk of the chamber is manufactured from Grade 304 and 316 stainless steel; however, stainless steel is susceptible to hydrogen diffusion and adsorption~\cite{Zajec2001} which also contributes hydrogen to the overall background pressure. In the QSCOUT chambers, we focused on eliminating these organics and reducing hydrogen within the stainless steel in an effort to improve our vacuum pressure.

\subsection{Hydrogen Mitigation }\label{subsec:Hmit}
In an effort to decrease the amount of hydrogen within the stainless steel components, we subject all of the purely stainless steel components to a high-temperature bake process. This includes the trap platform, the experimental chamber, and all of the nipples, tees, crosses, adapters, and blanks. Prior to baking, the parts are cleaned with solvents and degreasers that include 3M\texttrademark~Novec\texttrademark~72DE, acetone, and isopropanol. Initially, they are baked in a dry H$_2$  atmosphere at 1000\degree C for 30-45 minutes to remove any organic material from the stainless steel. Afterwards, the parts are baked in a vacuum environment for 4 hours at 800\degree C to remove hydrogen adsorbed and diffused into the steel~\cite{Sasaki1991}. 

The stainless steel elements of other components, such as viewports and feedthrough flanges, were subjected to a vacuum bake at 800\degree C for 10 hours by the manufacturer (MPF Inc.) prior to assembly with the feedthrough components or windows. Other elements including the ion pump, TSP, ion gauge, and bakeable valve were unable to be subjected to any aggressive bake process due to inherent elements that might be detrimentally affected (copper, filaments of tungsten, thoriated iridium, or titanium). 

After assembling the chamber, we bake it without the trap for several weeks at 200\degree C. After the bake, we slowly vent (over the course of several hours) to minimize particle disturbance in the trap.  Within 48 hours of completing the bake, we install the trap into the chamber in a cleanroom environment. Then a final bake is performed for 5-7 days at 200\degree C. The second bake is limited to a shorter time to prevent the accumulation of Al-Au intermetallics at the site of our wirebonds (i.e. purple plague)~\cite{Maunz2016, Chen1967, Selikson1964}. As the final step, we fire the TSP every 32 hours until the pressure as measured on the ion gauge does not improve at all with further firings of the TSP, usually after 5-7 days.

\subsection{Elimination of Organics }\label{elimorg}
For every element in the system in which the typical component contains organic material, we developed a suitable replacement using ceramic materials, such as the machinable ceramic MACOR\textsuperscript{\textregistered} (Corning), aluminum nitride (AlN), and aluminum oxide (Al$_2$O$_3$). Due to the brittleness of ceramic materials, additional design modifications were implemented to eliminate potential stresses on these ceramics. Each of these elements is described below.

\subsubsection{DC Delivery}

\begin{figure*}[htp]
\includegraphics[width=0.55\textwidth]{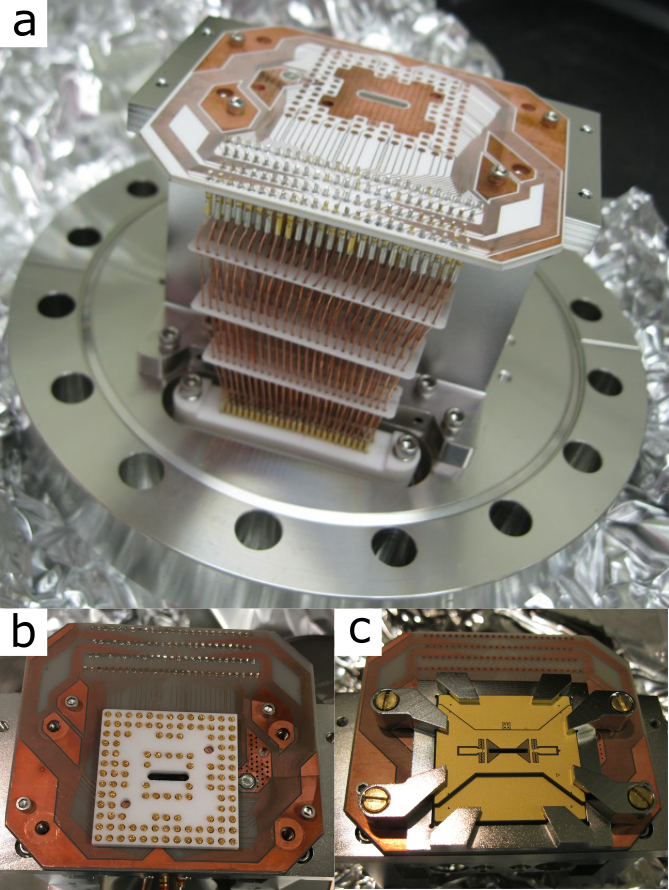}
\caption{a) Trap platform attached to the feedthrough flange. The trap platform, on which the trap will be mounted, is machined from stainless steel. The dc signals are routed via bare copper wires from a Micro-D feedthrough to an Al$_2$O$_3$ ceramic signal routing board. These wires are separated with Al$_2$O$_3$ spacers to prevent shorting. b) The signal routing board is a ceramic (AlN pictured here) circuit board with bare copper traces. A MACOR spacer is placed on top of the board with a series of \fuzz{}~placed inside to provide electrical contact with the trap package. c) The trap is secured to the board and ceramic spacer with a stainless steel clamp that preserves optical access to the ions. \label{internals}}
\end{figure*}

To bring the 100 dc signals to the trap control electrodes, we use a 100 pin Micro-D vacuum feedthrough. The standard commercial connector for this feedthrough on the vacuum side consists of a PEEK connector with 100 Kapton coated wires. Instead, we use a MACOR version of the connector (Wincheseter Interconnect) and a series of bare oxygen-free high-conductivity (OFHC) copper wires (AWG28) soldered into the receptacle pins of the connector. These wires are $\approx2"$ long and connect to the trap platform for distribution to the trap package as shown in Fig.~\ref{internals}a. The pitch of Micro-D pins is 0.05" with a row spacing of 0.043", each row offset by 0.025". We fan out the wires to a pitch of 0.07" with 0.16" row spacing and 0.035" row offset. In addition, to prevent shorting, we use three 0.05" thick Al$_2$O$_3$ spacers (laser machined with holes and grooves) to support and aid the separation of wires. The spacers remain solidly in place due to the tension of the wires. At the top of the trap platform sits a ceramic circuit board for signal delivery. Due to the high thermal conductivity of ceramic, the bare wires cannot be soldered directly to the board. Instead, BeCu pins are soldered directly to the board via a solder reflow oven process, using SAC305 (96.5\% Sn, 3\% Ag, 0.5\% Cu) solder. The BeCu pins are \#23 crimp contact pins (Glenair), without any markings, and are typically used in circular MIL spec connectors. The pins are small and thin enough to provide enough thermal isolation that the wires can be directly hand-soldered to the pins with SAC305 solder. All water-soluble flux from the solder is cleaned off through water baths, and then the entire assembly is cleaned with a series of ultrasonic baths with solvents and degreasers, discussed in Sec.~\ref{subsec:Hmit}.

\subsubsection{Signal Delivery Circuit Board}
The circuit board, manufactured by Millennium Circuits, consists of a 0.059" thick subtrate of an aluminum-based ceramic. The first chamber built was outfitted with an AlN circuit board (Fig.~\ref{internals}b), while the second has a Al$_2$O$_3$ board (Fig.~\ref{internals}a). The variation between the two chambers was part of an effort to reduce the board's thermal conductivity and allow for direct soldering of the wires to the board, but in the end, both board materials require the intermediate step of BeCu pins soldered via reflow. These substrates contain bare copper traces on the top for signal routing and a copper ground plane on the bottom, with the copper thickness corresponding to 1 oz./1~ft$^2$. Because the board is intended for a UHV environment, it contains no soldermask, no silkscreen, nor any other finish. The traces on the top of board are 0.005" thick and route the signal from the 100 dc wires to pads on the center of the board. These pads correspond to pads on the bottom of the trap package. In addition, the board contains routing for the trap rf. A thick AWG11 OFHC copper wire is soldered to the board (also via a solder reflow oven process) to carry the $\approx50$ MHz signal. The rf delivery is discussed in Sec.~\ref{rfdelivery}. The board also contains a series of holes for mounting to the trap platform inside the chamber, which are all copper plated to improve ground connection. The chamber itself acts as ground for both the rf and dc signals. 

\subsubsection{Trap Mounting}
Our trap package consists of a microfabricated surface-electrode trap attached to a ceramic package. For future iterations, we plan to use a trap that is attached via a solder process described in \cite{Revelle2020} as part of the effort to remove all organics. However, the trap currently in the system is attached to its package with an epoxy. 

Our package for these traps previously consisted of a 104-pin grid array (PGA), which contained Kovar pins that are then inserted into a zero-insertion-force (ZIF) socket, which is typically manufactured from PEEK. While the ZIF socket could potentially be manufactured out of a machinable ceramic such as MACOR, there was concern that the insertion and removal of traps may be hampered by brittleness of the ceramic with potential damage to the trap and the socket.  As such, we replaced the PGA package with a land grid array (LGA) package consisting of gold-coated pads on the backside of the package. 

The trap is then placed on a machined MACOR spacer. The 0.1" thick spacer sits on the circuit board and consists of an array of holes matching up with the array of pads, as well as additional holes for ground connections in the center of the package. Gold-plated beryllium copper 0.12" long \fuzz{}~(Custom Interconnects) are inserted into each slot in the MACOR spacer (Fig.~\ref{internals}b). The \fuzz{} provide both the path for the signal (or ground) and the elasticity necessary to ensure contact. In addition, the \fuzz{} eliminate the use of the magnetic Kovar pins found on the PGA packages, thus eliminating a potential source of stray magnetic fields near the ion. The trap is placed on top of the MACOR spacer and \fuzz, and a clamp machined from 316L stainless steel pushes down on the trap to ensure contact. The clamp is designed to ensure there is no loss of optical access around the trap. It consists of a series of fingers which clamp onto the package outside of designed laser path (Fig.~\ref{internals}c).

\subsubsection{Trap rf Delivery}\label{rfdelivery}
The rf is delivered to the circuit board via a AWG11 bare OFHC copper wire. This rf wire is connected through a barrel connector to another similarly gauged wire affixed to a feedthrough. The chamber serves as the rf ground. The air-side of the feedthrough also consists of a ground shield around the extruding wire, with an air gap (custom MPF P/N A19619-1). A helical resonator can is attached to the air side~\cite{Siverns2012}.

Given that the qubit splitting in \Yb{} is 12.642~GHz, it is not feasible to drive the qubit transition using an internal microwave antenna, as any signal would be attenuated drastically without the use of an internal coaxial cable. Because of our restriction on the use of organics inside the chamber, we instead use an external microwave horn to drive transitions. 

\subsection{Characterization of Vacuum}\label{Sec:VacuumChar}

\begin{figure*}
	\centering
	\includegraphics[width=\textwidth]{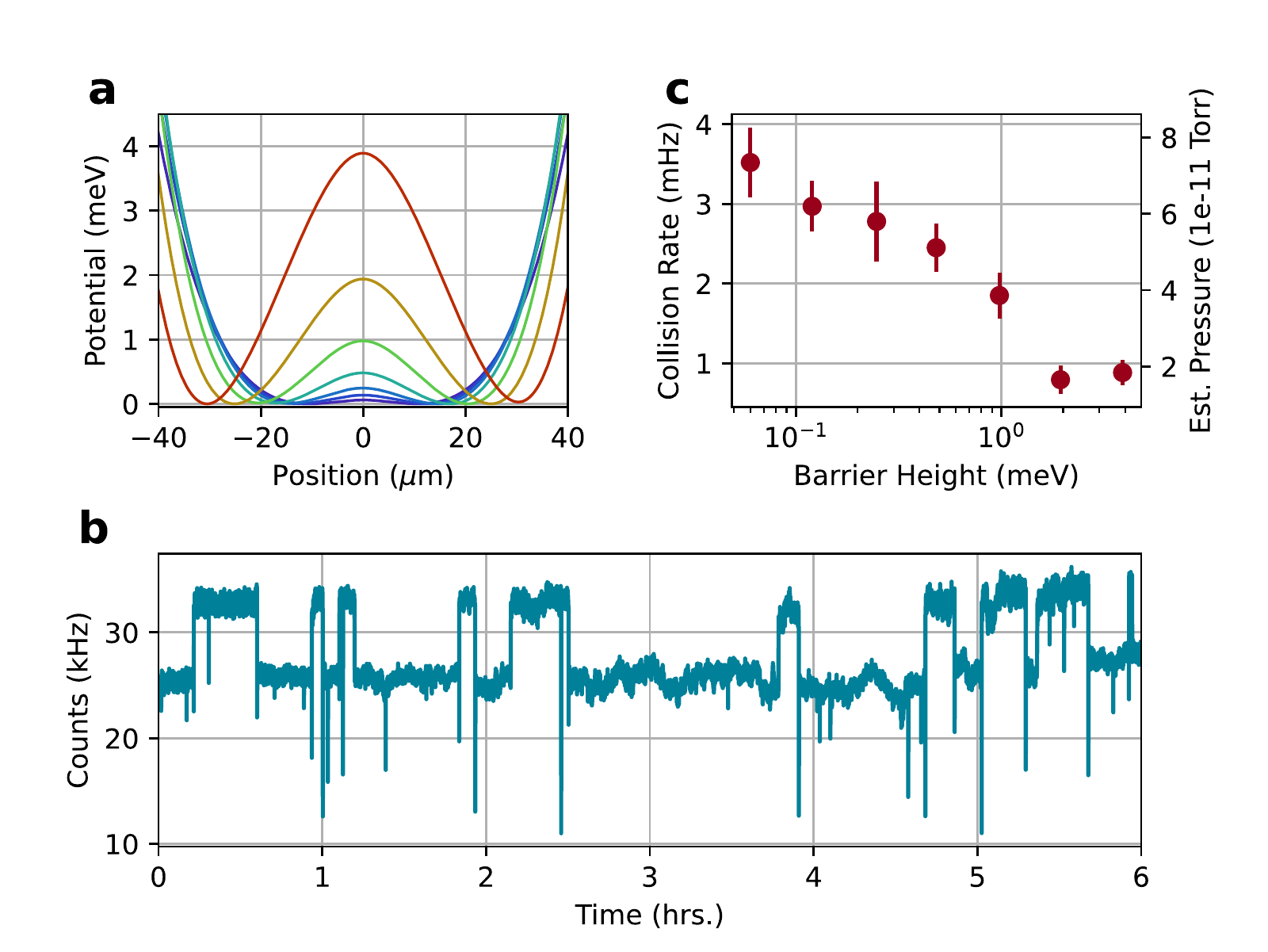}
	\caption{a) A series of slices of the axial potentials used to probe background gas collisions. The central barrier can be increased up to $\approx 4~$meV before reaching voltage limits on the trap dc electrodes (10 V). At each barrier height, we monitor the ion and measure how many times it shifts between the two sites of the potential.  b) An example of the collisions at a 979 $\upmu$eV energy barrier. Each scan was taken for at least 6 hours. One site registers $\approx 32~$kHz in the PMT while the other registers $\approx 25~$kHz, so hops are visible as changes in received photon counts.  c) The collision rate as a function of barrier height. The collision rate (left axis), can be transformed into an estimated pressure (right axis), which assumes all collisions are with a specific species, in this case hydrogen. The error bars reflect standard Poissonian counting error.}
	\label{fig:doublewell}
\end{figure*}

We utilize a variety of measurements to characterize our background pressure. The most rudimentary approach, and least informative, is a measurement of the pressure via the ion gauge, as the reading varies considerably based on the controller used and its particular calibration. The two assembled chambers measured pressures of 6e-12 and 4e-12 Torr via their respective ion gauges, but with the same controller.  We measure pressures using this controller in other chambers used by our group that do not undergo the organic elimination nor the hydrogen mitigation. We find that our QSCOUT chambers outperform the others by factors of 2-5.

Another method we use to examine the background pressure is through background gas-ion collision measurements using a double-well potential~\cite{Aikyo2020}. We generate a series of double-well trapping potentials with a variable height barrier (Fig.~\ref{fig:doublewell}a). Because hydrogen is the dominant residual gas, we begin with a low enough energy barrier to capture a significant portion of all collisions, 60~$\upmu$eV~\cite{Aikyo2020}. The imaging system is intentionally misaligned such that one site appears brighter in the detector than the other. We apply dc voltages to our segmented electrodes to create a shimming field along the trap axis ($x$-axis) that can push the ion from one site to the other. The optimal shim field is set in between the values of the shim fields which cause the ion to occupy each site.

We monitor the number of distinct jumps that occur at several different barrier heights.  At each height, we wait at least six hours or 20 distinct jumps before increasing the barrier. Fig.~\ref{fig:doublewell}b is a typical response over six hour time frame. For each potential, we count the number of jumps a single ion makes between the two sites and estimate the collision rate as twice the observed jump rate to account for collisions in which the ion returned to its original site~\cite{Aikyo2020, Pagano2018}. This collision rate corresponds to all background gas collisions with the ion that have a kinetic energy along the trap axis that is greater than the barrier height. By raising the barrier height, we aim to understand more about the residual gas species remaining in the chamber. Using a Langevin collision model, we estimate partial pressures of these gases~\cite{Pagano2018}.  As we increase the barrier height the number of collisions decreases until it flattens out around 2~meV~( Fig.~\ref{fig:doublewell}c).  This behavior suggests that the collisions are dominated by hydrogen, but there are some residual collisions at higher energies. The pressure suggested by these collisions is suitable for the purposes of this experiment and shows an improvement of a factor of roughly 2 over other chambers in our group.

\section{Continuous-Wave Laser Delivery }\label{cwlasers}

The \Yb{} ion is one of the predominant ion qubits due to its magnetically insensitive hyperfine states for ultra-stable qubit transitions \cite{Fisk1997}, relatively simple hyperfine structure (nuclear magnetic spin $I = \nicefrac{1}{2}$), and optical transitions requiring trapping and cooling lasers in the near-visible UV.  Our scheme for cooling, optical pumping, and detection is very similar to a scheme outlined previously~\cite{Olmschenk2007}.  This section will describe the continuous-wave (cw) lasers needed for trapping and cooling \Yb{}, as well as our method for light delivery into the vacuum chamber.  

\subsection{Level Diagram and Required Lasers }\label{leveldiagram}
We load ions into our trap via a two-photon transition of neutral Yb atoms sublimated from a heated source of solid Yb.  The two-photon transition first requires a resonant 399~nm photon, which (with isotope selectively) drives the atom into an excited state.  If this excitation is followed by another high frequency photon ($<$~394~nm), the second photon can strip off an electron, resulting in \Yb{}~\cite{Balzer2006}.  Once Yb is ionized, the pseudo-potential created by the combination of rf and dc electric fields from the surface trap will localize it above the chip surface.

 \begin{figure*}
 \includegraphics[width=6in]{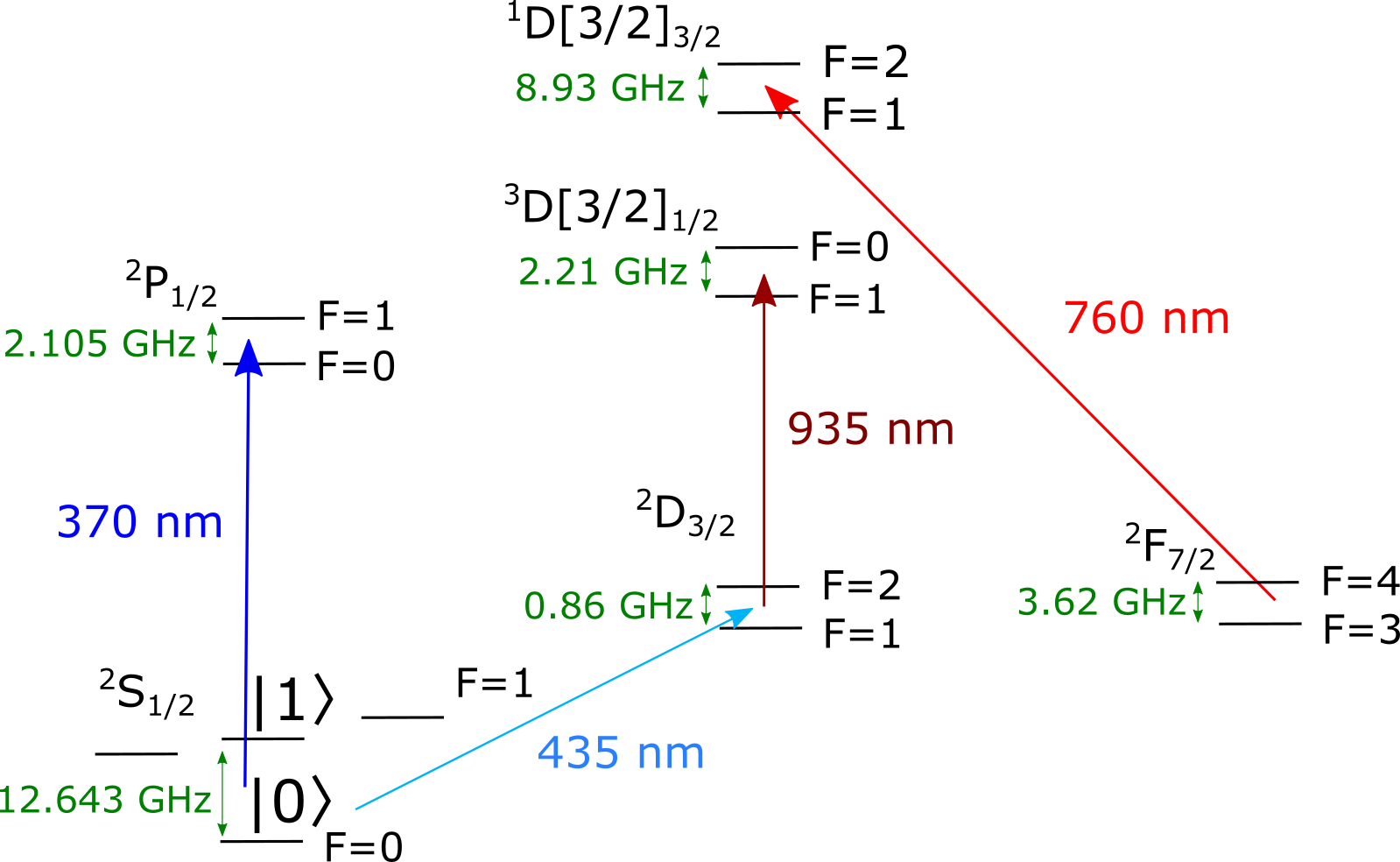}
 \caption{Energy-level diagram of \Yb{} (not to scale) showing lasers used to drive transitions and to keep the ion in the qubit manifold (solid arrows) .  The magnetic sublevels are shown only for the \state{2}{S}{1/2} state and the qubit energy levels (labeled \ket{0} and \ket{1}) are identified. The 370~nm laser is the laser used for Doppler cooling, detection, and state preparation (See Fig.~\ref{cdop}).  The 935~nm laser is used to pump ions out of the \state{2}{D}{3/2} state, while the 760~nm laser can pump out of the \state{2}{F}{7/2} state.  The 435~nm laser can be used for diagnostics, sideband cooling, or perhaps sympathetic cooling. \label{Ybsimple}}
 \end{figure*}

The lasers needed to keep the ion in the qubit subspace, (\state{2}{S}{1/2}~$|$F=0, $m_F =0\rangle$ and $|$F=1, $m_F=0\rangle$), are shown in Fig.~\ref{Ybsimple}~\cite{Olmschenk2007}.  Incoherent state manipulation (cooling, detection, state preparation), uses 370~nm light to excite transitions between the \state{2}{S}{1/2}~F=1 and the \state{2}{P}{1/2} F=0 states (with sidebands as described below for specific hyperfine state manipulation).  The 370~nm light performs Doppler cooling by removing energy through a series of photon absorption and emissions.  Since the pseudo-potential supplies a restoring force in all three-dimensions, only a single laser is needed to cool the ion; however, to do so, the propagation direction (\textbf{k}-vector) must have a projection along each axis of the ion's secular motion (axial and two radial directions).  The polarization needs to be a combination of linear ($\pi$) and both circular ($\sigma^+$, $\sigma^-$) directions in order to excite all the necessary transitions.  Additionally, the light is red-detuned from the resonant transition energy for optimal cooling~\cite{Olmschenk2007}.  

\begin{figure*}
\centering
\includegraphics[width=6in]{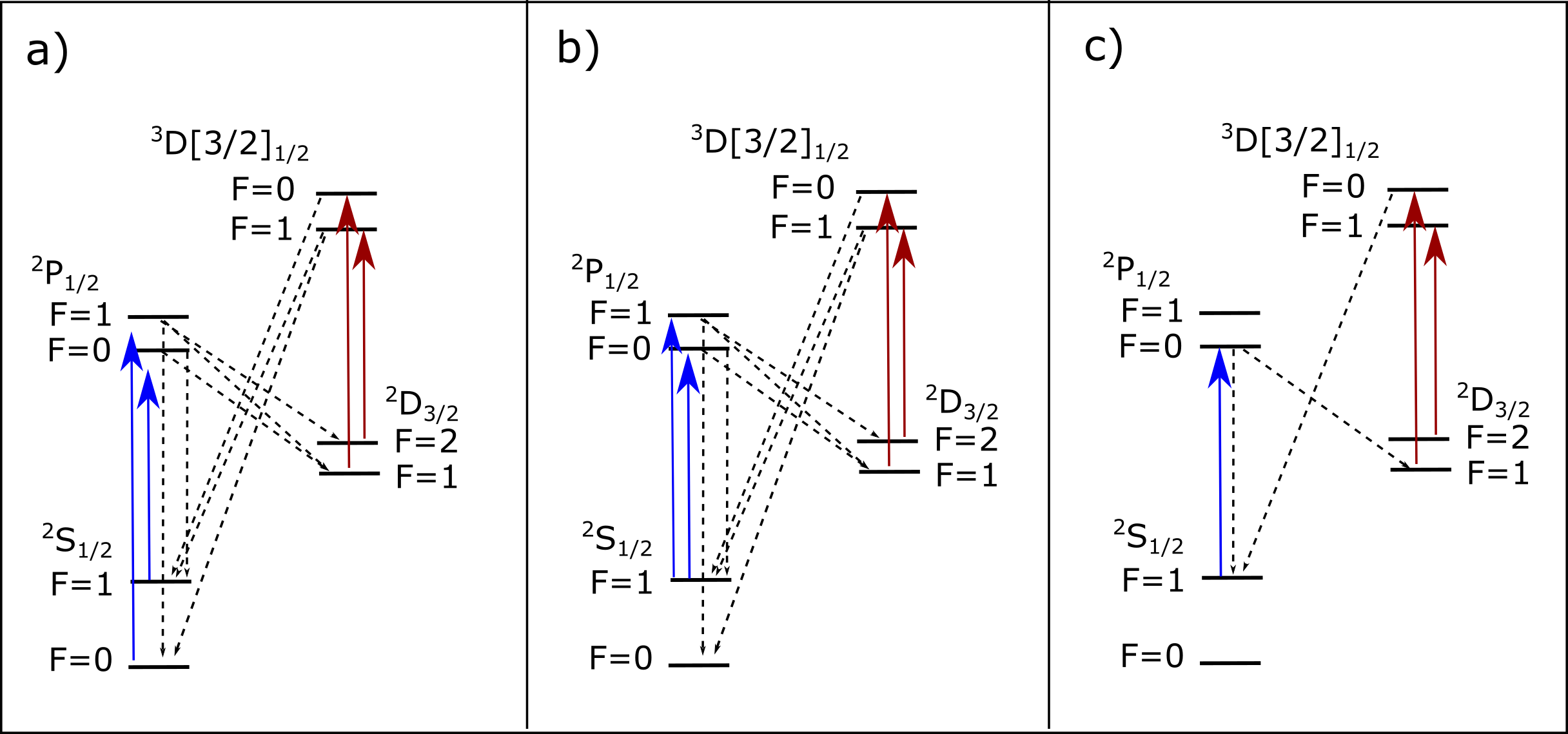}
\caption{Applied 370~nm light and sidebands for cooling, optical pumping, and detection.  a) For cooling, the 370~nm light is detuned -13~MHz from the resonant transition (transition linewidth, $\Gamma$ = 19.7~MHz) and 14.7~GHz sidebands are added to prevent the ion from becoming trapped in the \state{2}{S}{1/2}~F = 0 state.  b) For optical pumping, 2.105~GHz sidebands are applied, resulting in the ion becoming trapped in the  \state{2}{S}{1/2}~F = 0 state after several cycles.  c) For detection, sidebands on the 370~nm light are removed, causing the ion to emit many photons while the light is applied if it is in the F = 1 (bright) state, and none if in the F=0 (dark) state.}
\label{cdop}
\end{figure*}

During Doppler cooling, 14.7~GHz sidebands are added to the 370~nm beam via a Qubig free space electro-optic modulator (EOM) to prevent the ion from becoming trapped in \state{2}{S}{1/2}~F=0 hyperfine state, which can happen via off-resonant coupling to the  \state{2}{P}{1/2} F=1 states (See Fig.~\ref{cdop}a).  On the other hand, to prepare a single ion state, we remove the 14.7~GHz sidebands and add 2.105~GHz sidebands.  These sidebands excite to the \state{2}{P}{1/2}~F=1 state from the \state{2}{S}{1/2}~F=1 states, which can spontaneously emit to the \state{2}{S}{1/2}~F=0 state and become trapped there, as shown in Fig.~\ref{cdop}b.   

Finally, the 370~nm laser is also used for state detection, as shown in Fig.~\ref{cdop}c.  For state detection, in order to keep the ion in a cycling transition between the \state{2}{S}{1/2} F=1 and \state{2}{P}{1/2} F=0 states, no sidebands are used on the 370~nm beam.  Off-resonant coupling of the ion in the \state{2}{S}{1/2} F=1 state to the \state{2}{P}{1/2} F=1 state, which can result in an ion becoming dark if it was bright, is the leading source of detection error and varies depending on optical collection efficiency, detection time, and detection laser power~\cite{Olmschenk2007, Noek2013}.  More rarely, there is off-resonant coupling causing a dark ion to become bright, which also contributes to detection error (see Sec.~\ref{sec:imaging} for a more detailed discussion of detection error). 

The 370 nm light alone is not sufficient to prevent the ion from decaying into non-qubit states.  In particular, from the \state{2}{P}{1/2} levels, the ion relaxes to the low-lying \state{2}{D}{3/2} with a branching ratio of 0.5$\%$~\cite{Olmschenk2007}.  To mitigate leakage into this state, we use a 935 nm laser to resonantly excite the ion to the \state{3}{D[3/2]}{1/2} state, where it then decays back to the \state{2}{S}{1/2} manifold~\cite{Bell1991}.  During optical pumping, it is necessary to add 3.07~GHz sidebands to the 935~nm laser via a fiber EOM to ensure the ion is also pumped out of the F=2 state of the \state{2}{D}{3/2} manifold.  These sidebands are left on during cooling and detection.  

Collisions can cause the ion to relax to the \state{2}{F}{7/2} manifold, which has a lifetime of 3700 days~\cite{Roberts1997}.  A 760~nm laser is applied in order to pump out of the F state by exciting the ion to the \state{1}{D[3/2]}{1/2} state.  From there, it can decay back to the S manifold~\cite{Gerginov2011, Ransford2019}.  This repumping technique has yet to be tested on the QSCOUT system.  

Finally, another laser that is used for characterization of our system is the 435~nm laser.  It is used to perform transitions between the \state{2}{S}{1/2} state and the \state{2}{D}{3/2} state (as shown in Fig.~\ref{Ybsimple}).  This is a useful diagnostic tool, especially for sideband cooling of the axial modes (which are not accessible via the 355~nm pulsed laser (see Sec.~\ref{355laser})) or for potential sympathetic cooling by using co-trapped non-\Yb{} isotopes~\cite{Larson1986}.  The 435~nm laser has not yet been used extensively for QSCOUT, but we have designed to add this capability later.  

In summary, Table~\ref{usedLasers} lists the cw lasers used with \Yb{} and the sidebands applied for various stages  of ion initialization and readout.  The 399~nm, 935~nm, and 760~nm lasers are part of the Toptica MDL-PRO rack mountable system.  Fibers route light from these lasers to the experiment.  

 \begin{table*}
 \caption{Lasers, necessary modulations, and required linewidths for reliable trapping, state preparation, and detection of \Yb{}. \label{usedLasers}}
 \begin{ruledtabular}
 \begin{tabular}{ c c c c c c c }
 Laser & Linewidth & Cool & Optical Pump & Detect & 355 Experiment \\
370 nm & 1 MHz & + 14.7 GHz & + 2.105 GHz & On & Off &\\
935 nm & 5 MHz & + 3.07 GHz & + 3.07 GHz & + 3.07 GHz & + 3.07 GHz &\\
760 nm & 5 MHz & + 5.2 GHz &  + 5.2 GHz & + 5.2 GHz & +5.2 GHz &\\

 \end{tabular}
 \end{ruledtabular}
 \end{table*}

\subsection{The 370 Laser Path}
The 370~nm laser is the most complicated in terms of locking, modulation, and delivery requirements.  The light needs to be referenced to an atomic source and have a linewidth narrow enough ($<$~1~MHz) to efficiently excite the ion.  To meet these requirements, we have a multi-step approach consisting of four main modules: the transfer cavity, the laser breakout board with beat-note lock, modulation board, and beam delivery to the experiment.

\subsubsection{The Transfer Cavity Module}
The transfer cavity lock module consists of three separate locks.  All three objects being locked (two lasers and a transfer cavity) are mounted on a minus-K 100BM-8 benchtop vibration isolation system inside a Herzan acoustic enclosure to protect them from acoustic noise and provide passive temperature stability.  

\begin{figure*}
\includegraphics[width=6 in]{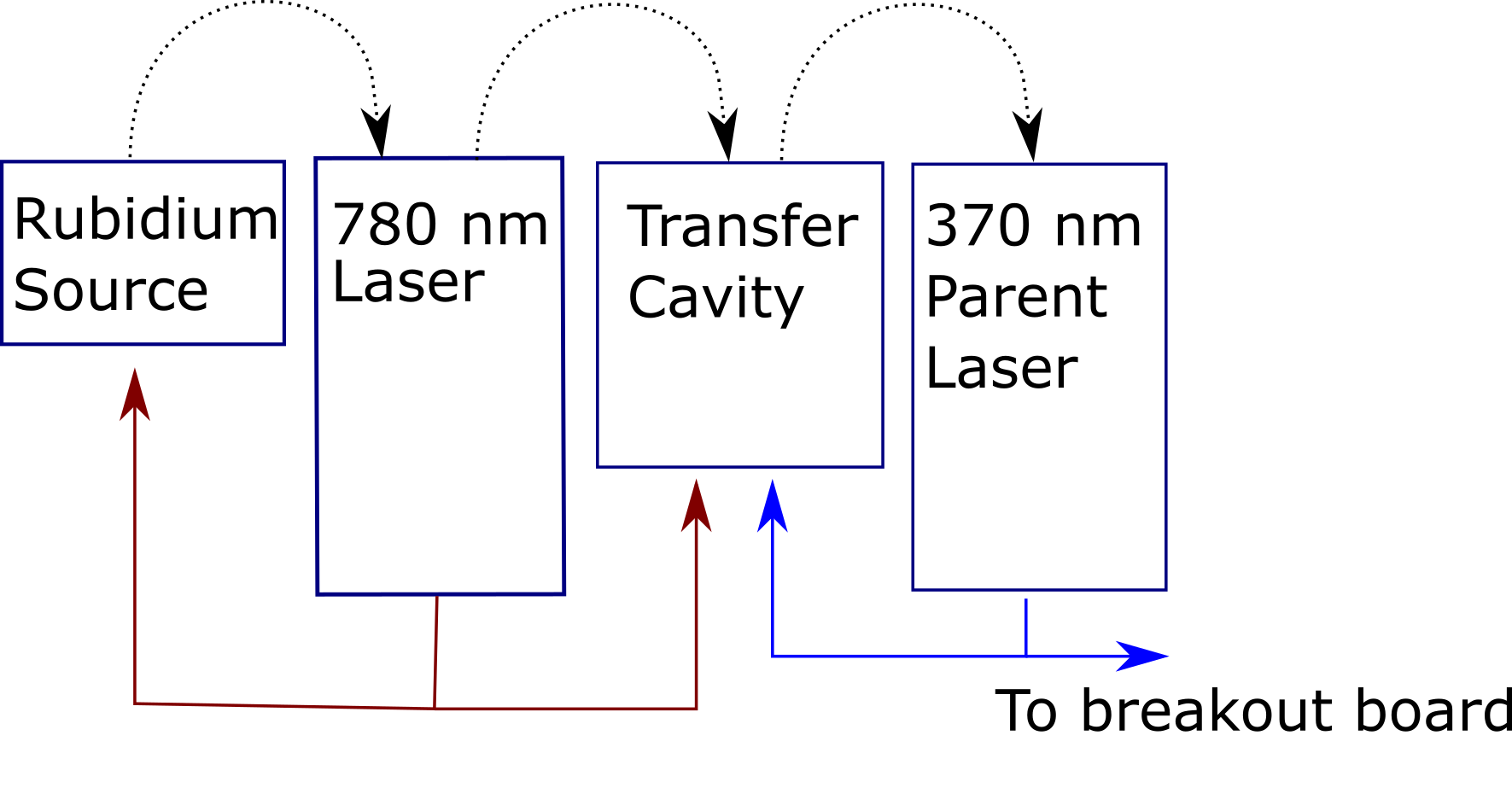}
\caption{Schematic of the transfer cavity module.  Red and Blue (solid) arrows represent optical connections and dotted black lines represent electrical feedback signals.\label{370locking}}
\end{figure*}

A schematic of the transfer cavity locking system is shown in Fig.~\ref{370locking}.  The Toptica DLC Pro 780~nm laser is first locked to a stable Rubidium source.  We use a side of fringe lock on the lower frequency side of the \state{5}{S}{1/2}~\ket{F = 2}~ to the \state{5}{P}{3/2}~$|$F$'$ = 2, $m_{F}$ = 2-3$\rangle$~crossover transition in \textsuperscript{87}Rb,  due to its relative height and peak sharpness.  The Rubidium source is a Toptica CoSy saturation spectroscopy module with a 5~MHz bandwidth.  A Toptica DLC-PRO Lock provides the feedback.  

\begin{figure*}
\includegraphics[width=0.9\textwidth]{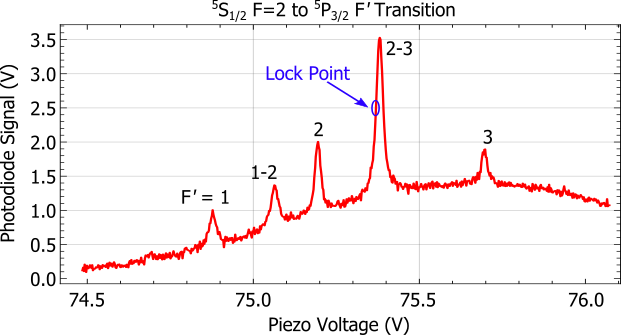}
\caption{Doppler free spectrum of \textsuperscript{87}Rb used to lock the 780 laser \label{Rbspec}}
\end{figure*}

\begin{figure*}
\includegraphics[width=6in]{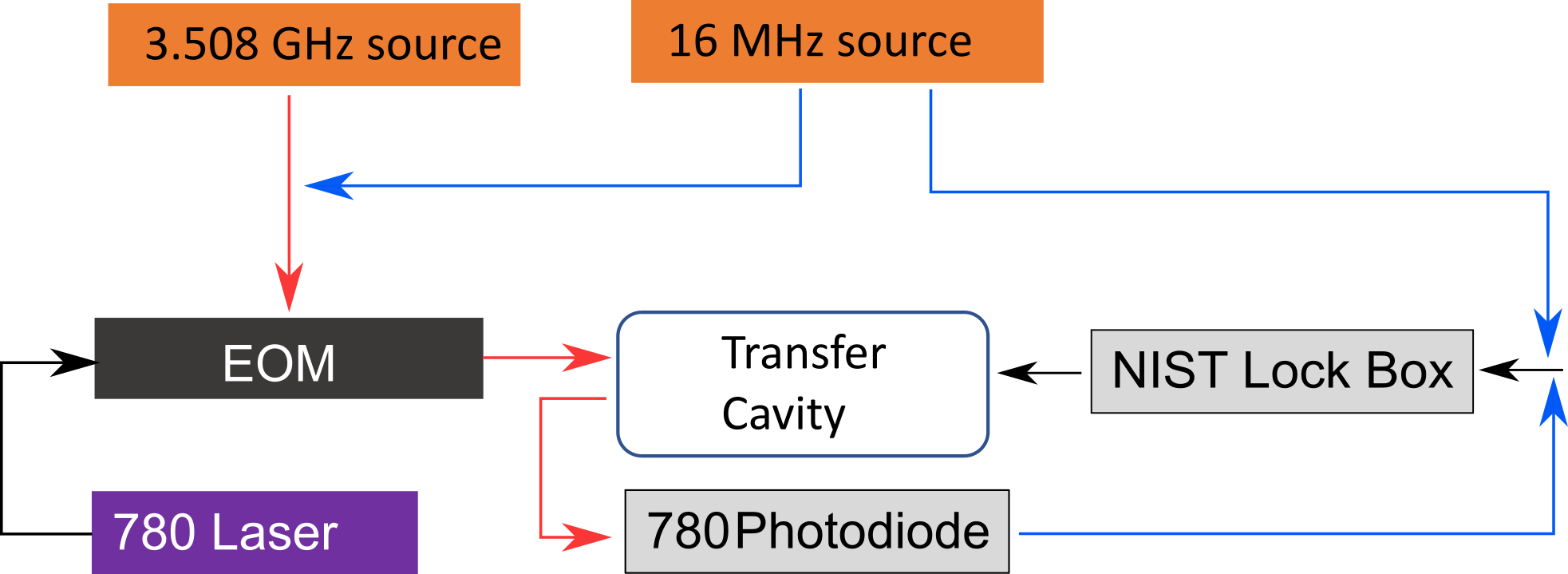}
\caption{Electronic components used to lock the transfer cavity to the stabilized 780 laser.  It is a Pound-Drever-Hall lock~\cite{Drever1983, Black2001}, where 16~MHz sidebands are added and then demodulated from a signal to yield a sharp locking feature.\label{TCelectronics}}
\end{figure*}

The stabilized light from the 780~nm laser is used to lock a SLS-PZT cavity from Stable Laser Systems.  The mirror substrates were coated by FiveNine Optics Inc. to have a finesse of 1000-3000 for 370~nm and 780~nm; the cavity length is scanned with a piezo.  We use a standard Pound-Drever-Hall (PDH) technique~\cite{Drever1983, Black2001} to lock the cavity by adding $\sim$16.0~MHz sidebands (plus a 3.508~GHz offset for the light to match the cavity at the desired frequency) to the 780~nm light using a fiber EOM from EOSpace. The cavity reflection is used as input to our cavity lock.  We use a Superlaserland digital servo, designed by NIST~\cite{Leibrandt2015}, to feedback to the cavity piezo (see Fig.~\ref{TCelectronics}).   

Next, the stablized cavity is used to lock the ``parent'' 370~nm laser (Toptica DL PRO), also using a PDH lock.  19.76~MHz sidebands are added to the light using a free space Qubig EOM.  The cavity reflection is mixed with a stable frequency source equal to the applied sideband frequency, and the resulting signal is used to lock the laser via a Toptica DLC-PRO Lock.  The frequency of the parent laser is chosen to be roughly centered between the \Yb{} and \Ybnext{} Doppler cooling transitions.  Therefore, the same locking electronics can be used to lock 370~nm experiment lasers to either isotope for diagnostics, calibration, or perhaps sympathetic cooling~\cite{Cetina2020,Revelle2019}.  Light from the parent 370~nm laser is split in free space and then coupled into fibers that are sent to various experiments and a Toptica High Finesse wavemeter (HF-ANGS WS8-2+1X8PCS) for monitoring.  

\subsubsection{Laser breakout board with beatnote lock module}
\begin{figure*}[!hbtp]
\centering
\begin{subfigure}{.5\textwidth}
    \includegraphics[width=3.0 in]{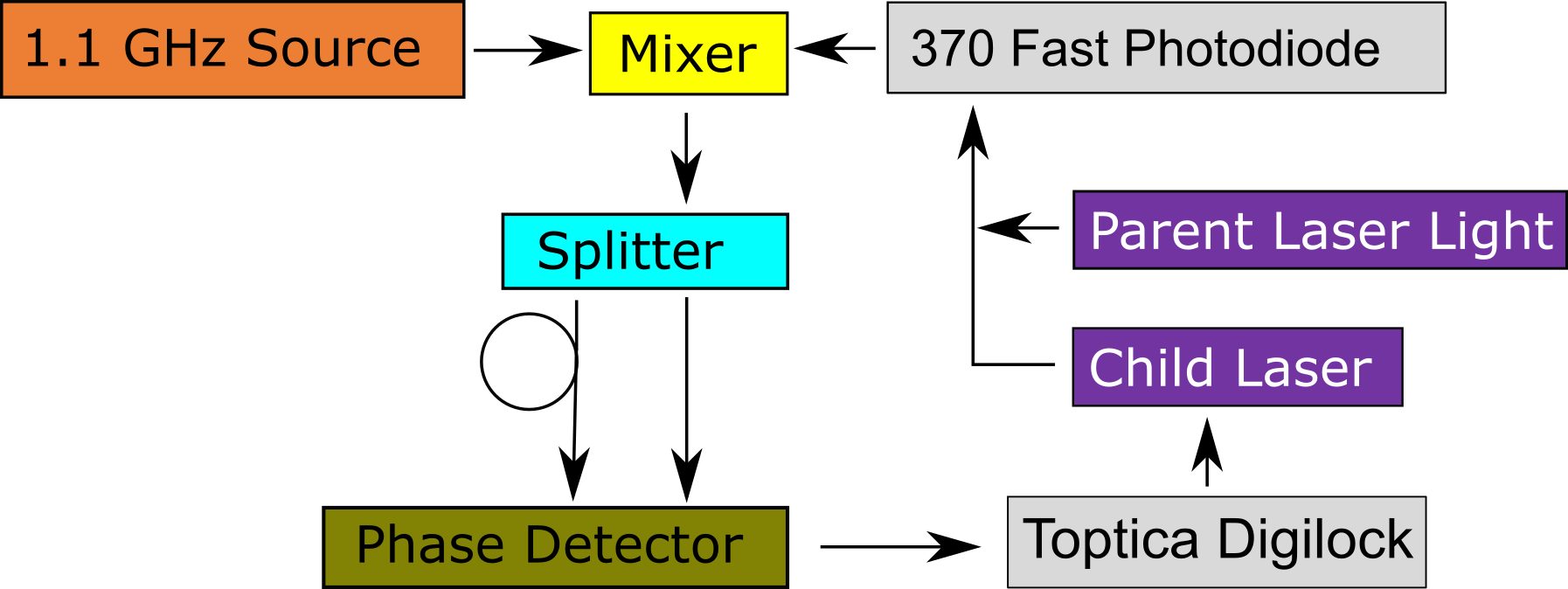}
    \caption{Beatnote lock schematic for stabilizing child laser frequency relative to parent laser (amplifiers and filters not shown).}
\end{subfigure}
\begin{subfigure}{.5\textwidth}
\includegraphics[width=3 in]{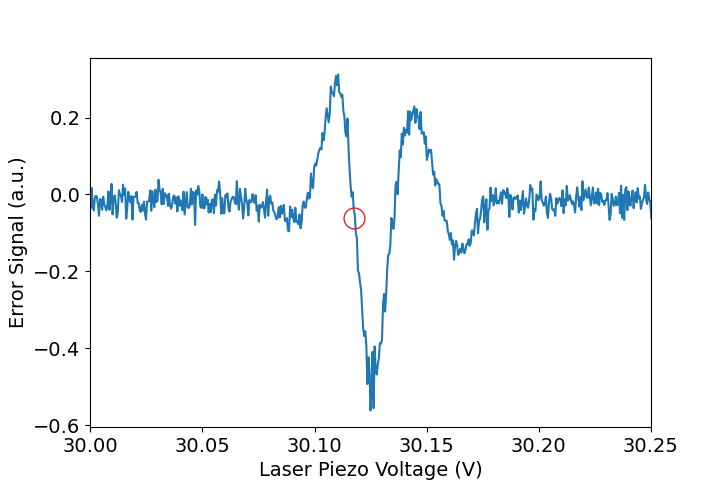}
\caption{Signal used for locking the child 370~nm laser generated from the electronics in part a).  We lock to the downslope of the dip with the highest contrast, as marked with the red circle. \label{beatnote}}
\end{subfigure}
\caption{Locking electronics and signal for 370~nm child laser beatnote lock.} 
\end{figure*}

The final laser is the ``child'' 370~nm laser (Toptica DL-PRO), which is shown on the breakout board in Fig.~\ref{breakout}.  A small portion of the light from this laser ($\sim 50~\mu$W) is split off and sent to a 50/50 beamsplitter, where it is combined with light from the parent 370~nm laser.  The combined light is coupled into a fiber and sent to an AlphaLas Si Photodetector (UPD-50-UP-FR-F).  The beatnote from the light is mixed with a stable 1.125~GHz reference frequency provided by a Vaunix lab brick and filtered to produce a beatnote signal~\cite{Schunemann1999}.  The Toptica DLC-PRO Lock electronics perform a side-of-fringe lock on the beatnote signal to set the frequency (Fig.~\ref{beatnote}).   

The breakout board also sends the light through an IntraAction AOM, which serves as a switch to add extra extinction to the 370~nm light on the ion as needed.  The first-order output from the AOM is coupled into a fiber and sent to a modulation board.  The zeroth-order output is sent to the wavemeter for monitoring.  

\subsubsection{Modulation board}
\begin{figure*}
\includegraphics[width=6 in]{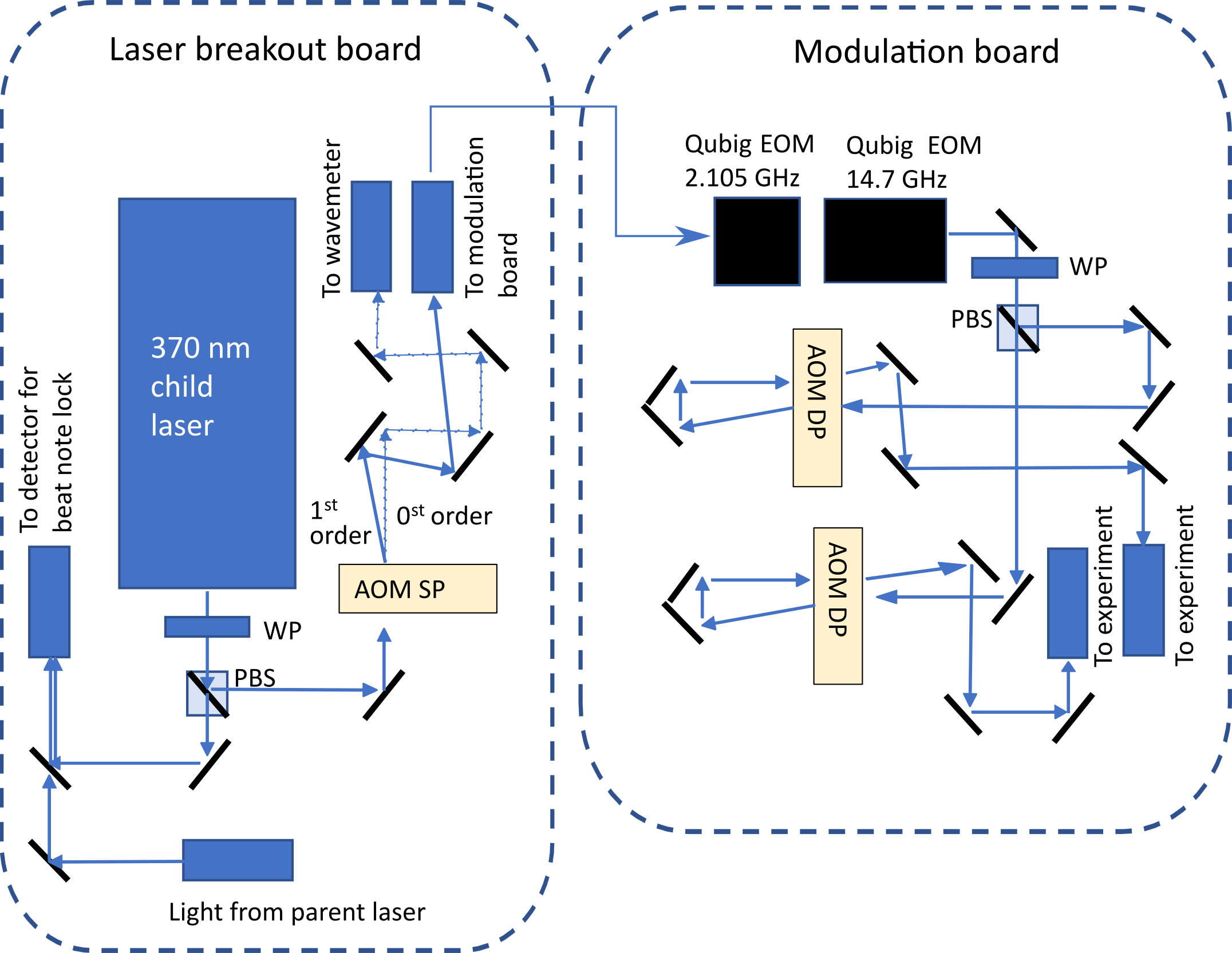}
\caption{Schematic of 370~nm breakout board (left) and modulation board (right).  PBS indicates a polarizing beam splitter, WP is waveplate.  AOM SP indicates an acousto-optic modulator in the single pass configureation, while AOM DP is in the double pass configuration (also apparent from the beam path arrows).   \label{breakout}}
\end{figure*}

At the modulation board, the 14.7~GHz and 2.105~GHz signals are added to the 370~nm light as outlined in Table~\ref{usedLasers}.  As shown in Fig.~\ref{breakout}, the light is first directed to the free space Qubiq EOMs~\cite{Qubig} after which it is split into two paths.  Each path goes through a double-passed AOM for wideband (up to $\approx$ 50 MHz) frequency control and then coupled into a fiber that is sent to the experiment.  There are two 370~nm beam paths available for diagnostics and loading assistance, but ultimately only one path is used for cooling, pumping, and detecting the ions.

\subsubsection{Light Delivery}
\begin{figure*}
\centering
	\includegraphics[width=5 in]{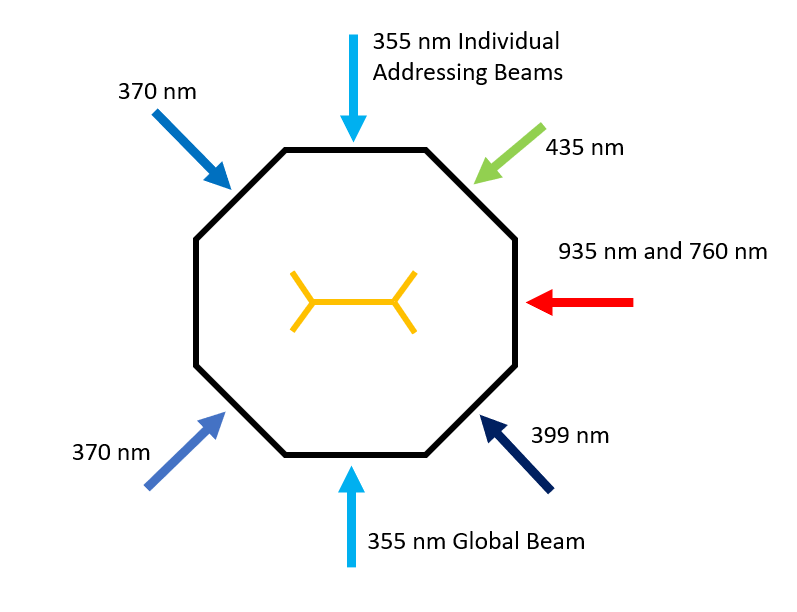}
	\caption{Schematic of chamber as seen from above, showing layout of lasers.  The gold bowtie in the center of the chamber shows the trap orientation.}
	\label{chambersch}
\end{figure*}

Once at the experiment, light exits the fiber and focuses on the ion.  Our current layout has two 370~nm beams each at a 45$^\circ$ angle with the axis of the trap.  The polarization of these beams is set so each path is able to cool all three secular frequency directions (Fig.~\ref{chambersch}).  The 399~nm light also makes a 45$^\circ$ angle with the trap axis, while the 935~nm and the 760~nm light are projected along the axis of the trap.  

We chose lenses to yield a specific spot size at the ion, such that some beams are elliptical while others are round based on their purpose and angle of incidence.  The table below (Table~\ref{tab:CWbeams}) sumarizes the lenses chosen for each beam and the expected beam size at the ion.  For the elliptical beams, the shorter focal length cylindrical lens determines the size of the beam in the direction perpendicular to the trap.  The longer lens is aligned so the ion is several millimeters behind the focus of the beam, and controls the beam size parallel to the trap.  This arrangement results in a beam with minimal scatter on the trap surface, but large enough extent to simultaneously address multiple ions.  

 \begin{table*}
 \caption{Optical components used for cw light delivery to the ion.  Waist is the $1/e^2$ radial waist. \label{tab:CWbeams}}
 \begin{ruledtabular}
 \begin{tabular}{ c c c c c c c }
Beam (nm) &Shape& Collimator & Focusing lens(es) (mm) & Waist at ion ($\mu$m)  \\
370 &Round &$\mu$laser FC10 & 150 & 6  \\
370 &Elliptical & $\mu$laser FC10 & 150 \& 200 & 6 \& 140\\
399 &Round& $\mu$laser FC5 & 250  & 32  \\
435 &Elliptical& $\mu$laser FC10 & 150 \& 200 & 7 \& 150  \\
935 and 760&Round & ThorLabs RC02FC-P01 & 250 & 74  \\
 \end{tabular}
 \end{ruledtabular}
 \end{table*}

\begin{figure*}[!hbtp]
    \includegraphics[width=2.5 in]{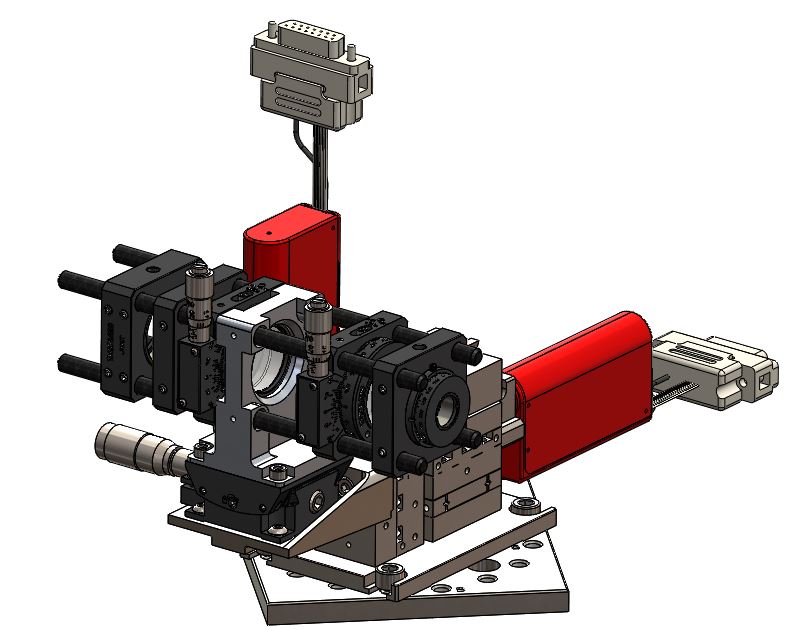}
    \caption{Example fiber launch assembly, showing custom pieces for attaching the translation stage to the breadboard, attaching the goniometer to the side of the translation stage, and for attaching the cage assembly above the goniometer.}
    \label{fig:cwlaunch}
\end{figure*}

We also created custom pieces to stably hold our beam launch optics, which can be seen in Fig.~\ref{fig:cwlaunch}.  The mounts each start with a Newport 561D-XYZ ULTRAlign\textsuperscript{TM} stage attached to a custom adapter plate that has a hole pattern allowing us to match the stage position to the holes of the optical breadboard (which differs depending on the position around the chamber).  Next a custom L-bracket is attached to the vertical short side of the translation stage, which holds a goniometer for adjusting the beam's angle (``tip'') as it is focused on the trap.  On top of the goniometer is another custom piece which serves as the adaptor to a Thorlabs cage mount system.  For the cage mount rods, we use carbon fiber instead of steel for its stiffness and low temperature sensitivity (though it is unclear if this has been beneficial).  The cage system holds and centers the collimators with the lenses and waveplates necessary for light delivery.  We use Newport picomotors for remote position adjustment in x and y.  

\section{Laser Based Qubit State Manipulation }\label{355laser}

\begin{figure*}
	\includegraphics[width=0.5\textwidth]{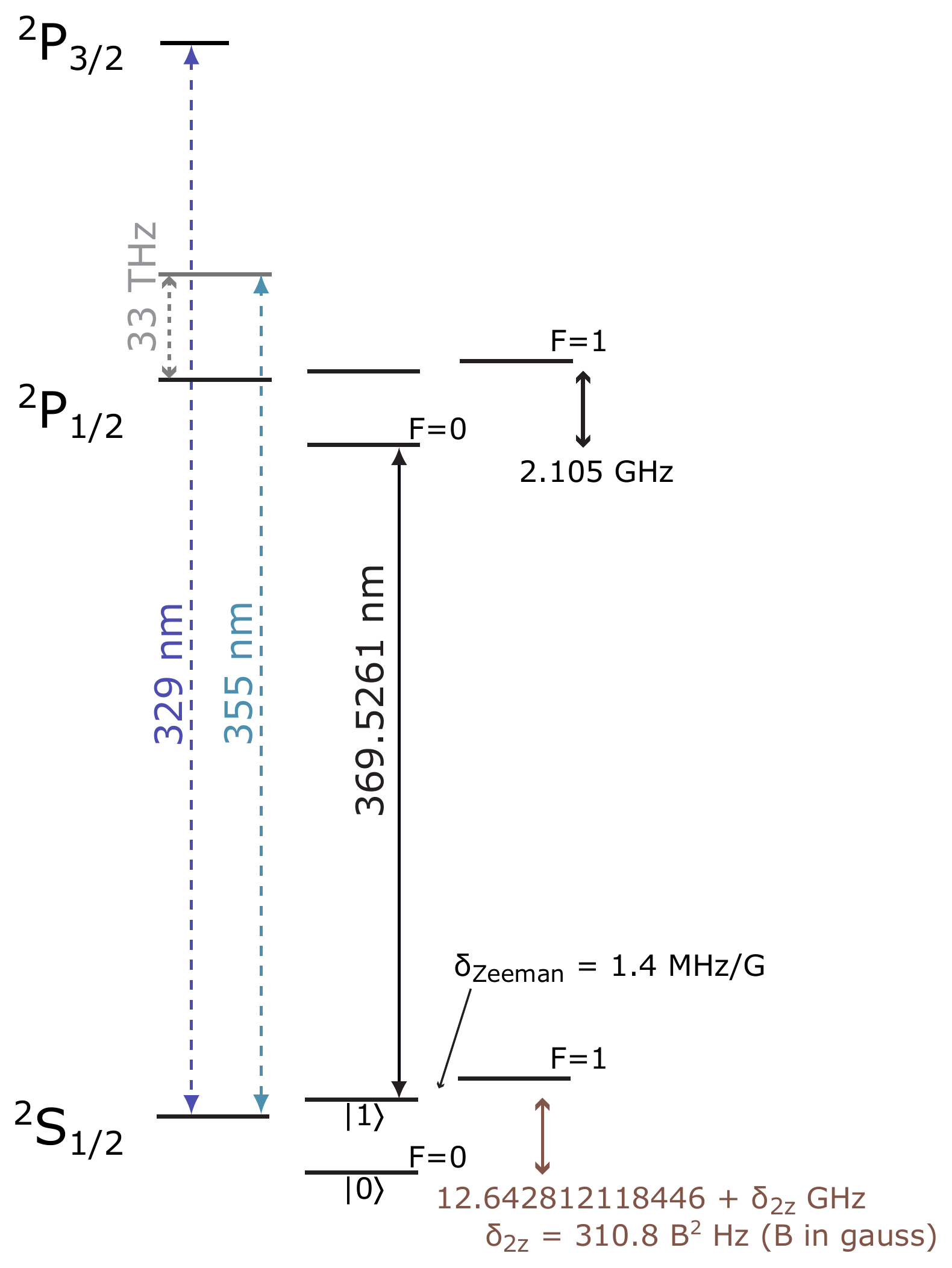} 
	\caption{\label{fig:355leveldiag}
	Cartoon of the relevant energy levels for the Raman transition in \Yb. 
	The 355~nm light used in the Raman beams spans from the $^2$S$_{1/2}$ state to a virtual state (grey) between the two P levels, approximately 33~THz from the $^2$P$_{1/2}$ state.
	By careful selection of the comb teeth used in the pulse laser, we span the 12.643~GHz offset between the \ket{0} and \ket{1} states of the qubit without high frequency electro-optical components or a second laser.}
\end{figure*}

While incoherent techniques for manipulating ions, as outlined in the previous chapter, are important for cooling, state preparation, and detection, quantum computing requires \emph{coherent} manipulation of qubit states.  As mentioned previously, we use the \Yb{} hyperfine clock transition ($^2S_{1/2}\,\ket{F=0,m_F=0}\rightarrow\ket{F=1,m_F=0}$) as the qubit states, depicted in Figs.~\ref{Ybsimple}~and~\ref{fig:355leveldiag}).  The frequency between the two qubit states is given by $f_{qubit} = 12.642812118466 + \delta_{2z}$~GHz, where $\delta_{2z} = (310.8)B^2$~Hz and B is the magnetic field in gauss~\cite{Fisk1997}, which to first order is insensitive to magnetic fields at small magnetic fields.   This magnetic field insensitivity and the fact that T1-type decay between states in the S manifold is extremely rare, leads to extremely long coherence times~\cite{Wang2017}, allowing for higher fidelity gate operations and a processor capable of greater circuit depth.  While these states can be directly manipulated with microwaves, individual addressing and two-ion interactions with microwaves require sophisticated techniques to create localized microwave fields and gradients~\cite{Ospelkaus2008,AudeCraik2014}.  Instead, we use a laser to perform single-ion and two-ion gates so we can take advantage of tight focusing and specific beam geometries to create the field gradients necessary for individual addressing and two-qubit gates mediated by a phonon bus~\cite{Molmer1999}.  Specifically, we use a 355~nm pulsed laser that drives two-photon Raman transitions~\cite{Wineland1998, Leibfried2003} via a virtual state, 33~THz detuned from the $^{2}P_{1/2}$ level (see Fig.~\ref{fig:355leveldiag}).  We chose to use a pulsed laser because the frequency comb associated with a train of pulses has sufficient bandwidth to span the qubit transition frequency~\cite{Islam2014}.

\subsection{Pulsed 355 nm Laser}
We use the Coherent Paladin Compact 355 as our pulsed Raman laser.  It is a commercial laser that has a repetition rate around 120~MHz~\cite{Coherent} with individual pulses roughly 10~ps long.  The pulse repetition rate was selected such that no pair of frequency comb teeth generated by a single pulse train would be resonant with any transition in \Yb.  This includes the carrier transitions (($^2S_{1/2}\,\ket{F=0,m_F=0}\rightarrow\ket{F=1,m_F=0}$), the estimated radial and axial sidebands, the Zeeman hyperfine transitions (\state{2}{S}{1/2}~$|$F=0;$m_F$=0$\rangle$ to \state{2}{S}{1/2}~$|$F=1;$m_F$=-1$\rangle$ or \state{2}{S}{1/2}~$|$F=0;$m_F$=1$\rangle$) and the so-called micromotion sidebands, which consist of any transition $\pm$ the trap rf drive frequency (usually 35-85~MHz for \Yb).  Additionally, the Paladin laser has an average output power of ~3.75~W, which is sufficient to split the beam multiple times (See section~\ref{sec:ImgPath}) and still have enough power at each ion to drive Raman carrier transitions at approximately 500~kHz.  The output power of the laser is internally stabilized against slow drift; thus, we only need to stabilize the beam intensity on fast times scales, which is acommplished by feeding back on a small portion of the beam.  However, the pulse repetition rate, which is critical for defining the frequency combs used to perform transitions, is not internally stabilized, and is expected to drift a few kHz over the course of the day. This can be corrected for by implementing a feed-forward system as described in the next section and in Sec.~\ref{hw:frequencyfb}. 

\subsection{Locking }
\label{sec:Locking}
\begin{figure*}
\centering
\includegraphics[width=\textwidth]{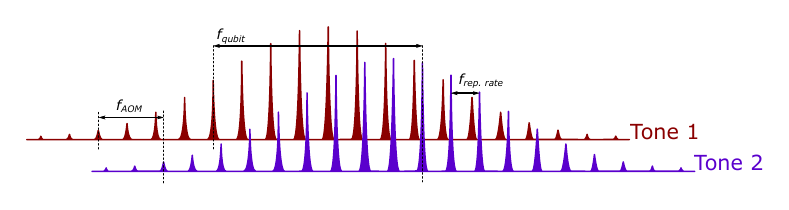}
\caption{Graphical representation of the AOM-shifted frequency comb and repetition rate feed forward scheme. 
To drive ion transitions, two offset frequency combs are applied to the ion, such that different teeth in those combs span the frequency separation of the desired transition, labeled $f_{qubit}$.
}\label{fig:355lock}
\end{figure*}

To drive transitions between the \ket{0} and \ket{1} states of the qubit, we need two frequency tones that are separated by $12.643$~GHz, the resonant frequency of the transition.  Because we have carefully selected our laser repetition rate to precisely \emph{not} overlap with any transitions, we will need to spatially overlap at least \emph{two} offset frequency combs in order to create the necessary frequency separation to drive transitions (see Fig.~\ref{fig:355lock}).  For example, assuming a repetition rate, $f_{rep}$, of exactly 120~MHz, the first and 105$^{th}$ comb lines nearly span the transition frequency. 
However, there is a 42.8~MHz offset, $f_0$, from the qubit transition to the comb separation.  To bridge this gap and drive the carrier transition, we apply a relative frequency shift, $f_0 = f_{qubit} - n\times f_{rep}$, between the two frequency combs using acoustic optic modulators (AOM), where $n$ is a fixed ratio (see Fig.~\ref{fig:355lock}).  Depending on the desired transition, the applied frequency offset of the combs may vary.  

Because the laser cavity length sets the repetition rate and thus the frequency comb spacing, it is sensitive to thermal drift, such that $f_{rep} \rightarrow f_{rep} + \delta f_{rep}$. 
In order to maintain a fixed frequency offset between the two tones, we actively track the variations in repetition rate and add an additional correction, $n\times\delta f_{rep}$, to the frequency offset~\cite{Islam2014}.  Specifically, we measure the 32$^{nd}$ harmonic of the pulsed laser repetition rate on an ultra-fast photodiode (Hamamatsu S9055-01) and mix the photodiode signal with a stable frequency reference (Wenzel 3.7~GHz MXO-PLD).  We chose the 32$^{nd}$ harmonic because of its increased sensitivity to frequency drift (as compared to the first harmonic), but still within the measurement range of our photodiode.  The resulting mixed frequency is cleaned up with a low pass filter and subsequently tracked by our rf control hardware (see Sec.~\ref{hw:frequencyfb}).
Any deviations from the expected repetition rate are scaled by $n = \nicefrac{105}{32}$ and added to (or subtracted from) one of the rf tones applied to the AOM.

\subsection{Optical Beam Paths}

As discussed in the previous sections, our qubit operations are performed using two-photon Raman transitions.  Since there are two photons involved, the beam propagation direction(s),  $\vec{k}$, determines how much total momentum is imparted on the ion, which can cause the ion to interact with particular motional modes of the ion trap.  For example, co-propagating beams are not as sensitive to the ion motion or temperature because $\Delta k=0$. 
Any other configuration yields $\Delta k \neq 0$, and results in a momentum kick imparted to the ion, which is necessary for sideband cooling and driving M\o{}lmer-S\o{}rensen (MS) gates~\cite{Molmer1999}.  We have chosen to use counter-propagating beams along the surface of the device, perpendicular to the ion chain axis, for driving gates that require a momentum kick.  In the counter-propagating configuration, momentum transfer is maximized ($\approx 2|\vec{k}|$) and since we are perpendicular to the trap access, momentum transfer is only possible to the radial modes of the ion (and to first order, gives no access to the axial modes).  

\begin{figure*}
\centering
\includegraphics[width=0.85\textwidth]{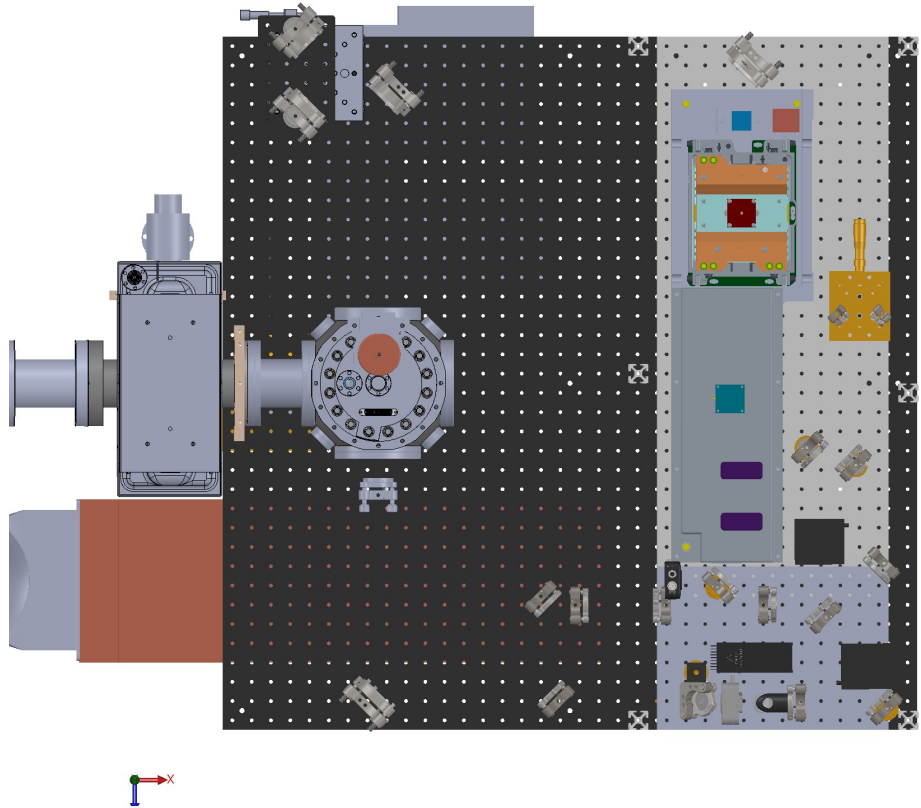}
\caption{3D rendering of the optical path for the Raman beams as seen from above.  The chamber is mounted to an optical breadboard, with the laser on the optical table below it (not shown).  The shared path between the global and individual setups is shown in green, where the light travels from the laser below  through the optical breadboard and then passes through polarization optics to set the total laser power and to an AOM which is used for power stabilization.  Then the light passes through a beamsplitter, where it is split into the two separate paths.  The global path (teal) goes to a mirror pair mounted on a stage allowing for path length matching of the beams, then its own AOM for frequency control. The individual addressing beam path (magenta) passes through the 32-channel Harris AOM, where it is first split into 33 beams by a diffractive element. Each beam goes to its own AOM crystal, which modifies the frequency, amplitude, and phase of that beam based on the driving rf signal. After the AOM in each path, there are optics to shape the beams and image all frequencies to the same locations at the ion chain.  Also shown are the placement of cw-laser beam delivery subsystems around the vacuum chamber.
}
\label{fig:355Globallayout}
\end{figure*}

The counter-propagating configuration requires two optical beam paths be directed to our ion chain.  When designing these optical beam paths, there are several considerations: 
\begin{enumerate}
\item{Individual pulses from the laser are 15~ps long, thus for the pulses to overlap at the ion from separate beam paths, the beam path lengths must match on the sub-millimeter scale.}
\item{Relative beam path lengths must not change due to vibrations or air currents.}
\item{We must be able to shift the frequency combs relative to each other in frequency space to address different transitions.}
\item{Co-propagating and MS gates require two frequency combs from the same direction, thus alignment needs to hold for a broad range of frequencies.}
\item{At least one of the beam paths much be able to individually address the ions.}
	\begin{itemize}
	\item{Each ion requires a laser with distinct frequency, amplitude, and phase control.}
	\item{Beams must be tightly focused to not overlap with nearby ions.}
	\end{itemize}
\end{enumerate}

Below, we will describe our design and how it allows us to achieve the above requirements.  A schematic of the beam paths is shown in Fig.~\ref{fig:355Globallayout}.  The beam paths include a method for: path length matching on the micron scale; carefully designed optical mounts to prevent vibrations; AOMs in both paths to shift the frequencies; optics designed to reimage a wide range of frequencies; one path with multiple beams with independent control of frequency, amplitude, and phase; and finally, optics to create tightly focused beams at the ions with low crosstalk and scatter.  To minimize air currents and other environmental disturbances, the entire system is enclosed.

\subsubsection{Global Beam Design}
The global beam (represented by the teal beam in Fig.~\ref{fig:355Globallayout}) is designed to work either alone, to drive motionally insensitive gates on the entire ion chain, or with individual beams to drive motionally sensitive gates on specific ions. 
To that end, the beam needs to have a nearly-uniform spatial profile, the alignment must be stable, and the timing must be aligned to the individual beams.

After splitting the individual path and the global bath at a beamsplitter, the global beam is diverted to a pair of mirrors on a linear translation stage to allow for fine-tuning of the path length without affecting downstream beam alignment.  Next, the beam passes through an AOM, allowing for full amplitude, phase, and frequency control of the beam. 

To select the global beam size, we needed to ensure we would not scatter on the surface of the trap and that we would be able to roughly equally illuminate 32 ions.  We calculated that a Gaussian horizontal beam with a waist of \um{160} spans 32 ions with a \um{4.5} ion spacing with less than 1\% power variation over the center three ions. 
Vertically, the ion is about \um{70} away from the trap surface, and the isthmus of the trap is 1.2~mm across, so a round beam with a waist of \um{160} (to match the desired horizontal beam waist) would scatter quite a bit.
This would result in trap charging, stray reflections, and degradation of beam quality at the ion.  Because the beam is Gaussian, the divergance of the beam is related to the focused beam waist, with more tightly focused beams diverging faster. Thus, in order to clear the trap edge with at least 4 beam radii and negligible aberrations, the vertical beam waist at the ion needs to be between about \um{5} and \um{13} with the maximum clearance from all surfaces occurring for with an \um{8} waist at the ion.  Thus, the Gaussian beam should be elliptical and $\um{8} \times \um{160}$ at the ion.

To achieve these beam waists, we use a cylindrical telescope to change the aspect ratio of the beam by using a concave/convex lens pair, which avoids putting the beam through any unnecessary focal planes.  A spherical lens is used to reimage the beam near the final focusing optics to minimize the effects of vibrations and diffraction. 

\subsubsection{Individual Addressing Beam Path}\label{sec:ImgPath}

The individual addressing beam path is more complicated than the global beam path, due to the extra requirements of splitting the beam many times, separately controlling each beam, and then imaging each beam on one and only one ion.  

We use an Illumination Module from Harris Corporation, which is a  multi-channel acousto-optic modulator AOM, that was specifically designed for this application~\cite{Harris}. 
It has integrated diffractive optics that splits the single beam into 33 equal and parallel beams, which are sent to 32 separate integrated miniature AOMs (the last beam is blocked internally).  Each AOM has an independent rf input from a dedicated rf amplifier.  The control signal to the amplifier is generated by an RFSoC (or an Octet, more details in Sec.~\ref{ControlHW}). It can generate multi-tone rf signals with arbitrary amplitude, frequency, and phase modulation capabilities, which allows us to exercise complete control over the light applied to each ion.  However, the AOM also deflects the beam depending on the applied rf frequency (in our case, vertically).  To prevent these deflections from causing different frequencies to have different overlap with an ion, light from the AOM needs to be carefully reimaged onto the ion.   

The output of the multi-channel AOM consists of 32 parallel beams with a \um{80} waist and a \um{450} pitch.  To image these onto the ions, the pitch of the AOM needs to be matched to the \um{4.5} pitch of the ions resulting in an \um{0.8} beam waist.  However, as discussed previously in the global beam design, due to the trap geometry, an \um{8} beam in the vertical direction is ideal for minimizing scatter from the surface of the trap.  Thus, we require an oval beam at each ion with a \um{0.8} axial (horizontal) waist and an \um{8} vertical waist.  The reimaging and beam shaping are accomplished using a combination of off-the shelf and custom optics, some of which are prealigned and bonded in place to make alignment of the entire beam path easier, as described in the next paragraph.  Additionally, the optical path was designed to keep the beams compact, to minimize the effect of air currents causing variations between the beams or along a single beam.  

After the AOM, the beams immediately go through a pair of cylindrical lenses, designed to change the aspect ratio of the beam to the desired 10:1.  These lenses were interferometrically aligned to each other to ensure that their optical axes were exactly aligned before being bonded into custom mounts.  The cylindrical lens assembly was mounted to the breadboard using a Newport LP-1A for $x$, $y$, tip, tilt, and roll position control.  The cylindrical pair is followed by a spherical lens, bringing the beams to a focus, that is reimaged by a spherical pair combined with a custom lens assembly from PhotonGear (PN:PG-18020-S)~\cite{Noek2013, PhotonGear}.  The PhotonGear lens relay has a 44.4~mm housing-to-image distance and a numerical aperture (NA) of 0.16 on the ion side and 0.04 on the input side. 
It has also been designed to account for the 3~mm thick vacuum window, additionally compensating for the aberrations caused by the window bowing while it is under vacuum.  To minimize aberrations at the ion, it is extremely important that the lens is aligned normal to the window surface, which is not necessarily parallel to the ion chain.  This alignment is particularly difficult because the relay lens is entirely inside the re-entrant bore window when it is the correct distance from the ion.  Additionally, the alignment of the spherical pair and the lens relay is crucial for setting the magnification.  Therefore, to hold and align these lenses relative to each other and to the ion, we designed and built a custom 5-axis flexure mount.  

\begin{figure*}
\centering
\includegraphics[width=0.75\textwidth]{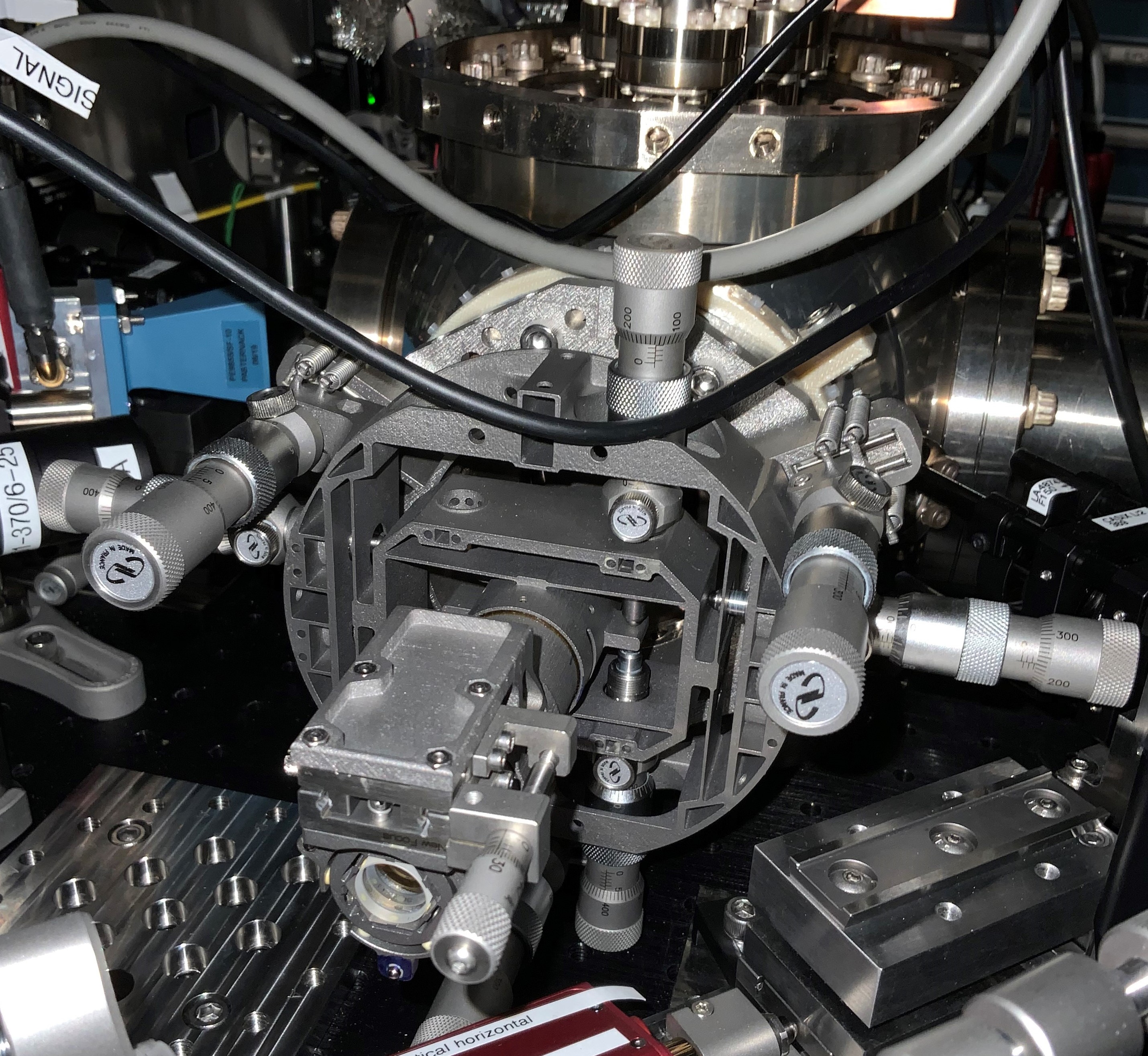}
\caption{Image of the 5 axis flexure mounted onto the individual addressing re-entrant flange.
The parts are 3D printed from titanium and steel, and then assembled before adding the lenses. 
The optics are aligned to the optical center of the lens (rather than the mechanical center) using external hardware and then bonded in place. 
By completing relative alignment of the optics, before mounting the flexure stage, we realize greater optical precision than we could achieve by hand.
This lends to the achievement of the desired beam waist with few aberrations.  
}
\label{fig:flexure}
\end{figure*}

The flexure stage is shown in Fig.~\ref{fig:flexure}.  It is bolted directly onto a mounting ring on the re-entrant window, registering the lenses to the center of the chamber.  Hidden from view in the photo is the relay lens itself, because it is entirely inside the re-entrant bore. The flexure design uses opposing micrometers to align the position in $x$ and $y$, which can then be brought together to lock the mount in place.  The range of motion is designed to cover the full length of the trap, if necessary, but currently, the individual addressing beams are aligned to the trap center. Tip and tilt are controlled by the three additional micrometers accessible on the front of the mount and allow us to reduce abberations by matching the tip and tilt of the lens to the tip and tilt of the vacuum window.  A pair of spherical lenses are interferometrically aligned to the relay lens.  They are bonded to a translation stage, which allows for small adjustments to the focal position, but they are mechanically linked to, and optically centered on, the lens relay.  By actively aligning the lenses to the lens relay on the same mount, we have created a single larger objective simplifying the alignment procedure. 

\subsection{Beam Characterization}

\begin{figure*}
\centering
\includegraphics[width=0.8\textwidth]{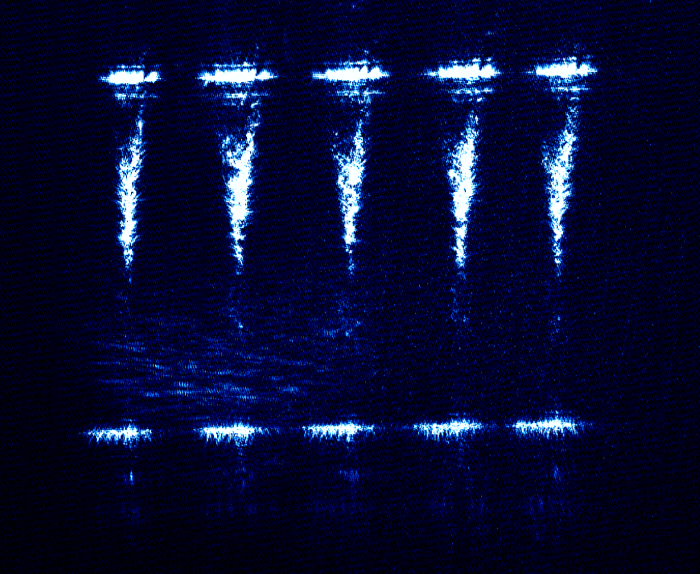}
\caption{Image of the individual addressing beams skimming across the trap surface. 
The horizontal features are where the reflected scatter is increased at the edge of trap electrodes. 
The bottom lines are at the edge of the slot on an HOA-2.0 trap and the top is the edge of the rf rail. 
The vertical cones are the beams reflecting from the surface of the trap slightly after the beam waist.
This shows beam 1, 8, 16, 24, and 32 (out of 32) from left to right. 
The imaging system is not designed to be fully diffraction limited at 355~nm and is limiting the resolution of the image, causing the beams to appear fuzzy and jagged, which is not a faithful representation of the beam profiles. }
\label{fig:alignindivid1}
\end{figure*}

To determine the quality of our optical alignment, we characterize the beam profile at the ions. 
For the global beam, a beam profiling camera at the ion location can be used (before the chamber is sealed and baked for vacuum). 
We measured a beam waist of \um{10.9} vertically and \um{182.8} horizontally, which is close to the desired \um{8} by \um{160} beam size.
However, the individual beams are much smaller than the pixel size of our beam profiling camera, so we must use the experiment itself to characterize those beams.
First, by taking advantage of the high resolution imaging system, we can align the beams to the trap surface and check each one for general shape and size (see Fig.~\ref{fig:alignindivid1}).
After aligning to the trap surface, it is straightforward to align to an ion by lifting the beams away from the trap surface using a micrometer.  

\begin{figure*}
\centering
\includegraphics[width=0.9\textwidth]{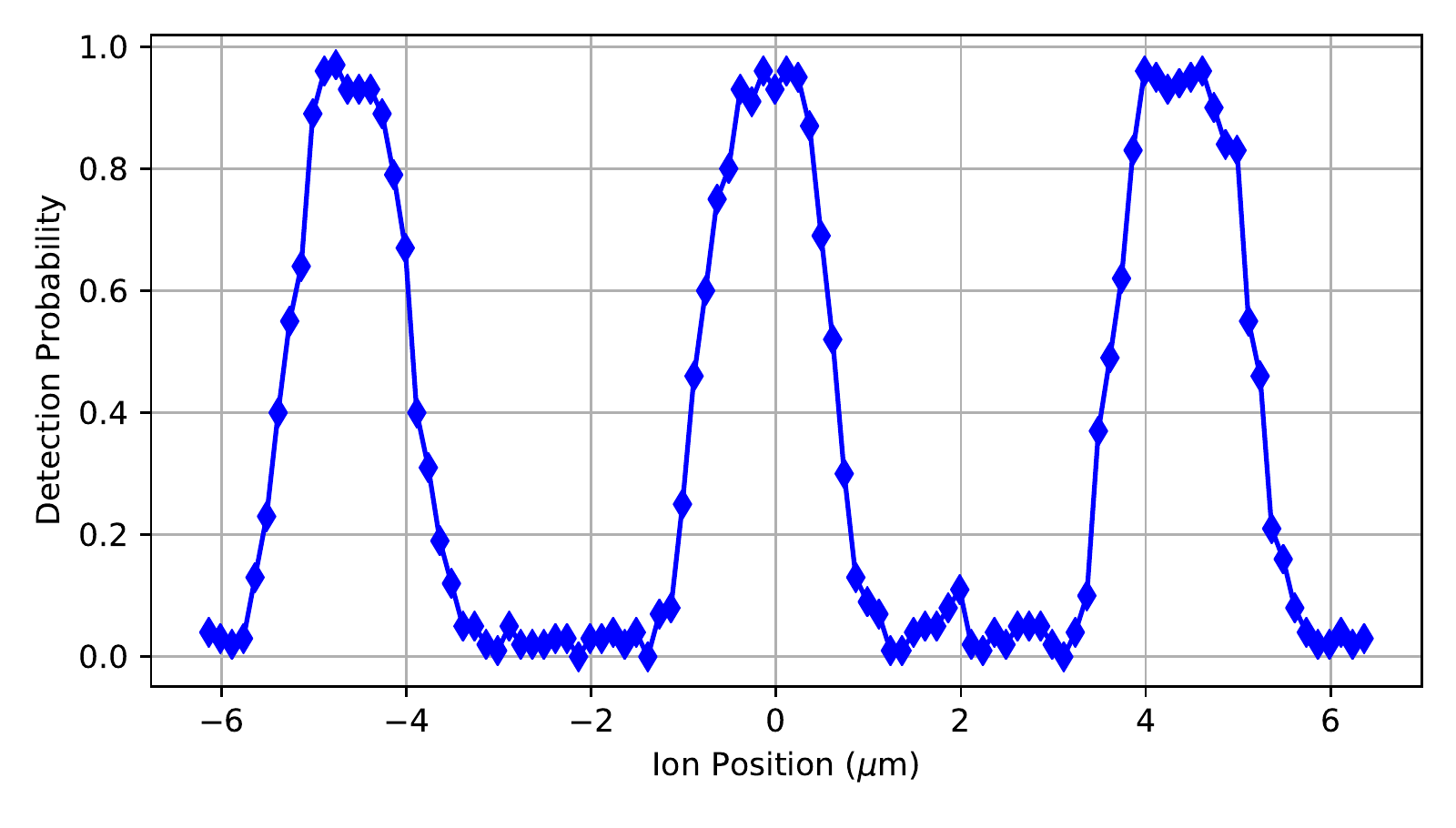}
\caption{For a fixed pulse duration and power, the probability of finding the ion in state \ket{1} while moving the ion through the individual addressing Raman beams. 
Fitting these profiles to a Gaussian, we determine the beam waist is approximately \um{0.8} with no obvious aberrations. 
We also confirm that the spacing between the beams is about \um{4.5} consistent with the \um{0.8} beam waist. 
}\label{fig:counterproppos}
\end{figure*}

We can shuttle the ion through the beams and extract horizontal beam profiles by driving qubit rotations and measuring the relative population as a function of ion position.
The ion position is controlled by voltage solutions that are precisely calculated to a sub-micron scale, resulting in a detailed and accurate measurement of the beam profiles along an ion chain, shown in Fig.~\ref{fig:counterproppos}.
The beam profiles confirm that the beam waists and spacings are consistent with the design and that there are no obvious aberrations.

\subsubsection{Crosstalk}\label{sec:355xtalk}
\begin{figure*}
\centering
\includegraphics[width=0.9\textwidth]{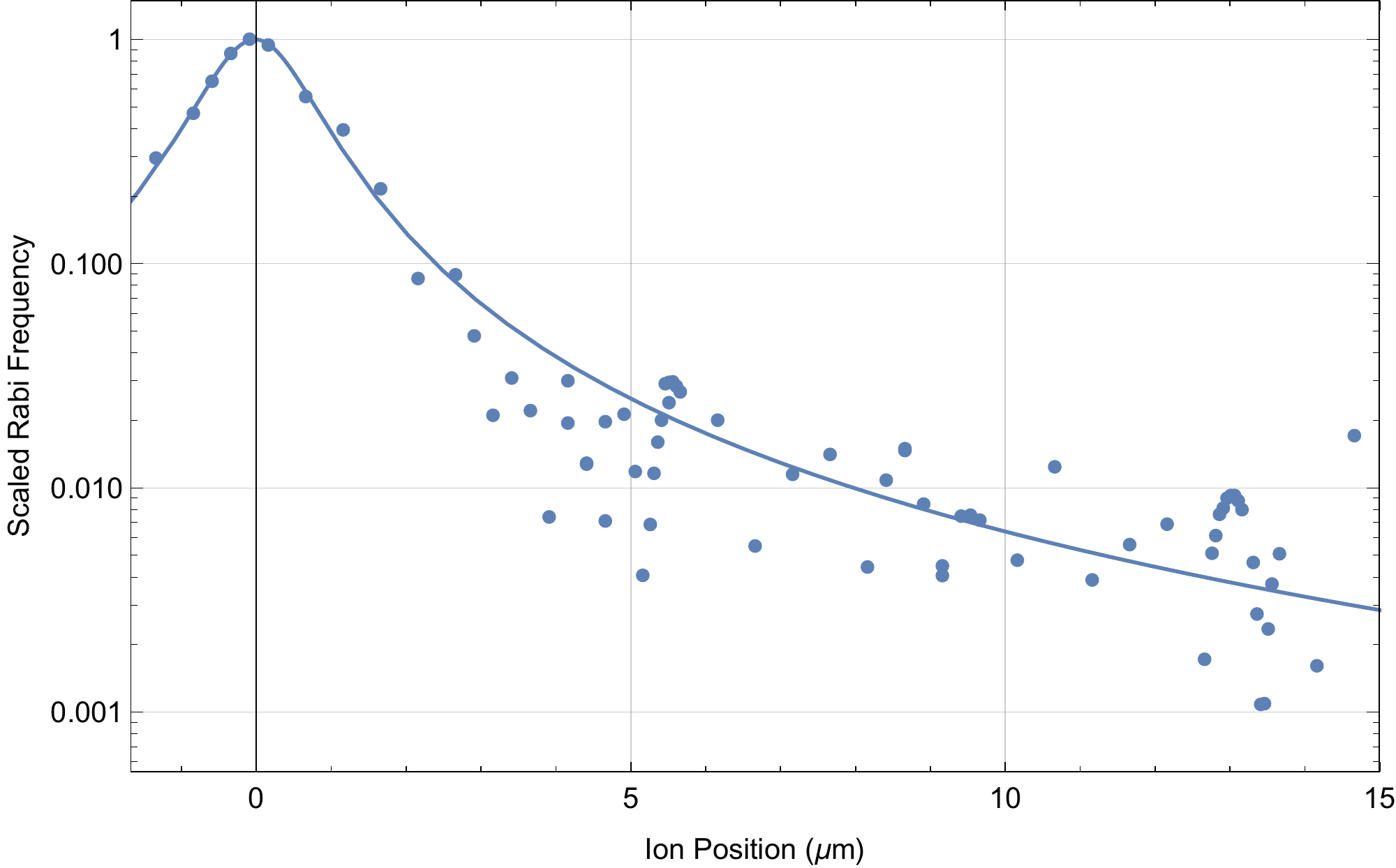}
\caption{Optical crosstalk is determined by the relative Rabi frequency measured when an individual addressing beam is interrogating a neighboring site. 
By measuring the Rabi frequency versus ion position using counterpropagating Raman beams, we determine the optical crosstalk value to be 2.3(6)\% at the neighboring beam location and 0.6(2)\% at the next nearest neighbor site. 
We do not see any clear evidence of electric or acoustic crosstalk on neighboring AOM channels causing an increase in the crosstalk at the ion. 
The solid line is a fit to a Lorentzian distribution with a full-width-half-max of \um{1.29}. }
\label{fig:rabivspos}
\end{figure*}

Linear chains of ions spaced closely together have stronger Coloumb interactions and thus faster two-qubit gates, but having neighboring sites close together increases the probability of optical crosstalk. 
We measure the impact of the neighboring beams at our \um{4.5} spacing by taking advantage of our ability to shuttle the ion. 
First, we drive Rabi oscillations using three consecutive individual addressing beams and measure the Rabi frequency at each beam position to ensure they are equal. 
Then, turning on only the center beam (centered at \um{0} in Fig.~\ref{fig:counterproppos} and~Fig.~\ref{fig:rabivspos}) we measure the Rabi frequency as a function of ion location.
Fig.~\ref{fig:rabivspos} shows the scaled Rabi frequency as a function of ion location, where the measured frequencies are normalized to the maximum observed Rabi rate at position \um{0}.
The decay of optical crosstalk from the center location is consistent with a Lorentzian distribution. 
We do not see any enhancement of the cross-talk above our baseline at the locations of the neighboring beams, suggesting minimal amounts of electrical crosstalk or acoustical crosstalk from the AOM.
We measure optical crosstalk of the neighboring site at about 2.3(6)\% and at the next nearest neighbor site = 0.6(2)\%.
To reduce the effects of the crosstalk, a cancellation tone can be applied to the neighboring channels (see  Sec.~\ref{CrosstalkComp}).

\section{Ion Imaging and Detection }\label{sec:imaging}
Another piece of hardware for our quantum system is the measurement and detection scheme.  
As described in Sec.~\ref{cwlasers}, detection is performed by illuminating ions on resonance with a cycling transition accessible to only one of the qubit states.  
The resulting ion florescence is collected, and depending on the number of photons captured in a particular time, we determine whether the ion is in the $\ket{0}$ (dark) or $\ket{1}$ (bright) state (see Fig.~\ref{Ybsimple}).
The bright and dark state histograms for repeated measurements ideally do not overlap, and we can distinguish the states with a high degree of accuracy~\cite{Acton2006}.  
Due to our detection efficiency of $\approx$~0.5$\%$ our detection time is set to 350~$\mu$s, during which we collect about 10 photons from an ion in the bright state.  
In this time, an ion in the bright state has a 1.7e-2 probability of off-resonant coupling to the incorrect P state followed by decay to a dark state (similarly there is a 1e-3 probability of a dark ion off-resonantly coupling to the bright state).
Depending on the number of photons collected before off-resonant coupling events, they do not necessarily result in a detection error. 

The thresholding technique is very good for a single ion.
However, when there are multiple ions, the histograms generated from repeated measurements of 2 bright ions and 1 bright ion overlap considerably.
At larger numbers of ions, distinguishing between the number of bright ions becomes even more challenging.  
Additionally for quantum algorithms, it is important to know \emph{which} ions are bright and dark, instead of just the pure number of bright ions.  
 
\begin{figure*}[t]
    \centering
    \begin{subfigure}[t]{0.59\textwidth}
        \centering
        \includegraphics[width=\textwidth]{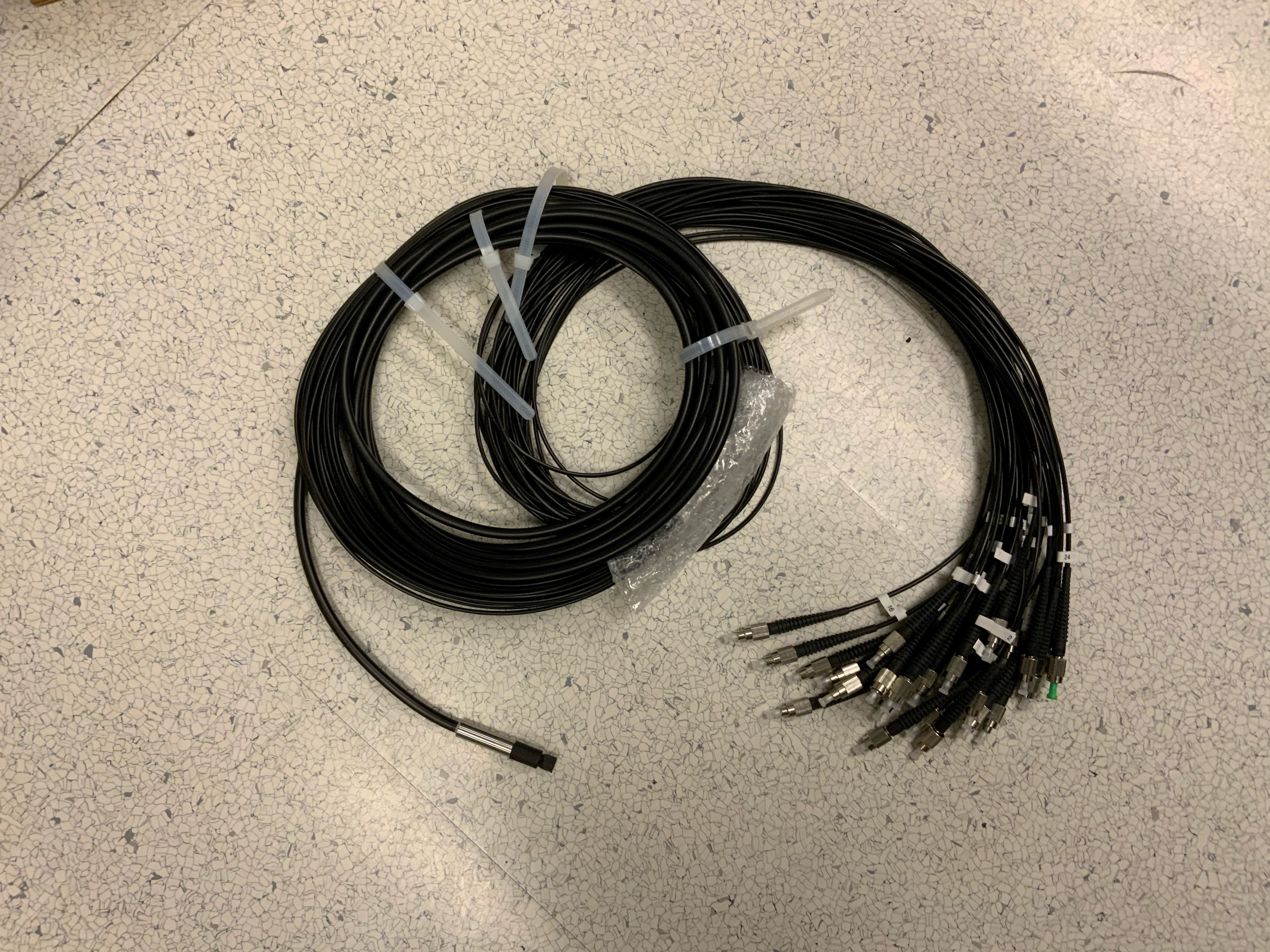}
        \caption{The entire multicore fiber.}
        \label{fig:MulticoreFiber}
    \end{subfigure}%
    ~ 
    \begin{subfigure}[t]{0.39\textwidth}
        \centering
        \includegraphics[width=\textwidth]{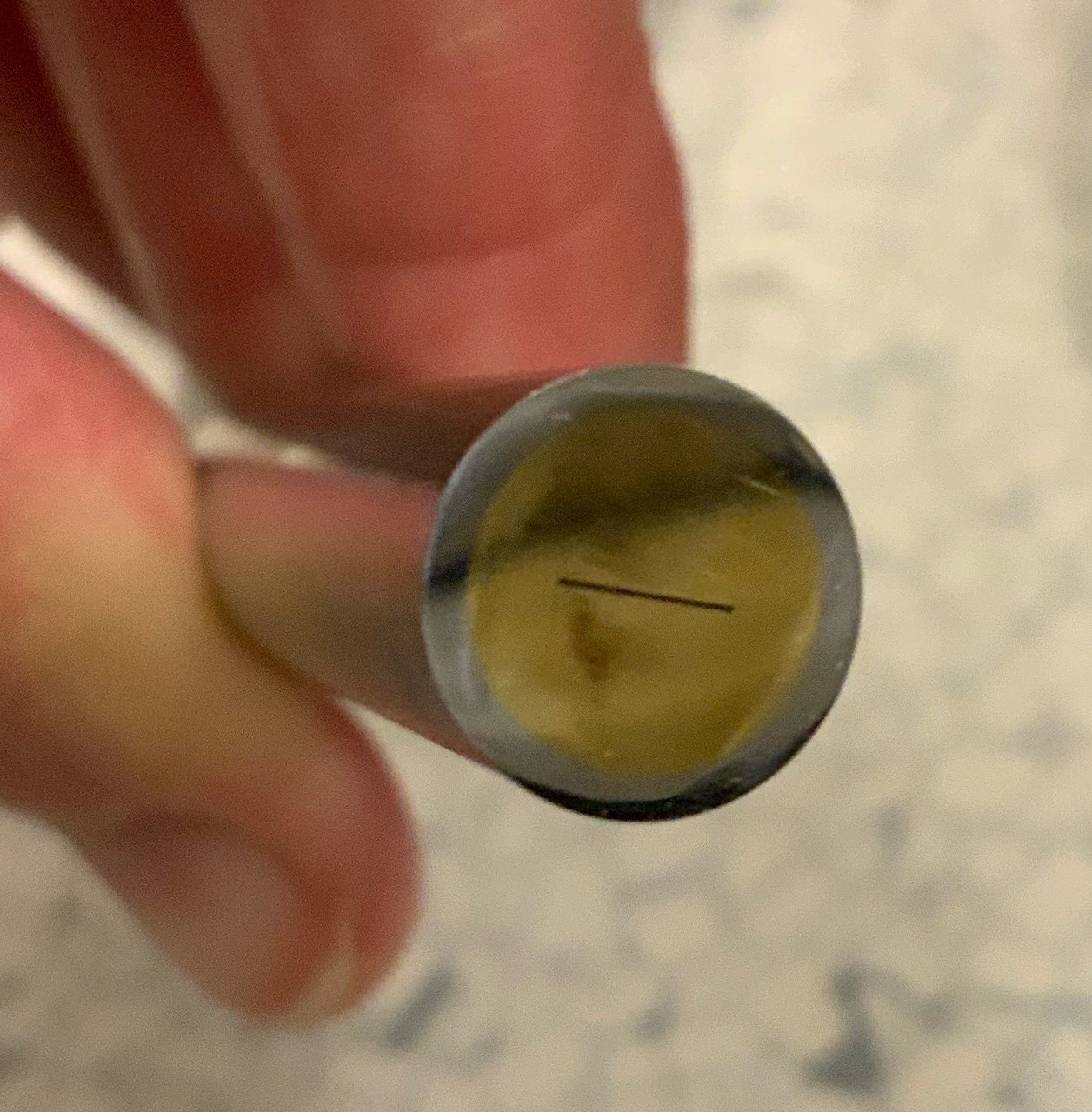}
        \caption{Photo of the multicore fiber tip. }    \label{fig:MulticoreFiberTip}
    \end{subfigure}
    \caption{The multicore fiber allows us to re-image each ion onto its own fiber and PMT. 
    The result is distinguishable detection with minimal loss and crosstalk.
    (a) A view of the single fiber tip fanning out into 32 individual fibers. These can be easily routed to separate PMTs.
    (b) A close view of the fiber tip. 
    The individual cores are indistinguishable in this photo, but the linear arrangement is apparent. }
	\label{fig:multicoreCombo}
\end{figure*}

To address these issues, each ion has a dedicated photomultiplier tube (PMT) to determine if that particular ion is bright or dark.
Dedicated PMTs provide increased sensitivity and readout speed compared to commonly available cameras and less crosstalk between channels as compared to a PMT array.  To get light to individual PMTs, we image the ion light into a multicore, multimode fiber (see Fig.~\ref{fig:multicoreCombo}).
This fiber has 32 separate \um{50} diameter cores spaced 125~$\mu$m apart (FiberTech Optica, custom), with light from each ion aligned to a distinct core.  
The cores are then broken out into individual fibers (see Fig.~\ref{fig:MulticoreFiber}), and each fiber plugs directly into a separate PMT (Hamamatsu H10682-210, dark count average 20 per second).    

\begin{figure*}
\includegraphics[width=\textwidth]{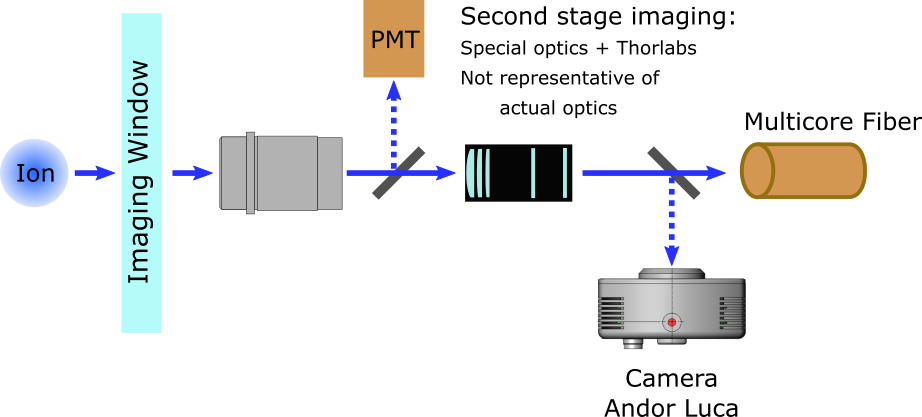}
\caption{Schematic of imaging system layout.
The motorized flip mirrors deflect the light (dashed arrows) when engaged, and pass light straight to the multicore fiber when disengaged. 
This way, the most sensitive optic is not prone to misalignment by actuating the mirrors.  \label{fig:ImagingSystemSch}}
\end{figure*}

For additional diagnostics, we have a camera (Andor Luca DL-604M-$\#$VP) and a free-space coupled global PMT (see Fig.~\ref{fig:ImagingSystemSch}).
The camera has an approximately \um{250} field of view and is typically used for alignment.
The global PMT also has a large field of view and works well for single ion diagnostics and rough alignment. 
These are each accessible via remote controlled motorized flip mirrors (also shown in Fig.~\ref{fig:ImagingSystemSch}).

Since the fiber cores are spaced by 125~$\mu$m and the ions are spaced by 4.5~$\mu$m, we need to reimage the light from the ions onto the cores.  
To do so, we use a two-stage imaging system.  
The first stage is a PhotonGear imager (PG-15470-S$\_$A) with a NA of 0.6 to collect light from the ions.  
There is a re-entrant window in the vacuum chamber (described in Sec.~\ref{vacuum}) in order to place the imager the proper distance from the ion (11~mm total, including a 4~mm fused silica window).
We adjust the tip and tilt of the lens to match the orientation of the chamber window. The lens is also mounted on linear translation stages to align precisely to ion chain.

After the PhotonGear lens, the light is focused and then reimaged using a combination of a Special Optics compound lens (54-17-29-370nm) and a pair of Thorlabs lenses (LB1294-A and LA4102-UV).  
This combination yields the desired 125/4.5 magnification for the ion light, with a final spot size of 2~$\mu$m radius per ion, which can easily couple into a 50~$\mu$m diameter multimode fiber core. 
With this arrangement, we achieve 10 counts per 350~$\mu$s from a \Yb{} ion on resonance, for an estimated total efficiency of 4e-3~\cite{Olmschenk2007}. 

\subsection{Detection Crosstalk }\label{subsec:crosstalk}
\begin{figure*}
\includegraphics[width=4 in]{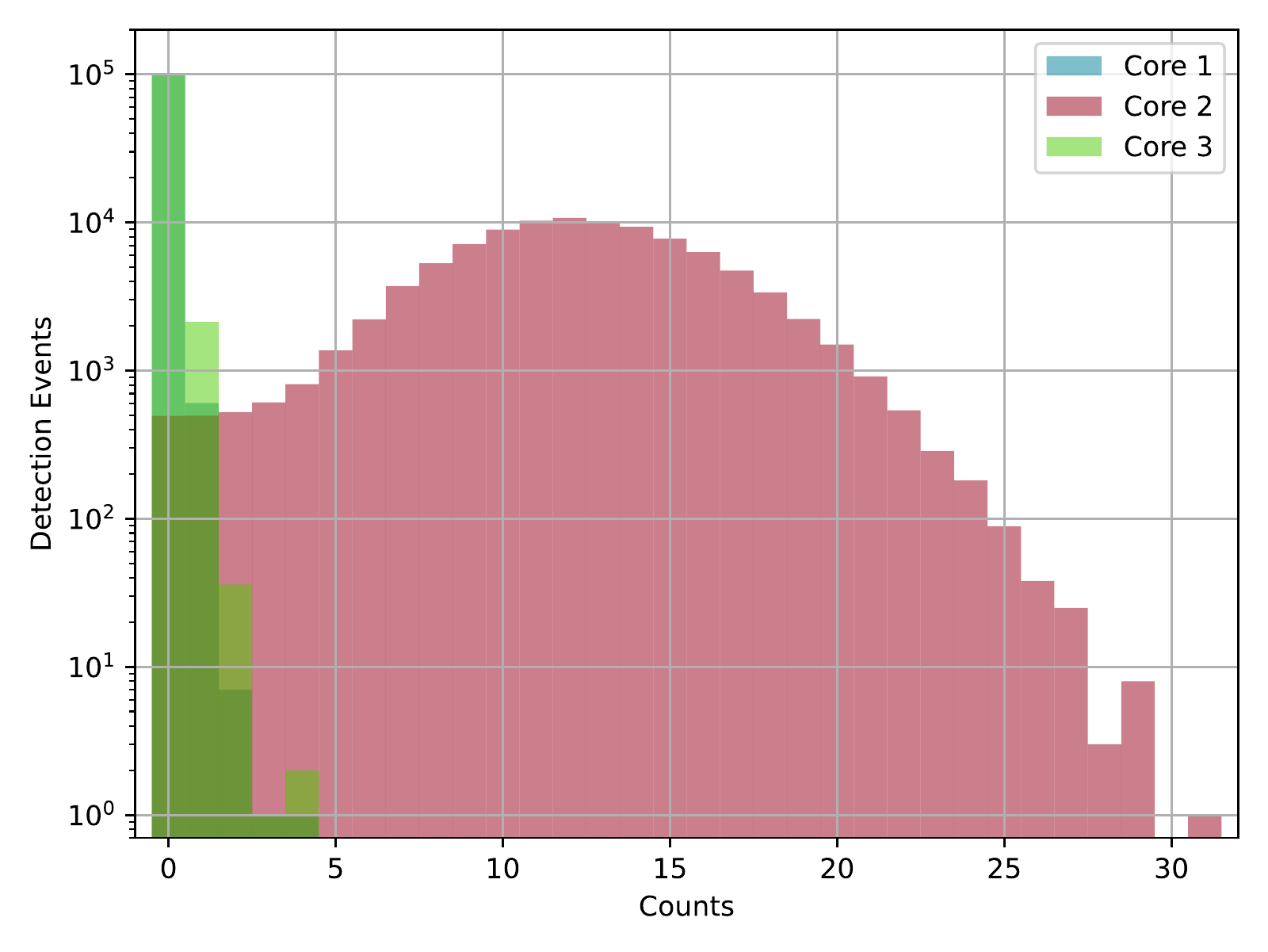}
\caption{Detection histogram displaying data from Table~\ref{tab:crosstalk}.  
Core 2 is centered on the ion and the corresponding detected events per detected photons shows a Poissonian distribution centered at 12 photons (or counts).  
Cores 1 and 3 are offset from the ion (imaging regions 4.5~$\mu$m on either side of the ion) and show mostly 0 count events.  \label{fig:DetectionHistogram}}
\end{figure*}

To determine the detection crosstalk of our system, we trap a single ion and measure the photons (or counts) on the PMTs connected to neighboring fiber cores.  
We Doppler cool the ion for 1~ms, optically pump for 10~$\mu$s as described in Sec.~\ref{cwlasers}, perform a $\pi$-pulse using microwaves, and then detect for 350~$\mu$s.
The light was collected into three PMTs, shown in Fig.~\ref{fig:DetectionHistogram}: one connected to the fiber core centered on the ion (core 2) and one connected on each side of the center ion (cores 1 and 3).
We repeat the measurement 101,100 times to look for light leakage from the bright ion on the empty cores.  
Detecting with a threshold of 1 (0 and 1 counts indicate a dark ion, $>$~1 count indicates bright), we realize an upper bound of fiber crosstalk error of 7e-4 (see Table~\ref{tab:crosstalk}).  
Since ion was prepared in state $\ket{1}$ for detecting the fiber crosstalk, error on the core 2 detection is indicative of a SPAM error. 

 \begin{table}
 \caption{Results of detecting a Doppler cooled bright ion on various detectors.  
 In fiber cores 1 and 3, we expect 0 counts, and indeed only see counts a small fraction of the time.  
 These extraneous counts could be due to light spillover from the bright ion in between the two cores, dark counts from the PMTs, or excess scatter of detection light off the trap.  
 For core 2, the ion was prepared in the bright state.  Thus, the fact that the ion is not perfectly bright is representative of a State Preparation and Measurement (SPAM) error.  \label{tab:crosstalk}}
 \begin{ruledtabular}
 \begin{tabular}{ c c c c }
Counts & Core 1 (no ion) & Core 2 (bright ion) & Core 3 (no ion)  \\
\hline
$\leq$ 1 &100191 & 991 & 100161  \\
$>$1 & 9 & 99209 & 39 \\
Error & 1e-4 & 1e-2  & 4e-4  \\
\hline
 \end{tabular}
 \end{ruledtabular}
 \end{table}

\section{Control Hardware }\label{ControlHW}

Low-level operation of the experiment, both in its quiescent state and when running experimental sequences, is predominantly controlled via custom FPGA-based circuits. 
Chief among these circuits are the following:
\begin{enumerate}
\item Master control system. 
\item Voltage control system. 
\item Coherent control system. 
\end{enumerate}
While these boards only comprise a subset of the electronics used throughout the system, they are the key interoperable components that execute timing-critical operations necessary for quantum algorithms.

The master control system's purpose is to orchestrate sequences involving signals that are used to control CW lasers, digital inputs and outputs (DIO), and analog inputs and outputs (AIO).
It is responsible for processes such as loading ions, doppler cooling, and state detection.
Certain steps in experimental sequences require the master control system to lend operational controls a dedicated subsystem, such as the voltage control system used for shuttling ions or the coherent control system for applying gates.
Only details of the coherent control system will be discussed in detail in order to elucidate the low-level pulse control needed to realize quantum gates.

\subsection{Coherent Control System}

The coherent control system uses a custom design, referred to as ``Octet'', implemented on a Xilinx rf system-on-chip (RFSoC).
It is responsible for generating the rf tones that drive the multi-channel AOM. Due to the nature of the pulsed laser system, the control electronics satisfy the following requirements:
\begin{enumerate}
\item Radio frequencies ranging from 0 MHz to 409.6 MHz.
\item Two tones per output channel.
\item Full waveform generation in the digital domain.
\item Global phase synchronization.
\item Gate sequencer, which does the following:
\begin{itemize}
\item Schedules pulse sequences used to realize gates.
\item Provides simultaneous control over frequency, phase, amplitude, and virtual Z-rotations for all tones and channels. All parameters support discrete and smooth modulation using an on-chip spline interpolator~\cite{Bowler2015}.
\end{itemize}
\item Dynamic correction for certain imperfections in the experimental hardware including the following:
\begin{itemize}
\item Frequency feedback to account for pulsed-laser cavity drift.
\item Cross-talk cancellation that adds output signals to nearest- and next-nearest-neighbor channels with tunable amplitude and delay.
\end{itemize}
\end{enumerate}

The details of these features and how they are used for pulse-level control differ from other standard rf delivery systems.
While many of these elements are separately optimized and are not exposed to the end user, several features are directly accessible.
Knowledge of how they work is vital to understanding how to write custom gates at the pulse level.
The latter features and how they are controlled are broken into separate categories.

\subsubsection{Global Phase Synchronization}
One of the main challenges of coherent control is the ability to have {\it absolute} control over the rf phase.
However, in most systems this is difficult; it either requires several independent frequency sources that are separately mixed or a lot of manual phase bookkeeping.
Using multiple frequency sources and mixing them externally does not scale well, and undesired phase shifts from changes in amplitude and frequency complicate calibration routines.
Manual phase bookkeeping allows one to re-purpose synthesizers by changing their frequency output. 
However, this comes with additional challenges and can add computational and data overhead.

\begin{figure*}[htp]
\includegraphics{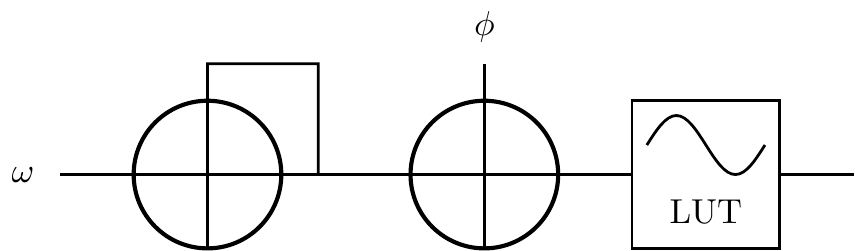}
\caption{
\label{fig:dds}
This diagram highlights the key elements for a simplified model of a DDS. The first component is a phase accumulator, which is represented as a summer that adds a frequency word, $\omega$, to the current output on each clock cycle (clock stimulus not shown). 
This is followed by a dedicated summer for changing the overall phase offset and a LUT that converts the final phase value to a digitized sinusoidal amplitude.
}
\end{figure*}

A direct digital synthesizer (DDS) can be represented as three main pieces: a phase accumulator~\footnote{The accumulator is represented as a summer with the output fed back into the second input, it should be assumed that all components are clocked}; a separate summer for shifting the phase accumulator output by a fixed phase value; and a lookup table (LUT) to convert the phase to a sinusoidal amplitude as shown in Fig.~\ref{fig:dds}. 
In the case where manual bookkeeping is used, either the external phase input, $\phi$, needs to be updated to rotate the phase accumulator output such that it is phase aligned to an earlier point in time after a frequency update, or the accumulator needs to be reset to a specific pre-calculated value.

\begin{figure*}[htp]
\includegraphics{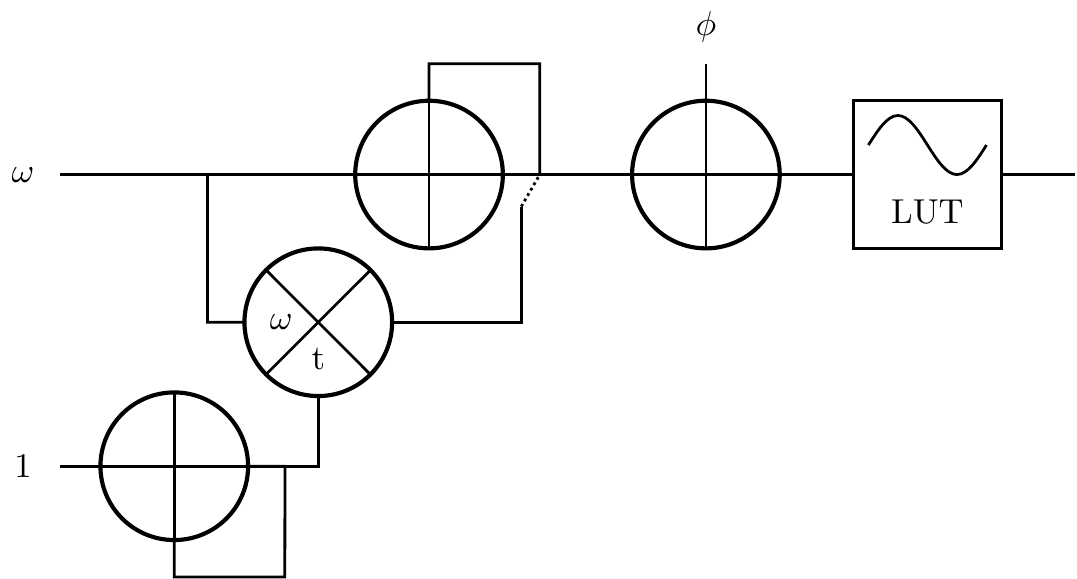}
\caption{
\label{fig:syncdds}
A modified form of the simplified DDS. A free-running counter tracks the global time, $t$, is constantly multiplied against the current frequency word, $\omega$, to calculate the global phase, $\phi' = \omega t$.
The global synchronization operation activates the switch at the output of the phase accumulator for a single clock cycle in order to overwrite the accumulated phase with the global phase.
}
\end{figure*}

In the QSCOUT Octet system, phase synchronization is handled using a separate paradigm, in which a global phase is constantly being calculated on chip as shown in Fig.~\ref{fig:syncdds}. 
In this case, a counter (represented as an accumulator with a unity frequency word) is multiplied by the DDS frequency word and updated on every clock cycle. 
This counter then tracks the global phase for any given frequency by calculating $\phi'=\omega t$ such that the global phase is zero when $t = 0$ regardless of frequency. 
Once a new frequency is applied, the accumulator can be aligned to the global phase for this frequency by overwriting the accumulator output with this global phase.
This simply requires sending a ``synchronization pulse'' that toggles the switch at the accumulator output for a single clock cycle. 
Although they are not shown in Fig.~\ref{fig:syncdds}, delay lines have been added at the appropriate places such that a simultaneous update of the frequency word and an application of a synchronization pulse will yield the global phase for the new frequency.

While each Octet board contains 16 custom DDS modules, resulting in 8 output channels that each comprise two tones, there is only a single global counter common to all DDSs. 
This counter is effectively nulled when the board is power cycled and initialized. 
Thus no assumptions should be made about the absolute value of the global counter. 
Proper usage of the synchronization mechanism requires that {\it every} pulse for which the rf frequency and phase are re-used should be applied with a synchronization operation. 
The absolute phase of the first pulse is effectively random in this case, as it depends on the value of the global counter. 
As long as the first pulse is synchronized, the phase of the Bloch vector will be set relative to the global phase. 
Subsequent pulses, for which phase synchronization is also applied, are guaranteed to be properly phase aligned not only to earlier pulses at the same frequency, but the phase alignment is identical across all tones and channels.

\begin{figure*}[htp]
\begin{center}
\includegraphics[width=4in]{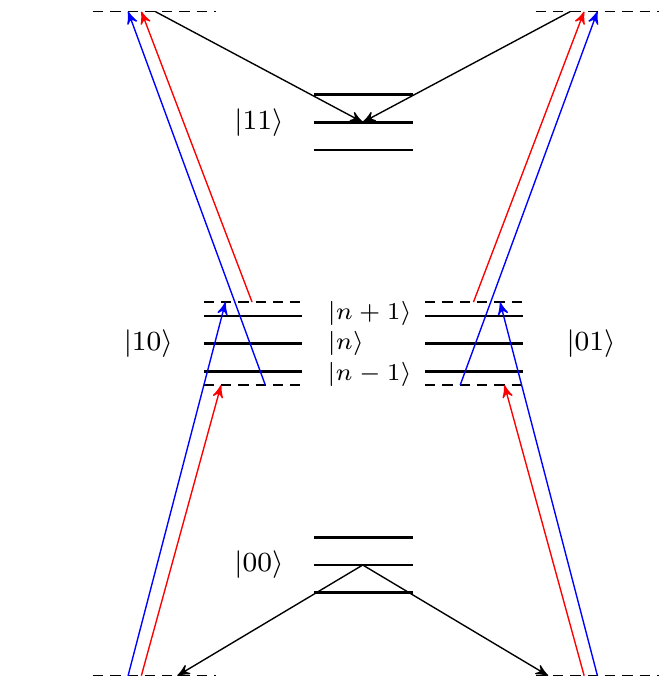}
\end{center}
\caption{
\label{fig:msgate}
Diagram of red and blue sideband transitions used in a two-qubit M\o lmer-S\o rensen gate.
}
\end{figure*}

This is particularly important for the two-qubit M\o lmer-S\o rensen (MS) gates (see Fig.~\ref{fig:msgate}), in which the red and blue sideband frequencies form a beat note.

\begin{figure*}[htp]
\begin{center}
\includegraphics[width=4in]{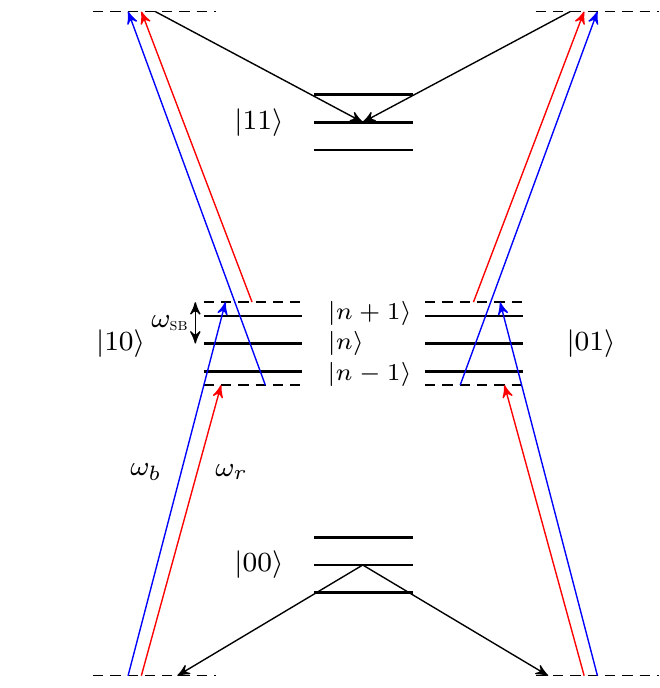}
\end{center}
\caption{
\label{fig:msgateraman}
Diagram of red and blue sideband transitions used in a two-qubit M\o lmer-S\o rensen gate realized via Raman transitions.
}
\end{figure*}

In this case, the global phase of the MS gate is set by the relationship between the phase of the beat note and the phase of the frequency resonant with the qubit transition.
Because the QSCOUT system qubit laser drives Raman transitions, any single transition is driven via a beat note given by the difference frequency of the two legs of the Raman transition.
This means the two-qubit gate picture is in reality a bit more complex, as shown in Fig.~\ref{fig:msgateraman}, and the phase of the beat note between the red and blue sidebands determines the global phase of the MS gate.

Fortunately, this imposes the requirement that the beat note formed by the sidebands needs to be phase aligned only to the relevant leg of the Raman transition.
This alignment is easily achieved by using the built-in the phase synchronization on the desired tones.
However, one of the caveats when working with any type of digital hardware is discretization effects, which can lead to subtle rounding errors because the frequency is limited to a depth of 40 bits.
The frequency word, $F$, which is sent to the hardware, is converted from a given frequency, $\nu$, by
\begin{equation}
F\left(\nu\right) = \textrm{round}(\nu/f_s\times 2^{40})
\label{eq:discrete}
\end{equation}
where $f_s \equiv 819.2\text{~MHz}$ is the effective sampling rate of the DDS.
The final value sent to the hardware is rounded to the nearest integer.

These rounding errors can play a larger role when the secondary beat note is involved.  For example, the beat frequency generated by the sideband tones is given by~\ref{eq:beatnote}.
\begin{equation*}
\begin{aligned}
f(t) &= \sin{(\omega_b t + \phi_b)} \sin{(\omega_r t + \phi_r)} \\
      &= \frac{1}{2}\left[\cos{\left((\omega_b-\omega_r\right)t + \phi_b-\phi_r)} - \cos{\left((\omega_b+\omega_r)t+\phi_b+\phi_r\right)}\right]
\end{aligned}
\label{eq:beatnote}
\end{equation*}
where $\omega_b$ ($\omega_r$) is the angular frenquency of the applied blue (red) sideband, and $\phi_b$ ($\phi_r$) is the corresponding phase.  These frequencies must obey the following condition for proper phase alignment:
\begin{equation}
2\omega_{carrier} = \omega_b + \omega_r.
\label{eq:freqRelation}
\end{equation}
We can define $\omega_b \equiv \omega_{carrier} + \omega_{SB}$ and $\omega_r \equiv \omega_{carrier} - \omega_{SB}$, and, as an example case, assign them the following values,
\begin{equation}
\begin{aligned}
\omega_{carrier} &= 2\pi\times 228732824.32571054 \text{~s}^{-1} \\
\omega_{SB} &= 2\pi\times 2235174.1793751717 \text{~s}^{-1}.
\end{aligned}
\label{eq:rawfreqs}
\end{equation}
We then use~(\ref{eq:discrete}) to convert these angular frequencies to the discretized carrier, $F_{carrier}$, red sideband, $F_{r}$, and blue sideband, $F_{b}$, frequencies, 
\begin{equation*}
\begin{aligned}
&F_{carrier} =  \textrm{round} \left(\frac{\omega_{carrier}}{2\pi f_s}\times 2^{40}\right) &=307000000000 \\
&F_{r} = \textrm{round}\left(\frac{(\omega_{carrier}-\omega_{SB})}{2\pi f_s}\times 2^{40}\right) &= 304000000000\\
&F_{b} = \textrm{round} \left(\frac{(\omega_{carrier}+\omega_{SB})}{2\pi f_s}\times 2^{40}\right) &= 310000000001\\
\end{aligned}
\label{eq:discreteunroundedfreqs}
\end{equation*}
Unfortunately, for the discretized frequencies,~(\ref{eq:freqRelation}) is no longer true and instead the two terms differ by one frequency epsilon, $f_{eps} \equiv f_s/2^{40} \approx 745~\mu\text{Hz}$.
This is because the discretization is performed {\it after} calculating the red and blue sideband frequencies ($\omega_{r}$ and $\omega_{b}$). 
While the frequency offset is small, the resulting phase then differs by $f_{eps}(t-t_0)$, where $t_0$ is arbitrary, leading to a large, unintended phase shift when applying phase synchronization.
This large error can be avoided by ensuring that~(\ref{eq:freqRelation}) is valid in the digital domain before applying phase synchronization. 

\subsubsection{Frequency Feedback}\label{hw:frequencyfb}

Drift in the cavity length of the pulsed laser that is used for driving quantum transitions is not actively corrected at the source.
Instead, we use a scheme similar to~\cite{Mount2016} to correct for frequency errors by feeding forward frequency corrections due to variations in the laser's repetition rate.
Variation in the repetition rate leads to a ``breathing'' of the frequency comb that is generated by the pulsed laser.
In order to bridge the 12.642 GHz qubit transition, the frequency difference between the Raman beams is set such that the closest {\it integer harmonic} is shifted into resonance with the qubit transition.
This means that the frequency error due to variations in the repetition rate must be amplified to account for the net frequency deviation at the target harmonic.

Details of how the repetition rate signal is monitored and converted into a useable signal for locking is described in Sec.~\ref{sec:Locking}.
This signal is passed into the RFSoC via one of the fast ADC inputs integrated on the chip.
The frequency lock involves a complex mixing stage between the ADC data and a dedicated DDS core in the firmware design.
The mixed output is sent to a PID module that feeds the error back onto the dedicated DDS to create a phase-locked loop that tracks the repetition rate.
The accumulated error is then {\it optionally} forwarded to the various output tones, and subsequently multiplied by the appropriate harmonic, depending on the specific details of the coherent operation or gate being applied.
Different locking configurations arise because the feed forward correction must add a {\it relative} shift to the tones that are used to realize the Raman transition.
In other words, one tone must be shifted by an additional amount to keep the frequency offset of the desired harmonic fixed relative to the other Raman beam.
The feed forward correction differs in sign, depending on which leg of the Raman transition the correction is being applied.

Since the correction depends on the integer harmonic and its sign, both of which are not dynamically configurable on a per-gate basis, the user need not worry about the specific details of the underlying lock settings.
It is instead more important to consider when the feed forward correction should be applied and to which tones.
There are two basic rules of thumb the user should follow.
\begin{enumerate}
\item Each Raman transition consists of two tones, where exactly one of those tones should have a feed forward correction applied.
\item The feed forward correction is applied to the higher frequency tone.
\end{enumerate}

Following the above conventions, frequency feedback for single qubit rotations is straightforward.  Whichever tone is higher in frequency receives the feedback, regardless of whether it is copropagating or counterpropagating.   
However, for the MS gate, red and blue sideband frequencies could be defined around $f_{upper}$ and thus the frequency feedback should be applied to both tones (meanwhile a counterpropagating global beam has a single lower frequency with no feedback).  
Conversely, the sidebands could be defined around $f_{lower}$, in which case $f_{upper}$ (as well as the feed forward correction) would be applied to the global beam channel.
However, other, more complicated configurations of the MS gate exist~\cite{InlekPhaseInsensitive} that impose different requirements on the lock parameters.  These can be handled by frequency locking the two Raman transitions to different harmonics of the frequency comb, which is possible using this hardware.  

\subsubsection{Frame Rotations}\label{sec:framerot}

Virtual Z-rotations are referred to as ``frame rotations'', since the functionality depends on the particular basis for the system. 
The ``frame'' nomenclature is used in the design simply for extensibility to future systems.
Frame rotations work on the same footing as the phase parameter, except that the phase values are accumulated such that all subsequent gates have an additional phase offset determined by the value in the frame rotation accumulator.
Thus, virtual Z-gates are realized by simply adding a phase offset to all subsequent gates, without the need for manual phase bookkeeping.  This phase offset can be defined within another gate or they can be a standalone operation.  
A standalone operation requires a minimum of 4 clock cycles, roughly 9.77 ns, to prevent underflows in a gate sequence.

Because the frame rotations are specific to the frame of the qubit itself, different configurations are required for how this phase is applied.
For example, the phase of a single-qubit gate is determined by a phase difference between the two legs of the Raman transition.
Applying a positive phase, $\phi$, to the tone used for $f_{upper}$ is equivalent to applying $-\phi$ to $f_{lower}$.
However, applying $\phi$ to both $f_{upper}$ and $f_{lower}$ will result in an overall phase shift of zero.

Copropagating gates must have the virtual Z-phase forwarded to the appropriate tone, and may require a change in sign.
On the other hand, the virtual Z-phase will only be correct for a M\o lmer-S\o rensen gate if the phase is forwarded to {\it both} tones.
Moreover, in alternate configurations, the sign of this phase might need to be inverted on one or both tones.
This functionality is handled at the hardware level, where the user only needs to specify which tones should have the virtual phase applied, and for which tones the phase should be inverted.

\subsubsection{Gate Sequencer}

The Octet design contains two separate modes of operation, which we refer to as {\it static} and {\it dynamic}.
These modes are not mutually exclusive, and are always running at the same time from the perspective of the DDS module. 
In other words the DDS is always taking into account separate inputs corresponding to both modes, and constantly combining the various input parameters based on their type. 
Static mode is meant for slow updates, where output channels can be set to a specific frequency, phase, and amplitude with an overall scale factor.
Static mode also includes settings for the frequency feed forward target harmonic and cross talk compensation settings (see Sec.~\ref{sec:CrossTalkCompensation}).
These settings, which are not exposed to the end user, are primarily used for other experimental operations, such as to assist with ionization when loading ions or adding slow drift corrections.

Dynamic mode is designed for running time-critical sequences of pulses that are used to realize quantum gates.
The final static and dynamic values for frequency, phase, and amplitude are simply added together, and the overall scale factor reduces the overall amplitude, each at the hardware level.
However, the user should assume that the static mode settings are all zero, except for a unity overall amplitude scaling.

Dynamic mode uses a firmware module, referred to as the ``Gate Sequencer'', that is fitted with fast LUTs for recycling pulse data and a set of spline interpolation modules (or ``spline engines'') for carrying out parameter modulation.
The spline interpolation scheme is based on a NIST design~\cite{Bowler2015} which maps the spline coefficients such that the interpolator can be modeled as a chain of accumulators.
This method uses third order polynomial B-splines, or one-dimensional cubic splines. 

The Octet uses spline engines for frequency, phase, amplitude, and frame rotations for both tones on all channels.
Because each set of coefficients is calculated for a precise number of time steps or clock cycles, each set of coefficients, as well as the number of clock cycles, is sent as a single word to the hardware.
Each word contains four 40-bit coefficients~\footnote{The 40 bit coefficient size is used since the frequency, phase, and frame rotation word size is 40 bits. However, the 16 bit amplitude words are zero-padded for uniform data size and to allow for additional precision in higher order coefficients where the least significant bits eventually accumulate and subtly affect the time-dependence of $u_0$.}, a 40-bit duration, and various other metadata bits used for controlling extra operations, such as phase synchronization, internal routing, and programming information.
For uniformity, and to match natural bus widths in the design, the total word size is 256 bits. 
It has the same format for all spline engines.

\begin{figure*}[htp]
\begin{center}
\includegraphics[width=\textwidth]{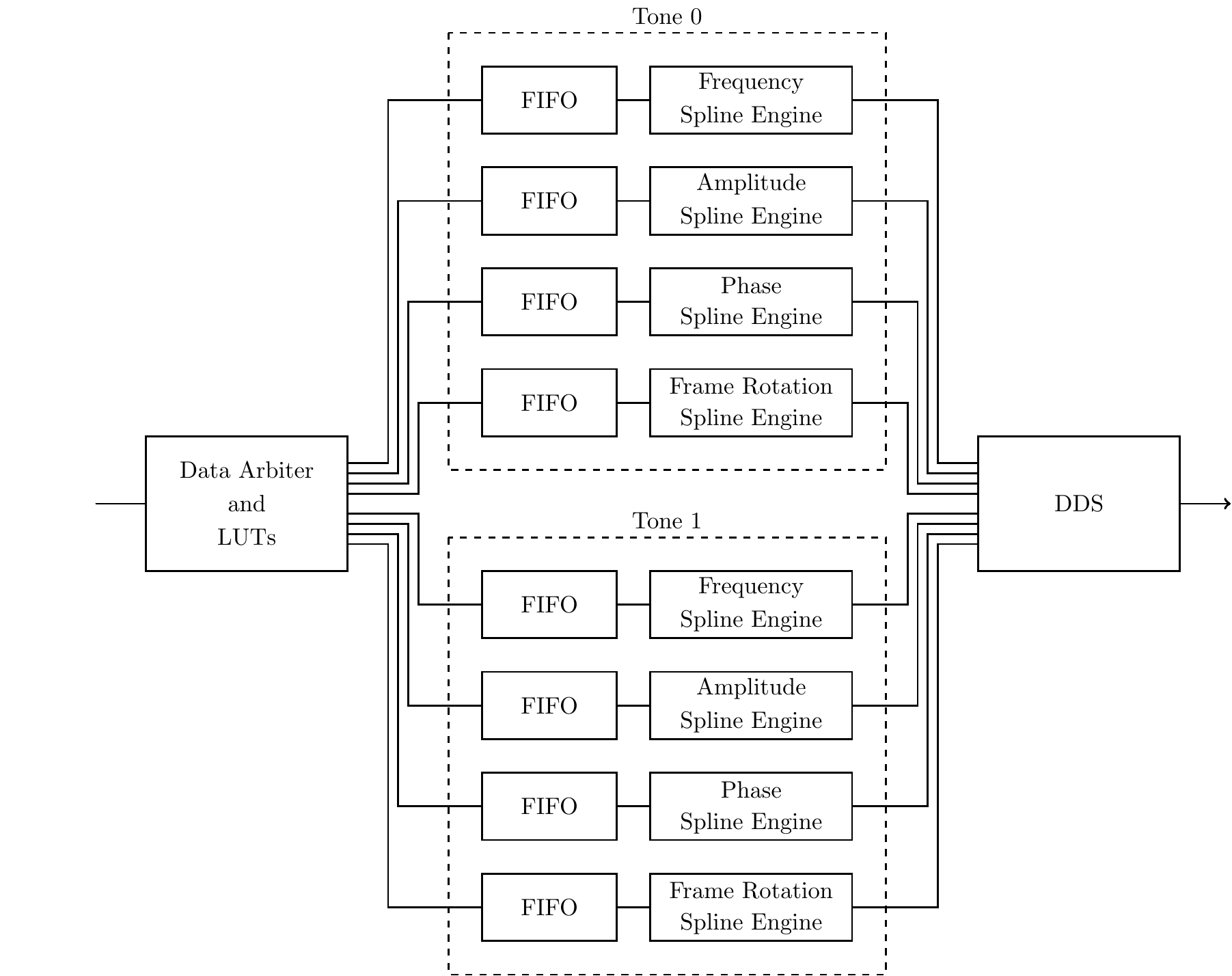}
\end{center}
\caption{\label{fig:gateseq}
Gate sequencer spline engine layout. Each parameter has a dedicated spline engine, each of which is fed by a 256 deep FIFO. Incoming data for the particular channel is routed through a data arbitration module and a series of LUTs for recycling data.
}
\end{figure*}

Each output channel has a dedicated gate sequencer module, each of which contains a data arbitration module, fast lookup tables (LUTs), and 8 spline engines with first-in first-out (FIFO) buffers, as shown in Fig.~\ref{fig:gateseq}.

Incoming data fills the FIFOs until the gate sequence is either exhausted, or the FIFOs are completely filled and present a blocking condition.
An external trigger simultaneously enables all spline engines, which will consume data until the FIFOs are empty or a {\it wait for trigger} flag is encountered in the metadata, at which point the spline engines will wait for another external trigger.

Because each data word carries its own timing information, each spline engine can consume data at different rates and runs independently from other spline engines.
As a result, the number of spline knots per parameter for a given pulse need not match. 
Rather the total sum of the duration arguments for each parameter should match at the gate level or, in the most extreme case, match the total elapsed timing between {\it wait for trigger} flags.
Asymmetry between number of spline points mainly imposes additional requirements on the order in which the data is sent. 
This is done in such a way that FIFO blocking for one parameter does not starve another FIFO.
This situation is avoided by properly interleaving data based on parameter type and time-sorting data.
Each output channel has a dedicated gate sequencer, where each gate sequencer is fed from a common arbiter. 
It has a structure almost identical to Fig.~\ref{fig:gateseq}, except that the spline engines would be replaced with gate sequencers, and each gate sequencer feeds a separate DDS.
Ultimately, 8 channels, each with 8 parameters, have to be run simultaneously.
They are, however, fed serially.
Thus, blocking conditions can arise from parameter FIFOs {\it and} channel FIFOs. 
The data sorting must also take into account ordering of all 64 parameter FIFOs across channels.

The overall model requires that {\it all} spline engines are consuming data during a circuit, even if the data is equivalent to a NOP (no operation), so that data is properly aligned at a later time when the value is potentially non-zero.  Specific channels and parameters can in fact be disabled to reduce overall data for smaller circuits and calibration routines, but this is generally not needed or used.
Ignoring potential optimizations, at least one word for each parameter must be provided to describe a gate, which is a minimum of 2 kib of data per gate.
This amount of data can add up for long data sets where latency between circuits needs to be minimal.
For an exhaustive protocol such as gate set tomography~\cite{nielsen2020gate}, the total number of gates used can easily be on the order of $10^6$, and with experimental averages and sideband cooling taken into account, the total amount of data associated with even the most simple gates can easily exceed 1 GB.
However, since the same basic set of gates is used, streaming the full data for each gate adds a lot of unnecessary overhead.

\begin{figure*}[htp]
\begin{center}
\includegraphics[width=\textwidth]{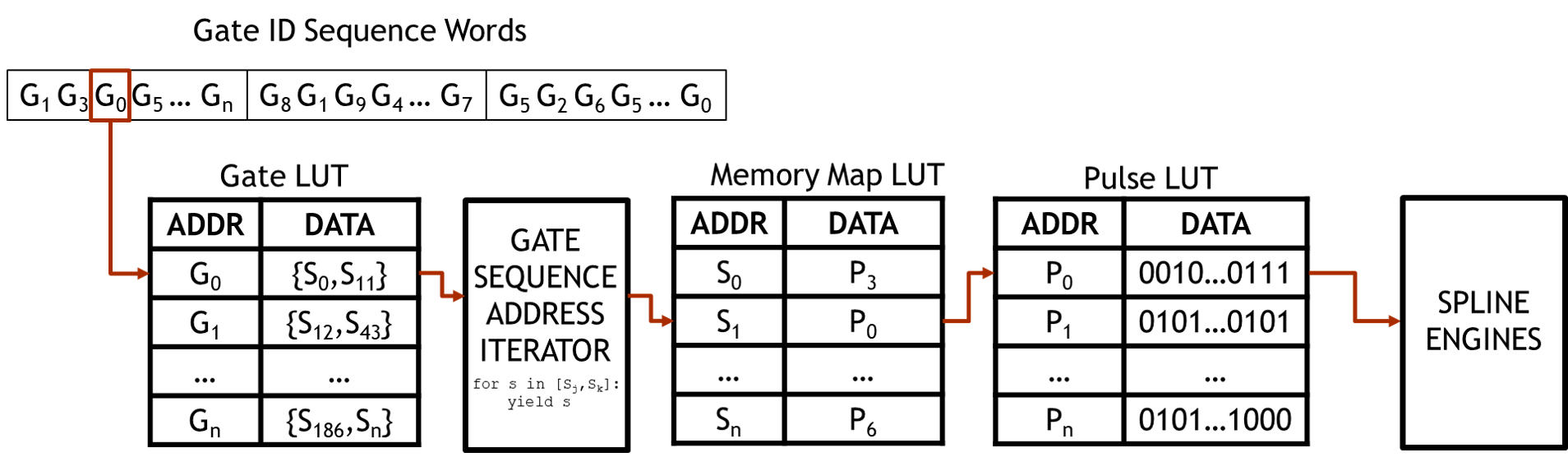}
\end{center}
\caption{\label{fig:gateseqLUTs}
Gate sequencer LUT layout.
Incoming words are, depending on metadata, optionally routed through a cascade of LUTs for reading out the associated pulse data on all parameters and subsequently routed to the correct spline engine input FIFO.
Gate IDs can be densely packed in single words to improve overall throughput.
These IDs are passed through an initial table to determine the memory bounds that need to be iterated over in a secondary LUT, which provides the final address associated with the final raw pulse data LUT.
}
\end{figure*}

Instead, the gate sequencer implements a series of LUTs (Fig.~\ref{fig:gateseqLUTs}) to store relevant pulse information for various gates.

The ``Pulse LUT'' (PLUT) contains raw data words, such as those used in the streaming case, that need to be forwarded to a particular spline engine FIFO.
However, a particular gate may contain a large number of PLUT entries, which all need to be sent out to their respective FIFOs.
Moreover, two gates might have common PLUT entries, and to reduce overall memory usage all PLUT entries are generally unique.
For example, a simple square pulse X-gate will contain 8 data words, but the equivalent type of Y-gate will be identical with the exception of a single phase word, thus a total of 9 PLUT entries is needed to describe both gates.
Because PLUT entries are unique, and often shared among different gates, a secondary ``Memory Map LUT'' (MLUT) is used to create dense arrays of pointers in a linear address space.
In other words, stepping through a particular set of MLUT addresses will return a series of address words of the PLUT entries needed to describe a particular gate.
Gates are then encoded using a set of MLUT address boundaries, packed into a single data word and stored in the ``Gate LUT'' (GLUT), the output of which is connected to an iterator module for reading out a sequence of MLUT addresses.
The GLUT address size is currently set to 6 bits, but it can easily be reconfigured in firmware, and sets the ultimate compression ratio for the data and an upper bound on the number of unique gates that can be stored locally.
Data still must be sent in 256-bit words and, because of metadata constraints, the number of 6-bit GLUT address words, or ``gate IDs'', that can be packed into a 256-bit word is 36.
This gives a compression ratio of $\approx 0.35\%$, neglecting the size of the data used to initially set the LUTs.

While it may seem that the additional overhead for programming the LUTs may, in some cases, encumber the data flow because of additional words needed to program the LUTs, the LUT approach can artificially increase FIFO depths and offers more flexibility.
Because programming data and sequence data can be packed into single 256-bit transfers, the total number of data words to program and stream a simple square pulse gate is $8 + 1 + 1 + 1 = 11$, where the GLUT, MLUT, and gate ID sequence require one word each.
However, there is no requirement that a gate entry incorporate all parameters for a pulse.
For example, a frequency modulated (FM) gate could be defined with a fixed amplitude, phase, frame rotation, and tone 0 frequency. 
The tone 1 frequencies could then be streamed directly after passing in the appropriate gate ID.
Additional gains come into play for large amounts of data, where LUTs are programmed at roughly the same update rate of the spline interpolation~\footnote{The LUTs operate on the 409.6 MHz clock domain, and data words are consumed at different rates depending on the type of programming data and the density of data packed into each programming word.}.
Hence, active use of the LUTs might have some additional data, but locally storing sequences in the LUTs can effectively expand the FIFO depth and offers additional gains if data can be partially recycled.
Another advantage of this method deals with the fact that data has to cross clock domains via a direct memory access (DMA) transfer, going from 300 MHz to 409.6 MHz.
Since the data is being transfered serially to the 300 MHz clock-domain-crossing FIFOs for 8 different channels, the effective frequency for feeding raw data on a single channel is $(300 \text{~MHz})/8 = 37.5 \text{~MHz}$.
Because the other channels need to be padded with NOPs, this leads to a total time of $\approx 213 \text{~ns}$ to stream in a single gate.
Minimizing FIFO filling on the 300 MHz domain, and maximizing the amount of data on the 409.6 MHz domain, will reduce potential data underflow conditions associated with fast spline operations or short gates.

The maximum LUT size is determined by constraints in fabric, and is dominated by the PLUT, which contains $2^{10}$ entries per channel, while the MLUT contains $2^{12}$ entries, and the GLUT contains $2^6$. 
Exceeding allocated LUT memory, such as when gates are defined by a large number of unique spline knots, requires either using a combination of directly streamed data with stored data or partially reprogramming of the LUTs between gates.
The latter has certain advantages owing to the clock domain crossings, where the buffer FIFOs (Fig.~\ref{fig:gateseqLUTs}) before the spline engines can store up to 256 words each. 

\subsubsection{Crosstalk Compensation }\label{sec:CrossTalkCompensation}
\label{CrosstalkComp}

Crosstalk effects induce undesirable rotations on nearest or next-nearest-neighbor qubits, and they can arise as a result of several possible sources.
Optical crosstalk can arise from larger individual addressing beam waists in which a small proportion of stray light is incident on the ions.
Other sources of crosstalk are due to unintended driving of neighboring AOM channels, either from electrical crosstalk between transducers or sympathetic vibrations between crystals.
Compensating for these effects requires applying cancellation tones on nearest- and next-nearest-neighbor channels.
Cancellation tones need to match the waveforms applied to the neighboring channels, but with a reduced amplitude and, for electrical or acoustic crosstalk, a delay.

To prevent frequency- and amplitude-dependent phase shifts associated with external rf components, crosstalk signals are added digitally in firmware.
Coarse delays are thus resolution limited by the 409.6 MHz clock used to transfer data.
However, fine tuning the delay can be approximated by applying an overall phase shift to the neighboring output before it is added onto the target channel's waveform.
Multiplying the waveform data in the complex domain gives full control over relative phase and amplitude.

Crosstalk compensation {\it could} be implemented to support infinite feedback, where the output of one DDS channel is added to its neighbor and the modified output from the neighbor is added back in to the original channel {\it ad infinitum}.
This implementation has some advantages, in the sense that it can be used to compensate crosstalk compensation signals themselves.
In other words, the compensation signal applied to channel 5 that accounts for the output on channel 7 can be further echoed to channel 3, which is compensating for the output on channel 5.
The downside of this approach is that the implicit delay is bounded by the latency of the digital signal processing (DSP) elements in the design that are used to scale and add signals from neighboring channels.
Instead, the initial version of the Octet design only applies signals from the {\it uncompensated} outputs onto neighboring channels. 
The uncompensated signals are passed into a delay line to match the latency of the DSP elements that are used to scale and add the compensation signals coming from neighboring channels.
This allows for optical crosstalk compensation, where the compensation signals must be perfectly synchronous with the original signal.

\section{Performance} \label{Results}
The previous sections outlined the hardware and design decisions that went into QSCOUT.  
Here we discuss some of the experiments where we tested the performance of the overall system. 
 
\subsection{Coherence Times}\label{sec:coherencetimes}
Ions are generally known for their long coherence times, and the QSCOUT system is no exception.  
Coherence times are typically limited by such factors as the reference clock, external magnetic fields, and noise (from vibrations and power supplies). 
To reduce the impacts of these factors, this experiment uses an ultra-stable Cesium clock CsIII Model 4310B from Microchip with a TSC 4145C OP01 quartz ultra clean-up oscillator  to provide a reference 10~MHz to all of our hardware.
The specified drift is 3e-13 from 1 to 100 seconds~\cite{MicroSemi1, MicroSemi2}, which allows our microwave coherence times to be long enough that they are heating limited on our current trap devices. 
Permanent SmCo magnets in a 3D printed ring mounted to the chamber produce the external magnetic field. 
These produce a field strength of approximately 4.37~G at the ions that is fairly insensitive to changes in temperature at \mbox{-0.04\% /K}~\cite{StanfordMagnets}. 
The optical table is floated to reduce the impact of vibrations from equipment and our sensitive equipment is run from non-switching power supplies. 

\begin{figure*}
 \includegraphics[width=0.7\textwidth]{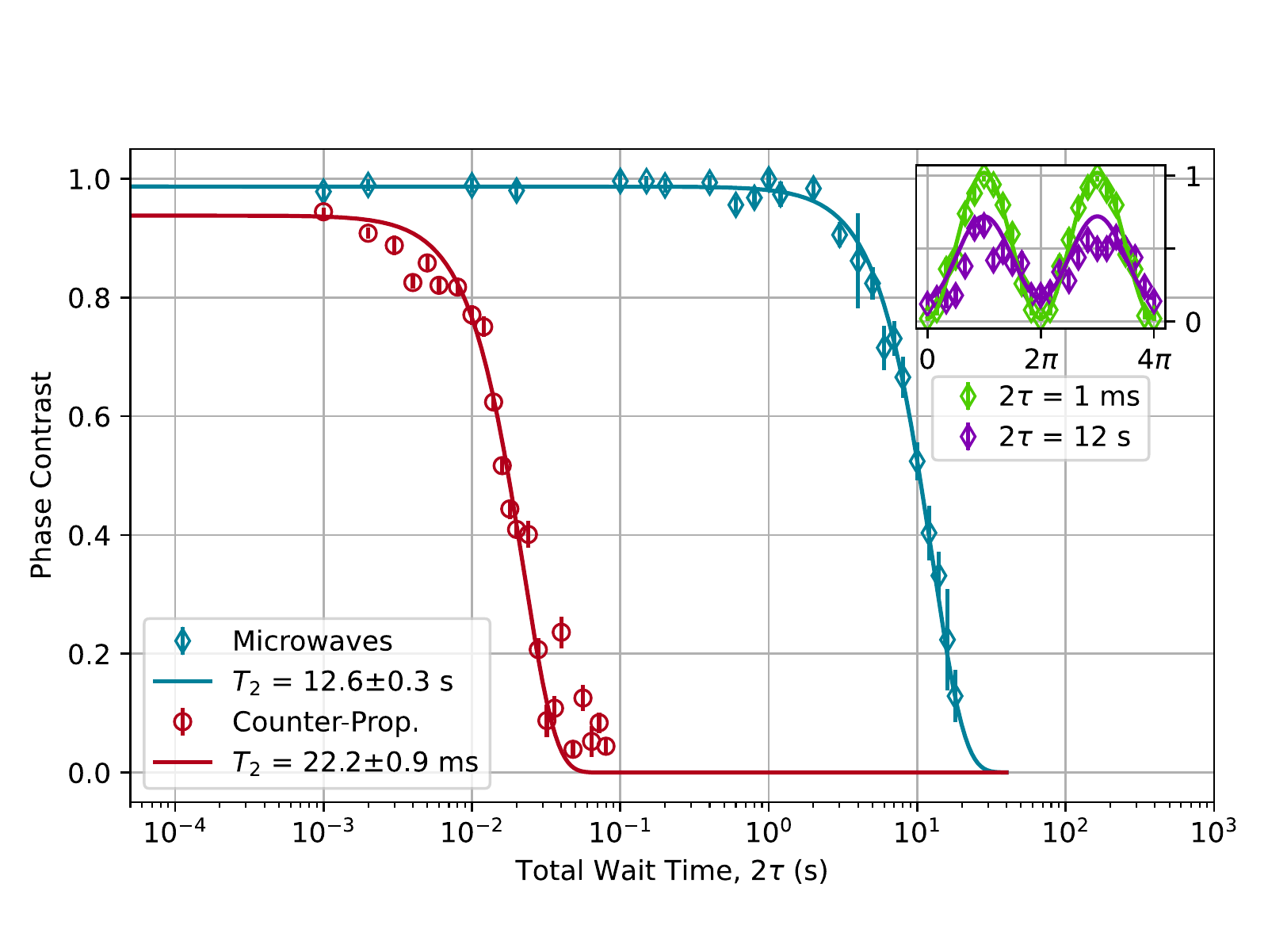}
 \caption{$T_2$ coherence measurements taken via a Hahn echo with microwaves reveals a T$_2$-time in excess of 10 s (blue). When the measurement is performed with counter-propagating pulses, the measurement is sensitive to motion and heating, and results in a heating-limited T$_2$ time of 22 ms (red).  Inset: Phase contrast measurements for two points on the microwave coherence measurement, total wait times of 1 ms (green) and 12 s (purple). The non-symmetric decay at 12 s indicates heating is a limiting factor to the coherence. \label{fig:coherences}}
 \end{figure*}

To characterize the coherence time, for each configuration we performed a Hahn Echo experiment to estimate the T$_{2}$~\cite{Hahn1950}.
We consecutively applied a $\nicefrac{\pi}{2}\,$-pulse, wait time, $\pi$-pulse, wait time, and a final $\nicefrac{\pi}{2}\,$-pulse of varying phase.
The final pulse is varied from 0 to 4$\pi$, and we fit the ion state projection to a sinusoid and extract the phase contrast (inset of Fig.~\ref{fig:coherences}). We increase the wait time and fit to a Gaussian decay revealing our T$_2$ coherence times, as shown in Fig.~\ref{fig:coherences}. As the wait time is increased, the sinusoid contrast should collapse symmetrically about 0.5 probability when coherence is lost. 
Cases where that decay is not symmetric, and the bright states appear to decay faster, may indicate heating as a coherence limiter.  
This in particular will be visible when measuring the coherence in a counter-propagating configuration, but can also be seen in the microwave measurements (inset of Fig.~\ref{fig:coherences}).

\subsubsection{Microwaves}\label{sec:uwaves}

Coherent microwaves can be used to drive the qubit frequency directly. 
To deliver the microwaves to the ion, we use an external microwave horn (Pasternack PE9855/SF-10). 
This horn is mounted externally to the chamber and the alignment is optimized to achieve the highest Rabi frequency. 
Typical microwave Rabi oscillations are shown in Fig.~\ref{fig:rabi}a.

The microwave frequency is generated by single-sideband modulation of a 12.6~GHz oscillator by an approximately 42.8~MHz signal from the RFSoC (see Sec.~\ref{ControlHW}), which allows for full frequency, phase and amplitude control.
The detailed description of the microwave frequency generation can be found in reference~\cite{Blume2017}.

\subsubsection{Laser Gates}\label{sec:lasergates} 

\begin{figure*}
	\centering
	\includegraphics[width=0.7\textwidth]{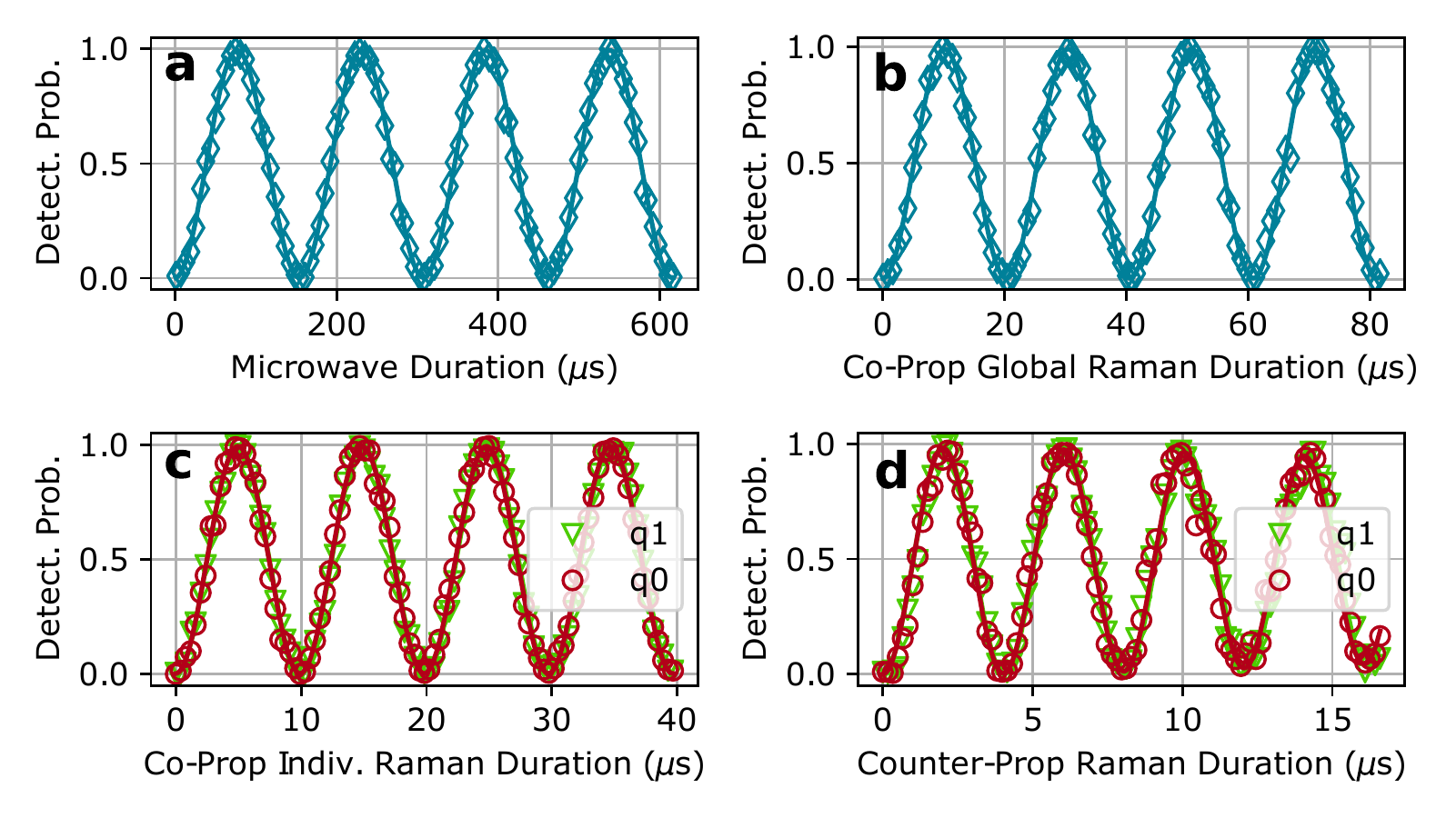}
	\caption{Typical Rabi oscillations using a) the microwave horn with a typical $\pi$-time of 80 $\mu$s, b) the 355 Raman laser in co-propagating global configuration with a typical $\pi$-time of 10 $\mu$s, c) the 355 Raman laser in co-propagating individual beam configuration with a typical $\pi$-time of 5 $\mu$s, and d) the 355 Raman laser in counter-propagating configuration with a typical $\pi$-time of 2 $\mu$s. Note, the oscillations in c) and d) respectively are performed on two ions simultaneously and detection is distinguished in their respective fiber cores. }
	\label{fig:rabi}
\end{figure*}

Our method for using a pulsed laser to drive Raman transitions for qubit manipulation is described in Sec.~\ref{355laser}. 
Just as in the case of the microwaves, the phase, frequency, and amplitude of an rf tone applied to an AOM is controlled by the RFSoC. 
Examples of Rabi oscillations from the 355 laser are shown in Fig.~\ref{fig:rabi}.
Typical $\pi$-times for the Raman transitions are 10 $\mu$s, 5 $\mu$s, and 3 $\mu$s for the co-propagating global beam, the co-propagating individual beams, and the counter-propagating gates respectively.
	
\subsection{Gate Fidelities}\label{sec:GST}
Gate fidelities are determined by using Gate Set Tomography (GST).
This is a characterization method developed at Sandia which gives a full tomographic description of gates, by performing pulse sequences to efficiently determine errors.  For a full explination of GST, see reference~\cite{Blume2017}.
For a single ion, GST probes $G_{I}$ (the identity gate), $G_{X}$ ( a $\nicefrac{\pi}{2}$ $X$ rotation), $G_{Y}$ ( a $\nicefrac{\pi}{2}$ $Y$ rotation).
The implementation of these gates was described in Sec.s~\ref{sec:uwaves} and~\ref{sec:lasergates}.

Additionally, we implemented GST using BB1 gates for our $G_{X}$ and $G_{Y}$ rotations~\cite{Shappert2013, Wimperis1994}.
These are a sequence of gates(time, phase) as follows: $G(\nicefrac{\pi}{2},\theta_{0}) - G(\pi,\theta_{1}) - G(2\pi,\theta_{2}) - G(\nicefrac{\pi}{2},\theta_{1})$.
This gate sequence corrects for pulse length errors, but the precise phase of the compensated pulse depends on the gate being applied. 
We also use a modified idle gate which has a gate time similar to the BB1 compensated gates. 
The compensated idle gate is a sequence of $G_{X}(2\pi,0) - G_{Y}(2\pi,90) - G_{X}(2\pi,180) - G_{Y}(2\pi,270)$.

The results from our gates are compared to an over-complete basis set from a ``black box'' model to determine the gate errors. 
As a result, we get a wealth of information, including gate decompositions, gate error generators, and other traditional fidelity metrics, which can be used to determine the sources of some gate infidelity. 
A summary of our typical GST result showing gate fidelity and rotations is in table~\ref{tab:GST}.

 \begin{table*}[hbt!]
 \caption{Single Qubit Gate Fidelities Determined by Gate Set Tomography}
% \begin{ruledtabular}
 \begin{tabular}{ L{0.38\textwidth}  C{0.14\textwidth} C{0.14\textwidth} C{0.14\textwidth} C{0.14\textwidth} }
\textbf{Parameter}	 						& 	\textbf{Rotation Angle ($\degree$)} 	&   \textbf{Infidelity} 	& \textbf{$\frac{1}{2}$-Trace Distance} 	& \textbf{Sequence Length} \\
\specialrule{.15em}{.2em}{.2em}
Microwave gates 						& 						& 			& 			& 		 	\\
\hspace{0.5cm}Bare gates 				& 	0.50353$\pi$		&  6.86e-4	&  6.49e-3	& 2048	 	\\
\hspace{0.5cm}BB1 gates					&	0.49997$\pi$		&  3.90e-5	&  1.78e-4	& 128		\\
\hline
Global Raman Beam						& 						& 			& 			& 		 	\\
\hspace{0.5cm}Bare gate					&	0.48989$\pi$ 		&  2.11e-3	&  1.61e-2	& 1024		\\
\hspace{0.5cm}BB1 gate					&	0.49713$\pi$		&  6.56e-4	&  4.69e-3	& 128		\\
\hline
Individual Coprop. Raman Beam 			& 						& 			& 			& 		 	\\
\hspace{0.5cm}Bare gates				&  	0.50749$\pi$		&  4.10e-3	&  1.24e-2 	& 1024		\\
\hspace{0.5cm}BB1 gates					&  	0.50015$\pi$	 	&  1.23e-3 	&  1.70e-3 	& 512		\\
 \end{tabular}
% \end{ruledtabular}
 \label{tab:GST}
 \end{table*}

\subsection{RF Stability}

\begin{figure*}
 \includegraphics[width=0.7\textwidth]{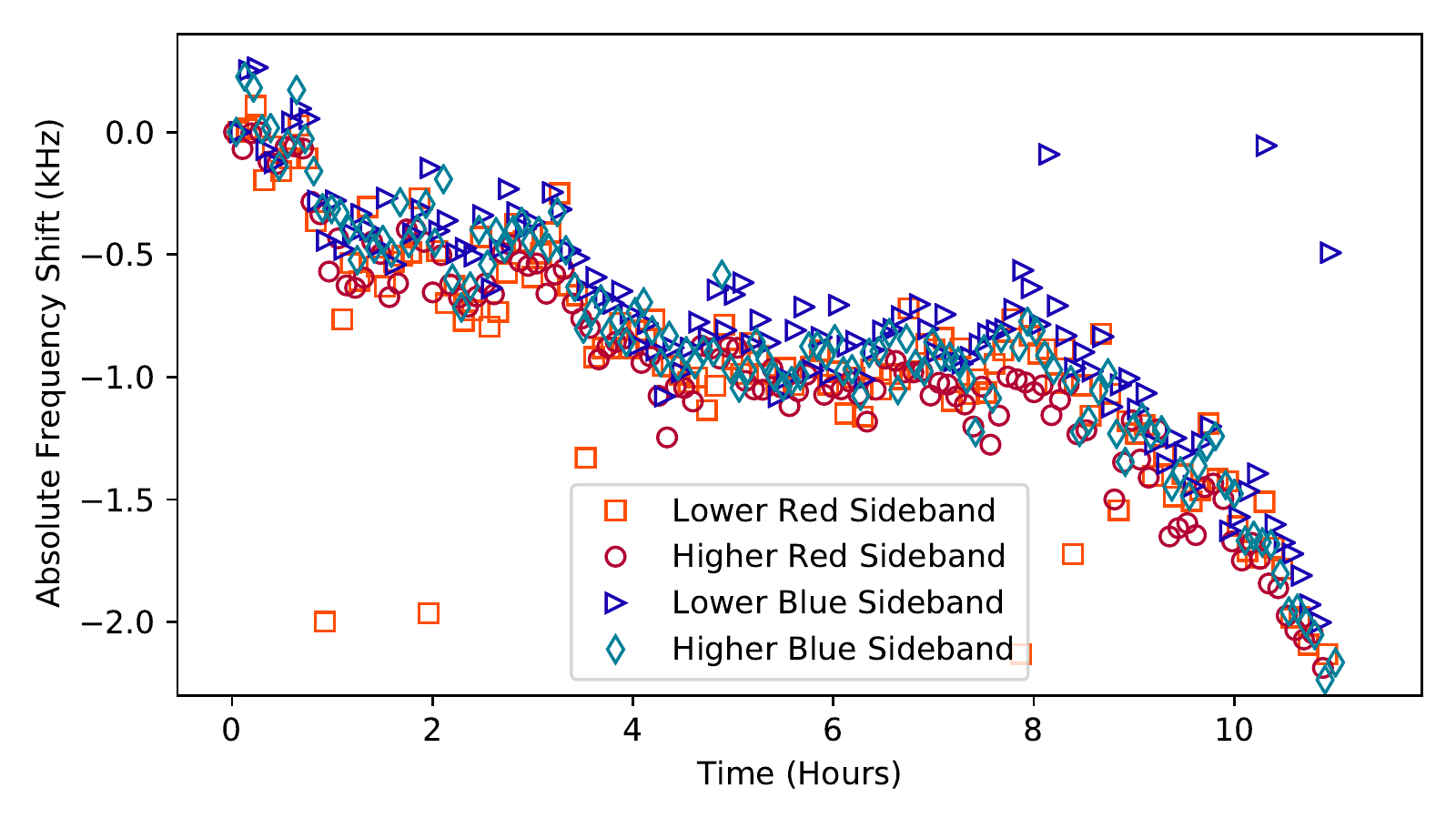}
 \caption{Measurements of radial secular frequencies for a single ion over the course of 10+ hours. 
 Both the red and blue sideband of the horizontal and vertical radial modes are tracked. 
 The bandwidth suggests a short-term stability of 200-300~Hz, and a long-term drift of a few kHz. 
 These drifts are small enough to achieve our gate fidelity goals.  
 \label{RFStability}}
 \end{figure*}

To achieve consecutive high fidelity gates, the motional modes of the ion need to be consistent from one gate to the next. 
Even small variations in the rf power can lead to destructive changes in the mode frequencies. 
We track the stability of our radial modes over time to determine the stability of the system (Fig.~\ref{RFStability}).
Short-term variation ($\approx$ 1~hour) is on the order of 200-300~Hz. 
The long term drift (several hours to days) is typically a few kHz. 
These drifts are within the range necessary to achieve our experimental goals. 

 \subsection{Crosstalk Compensation}
 
\begin{figure*}
	\centering
	\includegraphics[width=0.7\textwidth]{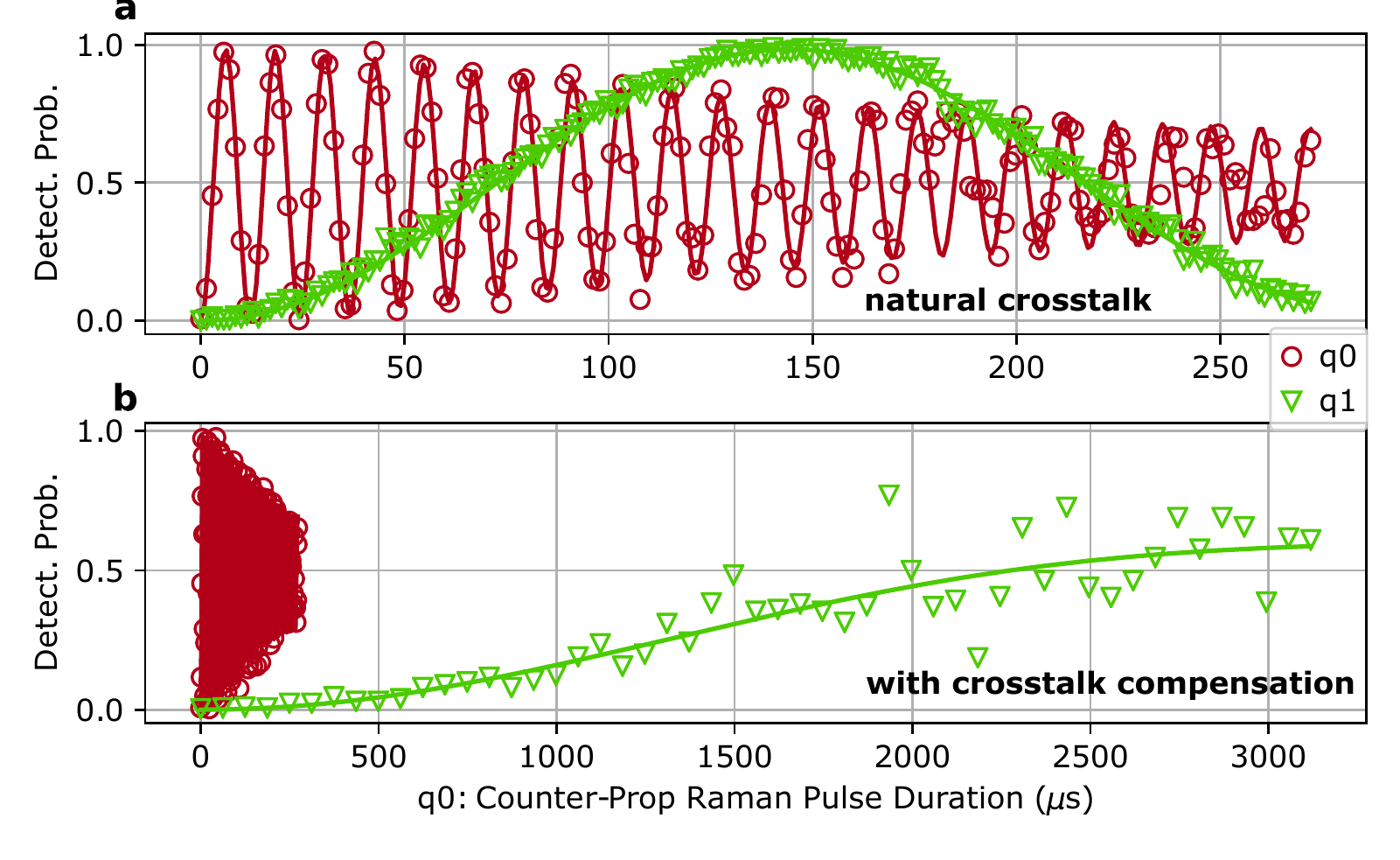}
	\caption{a) Rabi oscillations driven on qubit 0 (red) and corresponding natural crosstalk-driven Rabi oscillations seen on qubit 1 (green). b) After compensation through the Octet hardware (original signal with an amplitude of 0.034 and a phase shift of 68$\degree$, the resultant compensated crosstalk-driven Rabi oscillations seen on qubit 1 are now 0.17\% of the original signal on qubit 0.}
	\label{fig:crosstalkcomp}
\end{figure*}

 As discussed in Sec.s~\ref{sec:355xtalk} and~\ref{sec:CrossTalkCompensation}, we have the ability to compensate optical crosstalk from a gate being driven on one ion onto its nearest neighbor and next nearest neighbor ion. As a demonstration, we apply a counter-propagating pulse of varying duration on one ion, qubit 0, and measure its effect on its nearest neighbor, qubit 1, sitting 4.5 $\mu$m away. For example, as shown in Fig.~\ref{fig:crosstalkcomp}, the power necessary to drive a $\approx~6$ $\mu$s $\pi$-pulse on qubit 0, induces an  $\approx~140$ $\mu$s $\pi$-pulse on qubit 1. When we apply a crosstalk compensation on qubit 1, with an amplitude of 0.034 of the original signal on qubit 0 phase-shifted by 68$\degree$ (found empirically), we measure a resultant $\pi$-time of 3.6 ms, or 0.17\% crosstalk after compensation.
 
\subsection{Two-Qubit Gate}\label{sec:MSgate}

\begin{figure*}
	\centering
	\includegraphics[width=0.7\textwidth]{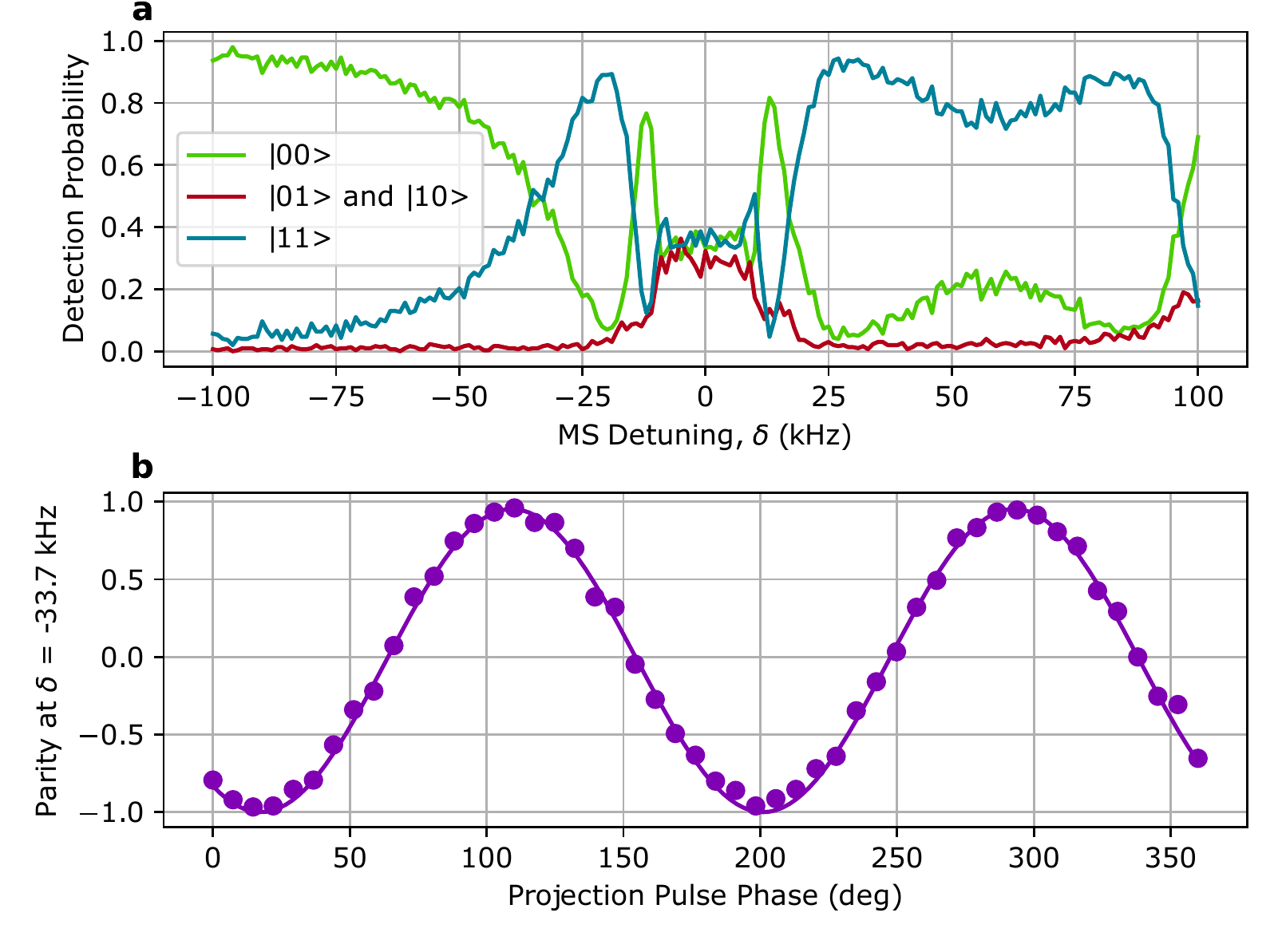}
	\caption{A M\o{}lmer-S\o{}rensen two-qubit gate is performed between two ions. a) For a pulse duration of 200 $\mu$s, a scan of the detuning from the red and blue sidebands with the populations is shown. The crossing at -33.7~kHz generates the entangled state, \ket{00} + \ket{11}. b) For a pulse duration of 200 $\mu$s and a detuning of -33.7 kHz, single-qubit projection pulses are then performed and their phase is swept. The parity is calculated from the populations of the different configurations, P$_{00}$ + P$_{11}$ - P$_{01}$ - P$_{10}$. An estimated parity of 0.955 suggests an entanglement fidelity of 0.978.}
	\label{fig:MS}
\end{figure*}

Two-qubit entangling gates are performed using a M\o{}lmer-S\o{}rensen gate~\cite{Molmer1999} and shown in Fig.~\ref{fig:MS}.  To verify the state is a Bell state~\cite{Sych2009}, we follow the M\o{}lmer-S\o{}rensen gate with a $\pi /2$-pulse of varying phase and measure the state parity  P$_{00}$ + P$_{11}$ - P$_{01}$ - P$_{10}$.  
We shape the pulses for the M\o{}lmer-S\o{}rensen gate to have a Gaussian envelope from the RFSoC, which empirically provides a higher fidelity than square pulses.  

\section{Summary}\label{conclusion}
In summary, this document has outlined the major decisions and implementations to build a quantum processor of up to 32 qubits based on trapped-ions.  It has created a blueprint for the current QSCOUT machine, with an emphasis on new technologies enabling moving from one or two qubits to many.  The details included are to assist QSCOUT users and potential users in evaluating the system for their algorithms, help them tailor programs for best chance of success, and help outside experimental trapped-ion groups build their own systems.  

\section{Acknowledgment}
We would like to thank Marko Cetina, Alan Bell, Ken Brown, Jungsang Kim, and Bert Tise for many useful discussions.  This material was funded by the U.S. Department of Energy, Office of Science, Office of Advanced Scientific Computing Research Quantum Testbed Program and by Sandia National Laboratories' Laboratory Directed Research and Development Program.  Sandia National Laboratories is a multimission laboratory managed and operated by National Technology \& Engineering Solutions of Sandia, LLC, a wholly owned subsidiary of Honeywell International Inc., for the U.S. Department of Energy’s National Nuclear Security Administration under contract DE-NA0003525.  This paper describes objective technical results and analysis. Any subjective views or opinions that might be expressed in the paper do not necessarily represent the views of the U.S. Department of Energy or the United States Government.

\section{QSCOUT Specifications Summary}
\label{App:Specs}
The below tables list general settings, parameters, infidelities, and approximate gate duration.  
 \begin{table*}[hbt!]%[H] add [H] placement to break table across pages
 \caption{General Settings }
 %\begin{ruledtabular}{0.6\texwidth}
 \begin{tabular}{  p{0.4\textwidth} C{0.3\textwidth}  }
 \hline\hline
\textbf{Setting}	 						& 	\textbf{Value} 	\ \\
\specialrule{.15em}{.2em}{.2em}
Rf drive 								&	46.888 MHz	\\
Magnetic Field							&	4.37 G		\\
Doppler Detuning						&	-13 MHz			\\
Doppler Cooling Duration				&		1 ms		\\
Detection Time							&	350 $\mu$s			\\
Optical Pumping Time 					&	10 $\mu$s			\\
Sideband Cooling Duration	&	1.7 ms 		\\
\hline \hline
 \end{tabular}
% \end{ruledtabular}
 \label{tab:Exptparams}
 \end{table*}

 \begin{table*}[hbt!]%[H] add [H] placement to break table across pages
 \caption{General Parameters }
 %\begin{ruledtabular}
 \begin{tabular}{ L{0.4\textwidth} C{0.28\textwidth} C{0.28\textwidth} }
 \hline\hline
\textbf{Parameter}	 							& 	\textbf{Value} 			& 	\textbf{Reference Section} 								\\
\specialrule{.15em}{.2em}{.2em}
Collision Rate									&	3.52 $\pm$ 0.44 mHz	 	& Sec.~\ref{Sec:VacuumChar}		 	\\ \hline
Coherence Times									&							& Sec.~\ref{sec:coherencetimes}		\\			
\hspace{0.8cm}Microwaves 						& 	12.6 s		& 								\\
\hspace{0.8cm}Global Raman Beam					&	11.5 s		& 		  						\\
\hspace{0.8cm}Individual Raman Beam 			&   7.0 s		&   		 					\\
\hspace{0.8cm}Counterprop. Raman Beam 			&	22 ms			&		 							\\
\hline\hline
 \end{tabular}
% \end{ruledtabular}
 \label{tab:generalSpecs}
 \end{table*}

 \begin{table*}[hbt!]%[H] add [H] placement to break table across pages
 \caption{Infidelities and Errors (Single Qubit, unless stated otherwise)}
% \begin{ruledtabular}
 \begin{tabular}{ L{0.4\textwidth} C{0.28\textwidth} C{0.28\textwidth} }
\textbf{Parameter}	 							& 	\textbf{Error} 	& 	\textbf{Reference} 				\\
\specialrule{.15em}{.2em}{.2em}
SPAM  											&	7e-3	 (need to check)				& 		 						\\
\hline
Crosstalk										&						&								\\
\hspace{0.8cm}Detection  						&	7e-4				& Sec.~\ref{subsec:crosstalk}	\\
\hspace{0.8cm}Coprop. Raman Beams				&	3e-4					& Sec.~\ref{sec:355xtalk}		\\
\hspace{0.8cm}Counterprop. Raman Beams 			&	4e-2					& Sec.~\ref{sec:355xtalk}		\\
\hline
Idle Gate 										&	7e-2					& 								\\
Idle Gate (Compensated)							&	8e-4					& 								\\
\hline	
Microwave Gate, bare 							& 	6.86e-4				& Sec.~\ref{sec:GST} 			\\
Microwave Gate, BB1  							&	3.9e-5				& Sec.~\ref{sec:GST}			\\
\hline
Raman Beam Single Qubit Gates  					&						&		\\
\hspace{0.8cm}Global, bare 						&	2.11e-3			 	& Sec.~\ref{sec:GST} 			\\
\hspace{0.8cm}Global, BB1 						&	6.56e-4			 	& Sec.~\ref{sec:GST} 			\\
\hspace{0.8cm}Individual Coprop., bare gates	&  	4.10e-3			 	& Sec.~\ref{sec:GST} 			\\
\hspace{0.8cm}Individual Coprop., BB1 gates		&   1.23e-3				& Sec.~\ref{sec:GST} 			\\
\hspace{0.8cm}Counterprop., bare gates	 		&	not measured					&		\\
\hspace{0.8cm}Counterprop., BB1		 			&	not measured				&		\\
\hspace{0.8cm}Z gate 							&	$360^{\textrm{o}}/2^{40}$ resolution 	&		\\
\hline
Two-Qubit MS gate 								&	2e-2				&	Sec.~\ref{sec:MSgate}	\\
Two-Qubit MS gate, Walsh 						&	2e-2					&	Sec.~\ref{sec:MSgate}	\\
 \end{tabular}
% \end{ruledtabular}
 \label{tab:Errors}
 \end{table*}

 \begin{table*}[hbt!]%[H] add [H] placement to break table across pages
 \caption{Approximate Gate Duration}
% \begin{ruledtabular}
 \begin{tabular}{ L{0.4\textwidth} C{0.28\textwidth} C{0.28\textwidth} }
\textbf{Parameter}	 						& 	\textbf{Time} 	& 	\textbf{Reference} \\
\specialrule{.15em}{.2em}{.2em}
Idle 										&	variable	&		\\
\hline
Microwave 									& 		& 		\\
\hspace{0.8cm}Bare gate 					& 	80 $\mu$s 	& 	Sec.~\ref{sec:uwaves}						\\
\hspace{0.8cm}BB1 gate						& 720 $\mu$s		&					 		\\
\hline
Global Raman Beam gate						&		&		\\
\hspace{0.8cm}Bare gate						& 10 $\mu$s	 	& Sec.~\ref{sec:uwaves}, Fig.~\ref{fig:rabi} 		\\
\hspace{0.8cm}BB1 gate						& 90 $\mu$s	 	& 		\\
\hline
Individual Coprop. Raman Beam gate			&	 	& 		\\
\hspace{0.8cm}Bare gate	 					&   5 $\mu$s	& Sec.~\ref{sec:uwaves}, Fig.~\ref{fig:rabi}   	\\
\hspace{0.8cm}BB1 gate			 			& 45 $\mu$s		&   	\\
\hline
Counterprop. Raman Beam gates				&	 	& 		\\
\hspace{0.8cm}Bare gate						& 2 $\mu$s		&Sec.~\ref{sec:uwaves}, Fig.~\ref{fig:rabi} 		\\
\hspace{0.8cm}BB1 gate		 				& 18 $\mu$s		&		\\
\hline
Z gate 										& 10 ns		& Sec.~\ref{sec:framerot}		\\
\hline
Two-Qubit MS gate 							&		&		\\
\hspace{0.8cm}Bare gate 					&	100-200 $\mu$s	&		\\
\hspace{0.8cm}Walsh gate	 				& 150-300 $\mu$s		&		\\
 \end{tabular}
% \end{ruledtabular}
 \label{tab:Times}
 \end{table*}


\clearpage

\begin{thebibliography}{00}
\bibitem{Arute2019} F. Arute~\emph{et al.}, ``Quantum supremacy using a programmable superconducting processor,'' \emph{Nature}, vol. 574, pp. 505-510, 2019, DOI https://doi.org/10.1038/s41586-019-1666-5.

\bibitem{Zhong2020} H.-S. Zhong~\emph{et al.}, ``Quantum Computing in the {NISQ} era and beyond,'' \emph{Science}, vol. 370, pp. 1460-1463, Dec. 2019, DOI 10.1126/science.abe8770.

\bibitem{Preskill2018} J. Preskill, ``Quantum computational advantage using photons,'' arXiv:1801.00862[quant-ph], 2018, DOI 10.22331/q-2018-08-06-79.

\bibitem{IBM} IBM Quantum Computing, IBM.
[Online]. Available:
\underline{https://www.ibm.com/quantum-computing/}. Accessed on: March 15, 2021.

\bibitem{Rigetti} Rigetti: Think Quantum, Rigetti Computing.
[Online]. Available:
\underline{https://www.rigetti.com/}. Accessed on: March 15, 2021.

\bibitem{IonQ} IonQ: The Future is Quantum, IonQ.
[Online]. Available:
\underline{https://ionq.com/}. Accessed on: March 15, 2021.

\bibitem{Fisk1997} P. H. Fisk, M. J. Sellars, M. A. Lawn, and G. Coles, ``Accurate measurement of the 12.6 {GHz} clock transition in trapped (171)Yb(+) ions,'' \emph{IEEE Trans Ultrason Ferroelectr Freq Control}, vol. 44(2), pp. 344-354, 1997, DOI 10.1109/58.585119.

\bibitem{Ballance2016} C. J. Ballance, T. P. Harty, N. M. Linke, M. A. Sepiol, and D. M. Lucas, ``High-Fidelity Quantum Logic Gates Using Trapped-Ion Hyperfine Qubits,'' \emph{Phys. Rev. Lett.}, vol. 117, pp. 060504, 2016, DOI 10.1103/PhysRevLett.117.060504.

\bibitem{Noek2013}
R.~Noek, G.~Vrijsen, D.~Gaultney, E.~Mount, T.~Kim, P.~Maunz, and J.~Kim,
  ``High speed, high fidelity detection of an atomic hyperfine qubit,''
  \emph{Opt. Lett.}, vol.~38, no.~22, pp. 4735--4738, Nov 2013. [Online].
  Available: \url{http://ol.osa.org/abstract.cfm?URI=ol-38-22-4735}


\bibitem{Landsman2019}
K.~A. Landsman, Y.~Wu, P.~H. Leung, D.~Zhu, N.~M. Linke, K.~R. Brown, L.~Duan,
  and C.~Monroe, ``Two-qubit entangling gates within arbitrarily long chains of
  trapped ions,'' \emph{Phys. Rev. A}, vol. 100, p. 022332, Aug 2019. [Online].
  Available: \url{https://link.aps.org/doi/10.1103/PhysRevA.100.022332}

\bibitem{Maunz2016}
P.~Maunz, ``High optical access trap 2.0,'' Sandia National Laboratories,
  Albuquerque, NM, Tech. Rep. SAND2016-0796R, 2016.

\bibitem{Islam2014}
R.~Islam, W.~C. Campbell, T.~Choi, S.~M. Clark, C.~W.~S. Conover, S.~Debnath,
  E.~E. Edwards, B.~Fields, D.~Hayes, D.~Hucul, I.~V. Inlek, K.~G. Johnson,
  S.~Korenblit, A.~Lee, K.~W. Lee, T.~A. Manning, D.~N. Matsukevich,
  J.~Mizrahi, Q.~Quraishi, C.~Senko, J.~Smith, and C.~Monroe, ``Beat note
  stabilization of mode-locked lasers for quantum information processing,''
  \emph{Opt. Lett.}, vol.~39, no.~11, pp. 3238--3241, Jun 2014. [Online].
  Available: \url{http://ol.osa.org/abstract.cfm?URI=ol-39-11-3238}

\bibitem{Landahl2020}
A.~J. Landahl, D.~S. Lobser, B.~C.~A. Morrison, K.~M. Rudinger, A.~E. Russo,
  J.~W. {Van Der Wall}, and P.~Maunz, ``Jaqal, the quantum assembly language
  for {QSCOUT},'' arXiv:2003.09382v1 [quant-ph] 2020.


\bibitem{Morrison2020}
B.~C.~A. Morrison, A.~J. Landahl, D.~S. Lobser, K.~M. Rudinger, A.~E. Russo,
  J.~W. {Van Der Wall}, and P.~Maunz, ``Just another quantum assembly language
  ({Jaqal}),'' arXiv:2008.08042v1 [quant-ph] 2020.

\bibitem{Pagano2018}
G.~Pagano, P.~W. Hess, H.~B. Kaplan, W.~L. Tan, P.~Richerme, P.~Becker,
  A.~Kyprianidis, J.~Zhang, E.~Birckelbaw, M.~R. Hernandez, Y.~Wu, and
  C.~Monroe, ``Cryogenic trapped-ion system for large scale quantum
  simulation,'' \emph{Quantum Science and Technology}, vol.~4, no.~1, p.
  014004, oct 2018. [Online]. Available:
  \url{https://doi.org/10.1088%2F2058-9565%2Faae0fe}

\bibitem{House2008}
M.~G. House, ``Analytic model for electrostatic fields in surface-electrode ion
  traps,'' \emph{Phys. Rev. A}, vol.~78, p. 033402, Sep 2008. [Online].
  Available: \url{https://link.aps.org/doi/10.1103/PhysRevA.78.033402}

\bibitem{Revelle2020}
M.~C. Revelle, ``Phoenix and peregrine ion traps,'' 2020.

\bibitem{OutgassingNASA}
 
``Outgassing data for selecting spacecraft materials system,'' 2020. [Online].
  Available: \url{https://outgassing.nasa.gov/}
 

\bibitem{Zajec2001}
 
B.~Zajec and V.~Nemanič, ``Hydrogen bulk states in stainless-steel related to
  hydrogen release kinetics and associated redistribution phenomena,''
  \emph{Vacuum}, vol.~61, no.~2, pp. 447 -- 452, 2001, proceedings of the 8th
  joint Vaccum Conference of Croatia, Austria, Slovenia and Hungary. [Online].
  Available:
  \url{http://www.sciencedirect.com/science/article/pii/S0042207X01001427}
 

\bibitem{Sasaki1991}
 
Y.~T. Sasaki, ``A survey of vacuum material cleaning procedures: A subcommittee
  report of the american vacuum society recommended practices committee,''
  \emph{Journal of Vacuum Science \& Technology A}, vol.~9, no.~3, pp.
  2025--2035, 1991. [Online]. Available: \url{https://doi.org/10.1116/1.577449}
 

\bibitem{Chen1967}
G.~{Chen}, ``On the physics of purple-plague formation, and the observation of
  purple plague in ultrasonically-joined gold-aluminum bonds,'' \emph{IEEE
  Transactions on Parts, Materials and Packaging}, vol.~3, no.~4, pp. 149--153,
  1967.

\bibitem{Selikson1964}
B.~{Selikson} and T.~A. {Longo}, ``A study of purple plague and its role in
  integrated circuits,'' \emph{Proceedings of the IEEE}, vol.~52, no.~12, pp.
  1638--1641, 1964.

\bibitem{Siverns2012}
 
J.~D. Siverns, L.~R. Simkins, S.~Weidt, and W.~K. Hensinger, ``On the
  application of radio frequency voltages to ion traps via helical
  resonators,'' \emph{Applied Physics B}, vol. 107, no.~4, pp. 921--934, Jun
  2012. [Online]. Available: \url{https://doi.org/10.1007/s00340-011-4837-0}
 

\bibitem{Aikyo2020}
 
Y.~Aikyo, G.~Vrijsen, T.~W. Noel, A.~Kato, M.~K. Ivory, and J.~Kim, ``Vacuum
  characterization of a compact room-temperature trapped ion system,''
  \emph{Applied Physics Letters}, vol. 117, no.~23, p. 234002, 2020. [Online].
  Available: \url{https://doi.org/10.1063/5.0029236}
 

\bibitem{Olmschenk2007}
S.~Olmschenk, K.~C. Younge, D.~L. Moehring, D.~N. Matsukevich, P.~Maunz, and
  C.~Monroe, ``Manipulation and detection of a trapped {Y}b$^+$ hyperfine
  qubit,'' \emph{Phys. Rev. A}, vol.~76, p. 052314, 2007.

\bibitem{Balzer2006}
C.~Balzer, A.~Braun, T.~Hannemann, C.~Paape, M.~Ettler, W.~Neuhauser, and
  C.~Wunderlich, ``Electrodynamically trapped {Y}b$^+$ ions for quantum
  information processing,'' \emph{Phys. Rev. A}, vol.~73, p. 041407(R), 2006.

\bibitem{Bell1991}
A.~S. Bell, P.~Gill, H.~A. Klein, and A.~P. Levick, ``Laser cooling of trapped
  ytterbium ions using a four-level optical-excitation scheme,'' \emph{Phys.
  Rev. A}, vol.~44, no.~1, p. R20, 1991.

\bibitem{Roberts1997}
M.~Roberts, P.~Taylor, G.~P. Barwood, P.~Gill, H.~A. Klein, and W.~R.~C.
  Rowley, ``Observation of an electric octupole transition in a single ion,''
  \emph{Phys. Rev. Lett.}, vol.~78, no.~10, p. 1876, 1997.

\bibitem{Gerginov2011}
 
V.~Gerginov, ``{Time and frequency metrology at PTB: recent results},'' in
  \emph{Time and Frequency Metrology III}, T.~Ido and T.~R. Schibli, Eds., vol.
  8132, International Society for Optics and Photonics.\hskip 1em plus 0.5em
  minus 0.4em\relax SPIE, 2011, pp. 84 -- 89. [Online]. Available:
  \url{https://doi.org/10.1117/12.892675}
 

\bibitem{Ransford2019}
A.~Ransford, personal communication, 2019.

\bibitem{Larson1986}
D.~J. Larson, J.~C. Bergquist, J.~J. Bollinger, W.~M. Itano, and D.~J.
  Wineland, ``Sympathetic cooling of trapped ions: A laser-cooled two-species
  nonneutral ion plasma,'' \emph{Phys. Rev. Lett.}, vol.~57, p.~70, 1986.

\bibitem{Drever1983}
R.~W.~P. Drever, J.~L. Hall, F.~V. Kowalski, J.~Hough, G.~M. Ford, A.~J.
  Munley, and H.~Ward, ``Laser phase and frequency stabilization using an
  optical resonator,'' \emph{Applied Physics B}, vol.~31, pp. 97--105, 1983.

\bibitem{Black2001}
E.~D. Black, ``An introduction to {Pound-Drever-Hall} laser frequency
  stabilization,'' \emph{American Journal of Physics}, vol.~69, p.~79, 2001.

\bibitem{Leibrandt2015}
D.~R. Leibrandt and J.~Heidecker, ``An open source digital servo for atomic,
  molecular, and optical physics experiments,'' \emph{Review of Scientific
  Instruments}, vol.~86, p. 123115, 2015.

\bibitem{Cetina2020}
M.~Cetina, L.~N. Egan, C.~A. Noel, M.~L. Goldman, A.~R. Risinger, D.~Zhu,
  D.~Biswas, and C.~Monroe, ``Quantum gates on individually-addressed atomic
  qubits subject to noisy transverse motion,'' 2020.

\bibitem{Revelle2019}
M.~Revelle, C.~W. Hogle, B.~Ruzic, P.~L.~W. Maunz, K.~Young, and D.~Lobser,
  ``Demonstration of sideband cooling on the 171yb+ quadrupole transition,''
  Sandia National Laboratories, Albuquerque, NM, Tech. Rep.
  SAND2019-0668C671715, 2019.

\bibitem{Schunemann1999}
U.~Sch{\"u}nemann, H.~Engler, R.~Grimm, M.~Weidem{\"u}ller, and
  M.~Zielonkowski, ``Simple scheme for tunable frequency offset locking of two
  lasers,'' \emph{Review of Scientific Instruments}, vol.~70, p. 242, 1999.

\bibitem{Qubig}
 
Q.~GmbH. (2021, Jan.) Pm -yb+ -qubig gmbh. [Online]. Available:
  \url{https://www.qubig.com/products/electro-optic-modulators-230/phase-modulators/ytterbium-ion.html}
 

\bibitem{Wang2017}
Y.~Wang, M.~Um, J.~Zhang, S.~An, M.~Lyu, J.-N. Zhang, L.-M. Duan, D.~Yum, and
  K.~Kim, ``Single-qubit quantum memory exceeding ten-minute coherence time,''
  \emph{Nature Photonics}, vol.~11, pp. 646--650, 2017.

\bibitem{Ospelkaus2008}
C.~Ospelkaus, C.~E. Langer, J.~M. Amini, K.~R. Brown, D.~Leibfried, and D.~J.
  Wineland, ``Trapped-ion quantum logic gates based on oscillating magnetic
  fields,'' \emph{Phys. Rev. Lett.}, vol. 101, p. 090502, 2008.

\bibitem{AudeCraik2014}
D.~P.~L. Aude~Craik, N.~M. Linke, T.~P. Harty, C.~J. Ballance, D.~M. Lucas,
  A.~M. Steane, and D.~T.~C. Allcock, ``Microwave control electrodes for
  scalable, parallel, single-qubit operations in a surface-electrode ion
  trap,'' \emph{App. Phys. B}, vol. 114, pp. 3--10, 2014.

\bibitem{Molmer1999}
K.~M\o{}lmer and A.~S\o{}rensen, ``Multipartite entanglement of hot trapped
  ions,'' \emph{Phys. Rev. Lett.}, vol.~82, pp. 1835--1838, 1999.

\bibitem{Wineland1998}
D.~J. Wineland, C.~Monroe, W.~M. Itano, D.~Leibfried, B.~E. King, and D.~M.
  Meekhof, ``Experimental issues in coherent quantum-state manipulation of
  trapped atomic ions,'' \emph{J. Res. Nat. Inst. Stand. Tech.}, vol. 103, p.
  259, 1998.

\bibitem{Leibfried2003}
D.~Leibfried, R.~Blatt, C.~Monroe, and D.~Wineland, ``Quantum dynamics of
  single trapped ions,'' \emph{Rev. Mod. Phys.}, vol.~75, pp. 281--324, 2003.

\bibitem{Coherent}
 
C.~Inc. (2020, Dec.) Paladin compact 355. [Online]. Available:
  \url{https://www.coherent.com/lasers/laser/paladin-compact-355}
 

\bibitem{Harris}
 
L.~Harris. (2020, Dec.) Acousto optic solutions. [Online]. Available:
  \url{https://www.l3harris.com/all-capabilities/acousto-optic-solutions}
 

\bibitem{PhotonGear}
 
P.~G. Inc. (2020, Dec.) Custom optics, assemblies, and interferometry: Photon
  gear. [Online]. Available: \url{https://photongear.com/}
 

\bibitem{Acton2006}
M.~Acton, K.-A. Brickman, P.~Haljan, P.~Lee, L.~Deslauriers, and C.~Monroe,
  ``Near-perfect simultaneous measurement of a qubit register,'' \emph{Quantum
  Inf. Comput.}, vol.~6, p. 465, 2006.

\bibitem{Bowler2015}
R.~Bowler, ``Coherent ion transport in a multi-electrode trap array,'' Ph.D.
  dissertation, University of Colorado, 2015.

\bibitem{Mount2016}
E.~Mount, D.~Gaultney, G.~Vrijsen, M.~Adams, S.-Y. Baek, K.~Hudek, L.~Isabella,
  S.~Crain, A.~{van Rynbach}, P.~Maunz, and J.~Kim, ``Scalable digital hardware
  for a trapped ion quantum computer,'' \emph{Quantum Inofrmation Processing},
  vol.~15, pp. 5281--5298, 2016.

\bibitem{InlekPhaseInsensitive}
 
I.~V. Inlek, G.~Vittorini, D.~Hucul, C.~Crocker, and C.~Monroe, ``Quantum gates
  with phase stability over space and time,'' \emph{Phys. Rev. A}, vol.~90, p.
  042316, Oct 2014. [Online]. Available:
  \url{https://link.aps.org/doi/10.1103/PhysRevA.90.042316}
 

\bibitem{nielsen2020gate}
E.~Nielsen, J.~K. Gamble, K.~Rudinger, T.~Scholten, K.~Young, and
  R.~Blume-Kohout, ``Gate set tomography,'' 2020.

\bibitem{MicroSemi1}
 
Microchip, (2021, March) Csiii model 4310b. [Online]. Available:
  \url{https://www.microsemi.com/product-directory/cesium-frequency-references/4114-csiii-model-4310b}
 

\bibitem{MicroSemi2}
 
------. (2021, March) Quartz ultra clean-up oscillator. [Online]. Available:
  \url{https://www.microsemi.com/existing-parts/parts/138085}
 

\bibitem{StanfordMagnets}
 
S.~Magnets. (2021, March) How does temperature affect samarium cobalt magnets.
  [Online]. Available:
  \url{https://www.stanfordmagnets.com/how-does-temperature-affect-samarium-cobalt-magnets.html}
 

\bibitem{Hahn1950}
E.~L. Hahn, ``Spin echos,'' \emph{Phys. Rev.}, vol.~80, p. 580, 1950.

\bibitem{Blume2017}
R.~Blume-Kohout, J.~K. Gamble, E.~Nielsen, K.~Rudinger, J.~Mizrahi, K.~Fortier,
  and P.~Maunz, ``Demonstration of qubit operations below a rigorous fault
  tolerance threshold with gate set tomography,'' \emph{Nat. Comm.}, vol.~8, p.
  14485, 2017.

\bibitem{Shappert2013}
C.~M. Shappert, J.~T. Merrill, K.~R. Brown, J.~M. Amini, C.~Volin, S.~C. Doret,
  H.~Hayden, C.-S. Pail, K.~R. Brown, and A.~W. Harter, ``Spatially uniform
  single-qubit gate operations with near-field microwaves and composite pulse
  compensation,'' \emph{New J. Phys.}, vol.~15, no. 083053, 2013.

\bibitem{Wimperis1994}
S.~{Wimperis}, ``{Broadband, narrowband, and passband composite pulses for use
  in Advanced NMR Experiments},'' \emph{Journal of Magnetic Resonance}, vol.
  109, no.~2, pp. 221--231, Jan. 1994.

\bibitem{Sych2009}
 
D.~Sych and G.~Leuchs, ``A complete basis of generalized Bell states,''
  \emph{New Journal of Physics}, vol.~11, no.~1, p. 013006, jan 2009. [Online].
  Available: \url{https://doi.org/10.1088/1367-2630/11/1/013006}
 

\end{thebibliography}
\end{document}